\documentclass[a4paper,longauth]{aa}  

\usepackage{subfig,graphicx}
\usepackage{threeparttable,booktabs}
\usepackage{txfonts}
\usepackage{xcolor}
\usepackage[export]{adjustbox}
\definecolor{CornflowerBlue}{rgb}{0.39,0.58,0.93}
\definecolor{forestgreen}{rgb}{0.13, 0.55, 0.13}
\usepackage{hyperref}
\usepackage{natbib}
\hypersetup{colorlinks=true,allcolors=CornflowerBlue}

\begin{document}

   \title{The eROSITA Final Equatorial-Depth Survey (eFEDS):}

   \subtitle{A complete census of X-ray properties of Subaru Hyper Suprime-Cam weak lensing shear-selected clusters in the eFEDS footprint}

   \author{M.~E.~Ramos-Ceja\inst{1} \thanks{e-mail: \href{mailto:mramos@mpe.mpg.de}{\tt mramos@mpe.mpg.de}}
          \and M.~Oguri\inst{2,3,4}
          \and S.~Miyazaki\inst{5,6}
          \and V.~Ghirardini\inst{1}
          \and I.~Chiu\inst{7,8,9}
          \and N.~Okabe\inst{10,11,12}
          \and A.~Liu\inst{1}
          \and T.~Schrabback\inst{13}
          \and D.~Akino\inst{10}
          \and Y.~E.~Bahar\inst{1}
          \and E.~Bulbul\inst{1}
          \and N.~Clerc\inst{14}
          \and J.~Comparat\inst{1}
          \and S.~Grandis\inst{15}
          \and M.~Klein\inst{15}
          \and Y.-T.~Lin\inst{9}
          \and A.~Merloni\inst{1}
          \and I.~Mitsuishi\inst{16}
          \and H.~Miyatake\inst{4,16,17,18}
          \and S.~More\inst{4,19}
          \and K.~Nandra\inst{1}
          \and A.~J.~Nishizawa\inst{18}
          \and N.~Ota\inst{13,20}
          \and F.~Pacaud\inst{13}
          \and T.~H.~Reiprich\inst{13}
          \and J.~S.~Sanders\inst{1}
          }

   \institute{Max Planck Institut f{\"{u}}r extraterrestrische Physik, Giessenbachstra{\ss}e 1, D-85748 Garching, Germany
         \and Research Center for the Early Universe, University of Tokyo, Tokyo 113-0033, Japan
         \and Department of Physics, University of Tokyo, Tokyo 113-0033, Japan
         \and Kavli Institute for the Physics and Mathematics of the Universe (Kavli IPMU, WPI), University of Tokyo, Chiba 277-8582, Japan
         \and National Astronomical Observatory of Japan, 2-21-1 Osawa, Mitaka, Tokyo 181-8588, Japan
         \and SOKENDAI (The Graduate University for Advanced Studies), Mitaka, Tokyo, 181-8588, Japan
         \and Tsung-Dao Lee Institute, and Key Laboratory for Particle Physics, Astrophysics and Cosmology, Ministry of Education, Shanghai Jiao Tong University, Shanghai 200240, China
         \and Department of Astronomy, School of Physics and Astronomy, and Shanghai Key Laboratory for Particle Physics and Cosmology, Shanghai Jiao Tong University, Shanghai 200240, China
         \and Academia Sinica Institute of Astronomy and Astrophysics (ASIAA), No. 1, Section 4, Roosevelt Road, Taipei 10617, Taiwan
         \and Physics Program, Graduate School of Advanced Science and Engineering, Hiroshima University, 1-3-1 Kagamiyama, Higashi-Hiroshima, Hiroshima 739-8526, Japan
         \and Hiroshima Astrophysical Science Center, Hiroshima University, 1-3-1 Kagamiyama, Higashi-Hiroshima, Hiroshima 739-8526, Japan
         \and Core Research for Energetic Universe, Hiroshima University, 1-3-1, Kagamiyama, Higashi-Hiroshima, Hiroshima 739-8526, Japan
         \and Argelander-Institut f{\"{u}}r Astronomie (AIfA), Universit{\"{a}}t Bonn, Auf dem H{\"{u}}gel 71, 53121 Bonn, Germany
         \and IRAP, Université de Toulouse, CNRS, UPS, CNES, Toulouse, France
         \and Faculty of Physics, Ludwig-Maximilians-Universit{\"a}t, Scheinerstr. 1, 81679, Munich, Germany
         \and Division of Physics and Astrophysical Science, Graduate School of Science, Nagoya University, Nagoya, Aichi 464-8602, Japan
         \and Kobayashi-Maskawa Institute for the Origin of Particles and theUniverse (KMI), Nagoya University, Nagoya, 464-8602, Japan
         \and Institute for Advanced Research, Nagoya University, Nagoya, 464-8601, Japan
         \and The Inter University Centre for Astronomy and Astrophysics, Ganeshkhind, Pune 411007, India
         \and Department of Physics, Nara Women's University, Kitauoyanishi-machi, Nara, 630-8506, Japan
             \\
             }

   \date{A\&A accepted 2021}

\titlerunning{X-ray properties of shear-selected clusters in the eFEDS footprint}

 
  \abstract
   {The eFEDS survey is a proof-of-concept mini-survey designed to demonstrate the survey science capabilities of SRG/eROSITA. It covers an area of $140$~deg$^2$ where $\sim 540$ galaxy clusters have been detected out to a redshift of $1.3$. The eFEDS field is partly embedded in the Hyper Suprime-Cam Subaru Strategic Program (HSC-SSP) S19A data release, which covers $\sim 510$~deg$^2$, containing approximately $36$ million galaxies. This galaxy catalogue has been used to construct a sample of $\sim 180$ shear-selected galaxy clusters. The common area to both surveys covers about $90$~deg$^2$, making it an ideal region to study galaxy clusters selected in different ways. }
   {The aim of this work is to investigate the effects of selection methods in the galaxy cluster detection by comparing the X-ray selected, eFEDS, and the shear-selected, HSC-SSP S19A, galaxy cluster samples. There are $25$ shear-selected clusters in the eFEDS fooprint.}
   {The relation between X-ray bolometric luminosity and weak-lensing mass is investigated ($L_{\rm bol}-M$ relation), comparing this relation derived from a shear-selected cluster sample to the relation obtained from an X-ray selected sample. Moreover, the dynamical state of the shear-selected clusters is investigated and compared to the X-ray selected sample using X-ray morphological parameters and galaxy distribution.}
   {The normalisation of the $L_{\rm bol}-M$ relation of the X-ray selected and shear-selected samples is consistent within $1\sigma$. Moreover, the dynamical state and merger fraction of the shear-selected clusters is not different from the X-ray selected ones. Four shear-selected clusters are undetected in X-rays. A close inspection reveals that one is the result of projection effects, while the other three have an X-ray flux below the ultimate eROSITA detection limit. Finally, $43\%$ of the shear-selected clusters lie in superclusters.}
   {Our results indicate that the scaling relation between X-ray bolometric luminosity and true cluster mass of the shear-selected cluster sample is consistent with the eFEDS sample. There is no significant population of X-ray underluminous clusters, indicating that X-ray selected cluster samples are complete and can be used as an accurate cosmological probe.}

   \keywords{Galaxies: clusters: general -- 
                 X-rays: galaxies: clusters --
                Shear-selection: clusters
               }

   \maketitle
%

\section{Introduction}
\label{sect:intro}

There are a number of completed and on-going galaxy cluster surveys across the electromagnetic spectrum ranging from sub-millimetre wavelengths, infrared and optical bands, to X-rays. While clusters are rare, constructions of large samples of clusters have been made possible thanks to recent technological advances of sensitive detectors and telescopes that allow us to collect huge amounts of data in a reasonable amount of time.
Clusters selected in different wavelengths are complementary with each other because they are sensitive to distinct cluster components. Broadly speaking, $\sim 80-85\%$ of the cluster mass is in the form of dark matter and the rest, the baryonic component, is divided into $\sim 12\%$ as a hot plasma called the intracluster medium (ICM), and $\sim 3-8\%$ cluster member galaxies and a relativistic population of electrons that are part of the ICM. 

In the optical and the near-infrared, galaxy clusters are identified as spatially localised and projected overdensities of galaxies \citep[e.g.][]{Wen2012, Rykoff2014,Oguri2014}. In X-rays and sub-millimetre wavelengths, galaxy clusters are identified via the ICM, whose angular extent is comparable to that of the galaxy member distribution \citep[e.g.][]{Boehringer2001,Pacaud2006,Reichardt2013,PlanckCollaboration2016}. While X-rays detect the integrated emission from the optically thin bremsstrahlung emission from the ICM, the sub-millimetre wavelengths detect the apparent decrement in the brightness of the cosmic microwave background (CMB) caused by the inverse Compton scattering of the CMB photons by energetic electrons in the ICM, that is the so-called Sunyaev-Zel'dovich (SZ) effect \citep[][]{Sunyaev1970}. Furthermore, it is possible to identify galaxy clusters directly using weak gravitational lensing, that is the distortion in the shapes of background galaxies caused by the total projected mass along the line of sight, presumably dominated by the cluster \citep[e.g.][]{Miyazaki2002, Hetterscheidt2005,Wittman2006,Schirmer2007,Shan2012,Liu2015,Miyazaki2018, Hamana2020}. Finally, in radio wavelengths spatially localised and projected overdensities of wide- and narrow-angle tail radio galaxies and/or spatially extended, low-frequency radio emission, the so-called radio relics and radio halos, can also be used to identify galaxy clusters \citep[e.g.][]{Kale2015}.
 
Two important aspects of the multi-wavelength study of galaxy cluster samples are worth emphasising. First, regardless of the galaxy cluster search technique, there is a selection bias, that is the distinct wavelengths tend to preferentially select a particular type of clusters with respect to the full population \citep[see][for a review on cluster survey biases]{Giodini2013}. Secondly, there are galaxy cluster correlations, the so-called scaling relations, that describe the relationship between different galaxy cluster properties \citep[e.g.][]{Kaiser1986,Giodini2013}. These relations are not exclusive of a given waveband and are very important because they relate easily observable quantities to other cluster properties that are difficult to determine by direct observations. Although the different cluster surveys have their strengths and challenges, the ultimate goal is to combine them and use their synergies to obtain a sufficiently complete view of the cluster populations and their place in the hierarchy of cosmic large-scale structure formation. Only if we succeed with this challenge, large galaxy cluster samples can be reliably used for competitive, precision cosmological constraints.

There are three main ways to have a better understanding of the selection bias in cluster samples and to understand their multi-wavelength properties: {\it i}) by comparing results obtained from different cluster samples \citep[e.g.][]{Gilbank2004, Rozo2014}, {\it ii}) by doing an exhaustive follow-up of a given cluster sample \citep[e.g.][]{Rossetti2017, Andrade-Santos2017}, and {\it iii}) by comparing cluster samples obtained over a common sky area \citep[e.g.][]{Sadibekova2014, Donahue2002, Willis2018, Willis2021}. This work focuses on the third method and compares an X-ray detected galaxy cluster sample and a shear-selected sample obtained over the same sky region.

Most of studies in this line of investigation have focused on determining the cluster masses from weak-lensing observations for X-ray selected galaxy cluster samples \citep[e.g.][]{Okabe2010, Hoekstra2012, Mahdavi2013, Israel2014}. Very few studies have performed an X-ray follow-up of shear-selected galaxy clusters \citep[e.g.][]{Giles2015, Deshpande2017}. \cite{Miyazaki2018} studied the X-ray properties of shear-selected clusters using archival X-ray data, which is shallow and incomplete. In general, all these studies lack having a concise X-ray or weak-lensing selected cluster sample, either because they are noisy, very small or are not completely followed up. This was a consequence of the lack of large area and deep optical data that is necessary to construct a large sample of shear-selected clusters and the difficulty of a uniform X-ray analysis. These issues are now alleviated thanks to the multi-wavelength coverage of the eROSITA \citep{Predehl2021} pilot survey by deep wide-field optical imaging enabled by Subaru Hyper Suprime-Cam \citep{Miyazaki_HSC}.
Using these data sets, we can for the first time conduct a complete census of X-ray properties for a large sample of shear-selected clusters. 

The detailed exploration of weak lensing shear-selected clusters is particularly important for advancing our understanding of the observational bias in cluster samples, because their selection can be well modelled in analyses of cosmological simulations. The selection bias of shear-selected clusters depends on the density profile of clusters as well as the density fluctuations along the line-of-sight, both of which are reasonably well understood from $N$-body simulations \citep{Hamana2004,Hamana2012,Chen2020}. In contrast, selections of clusters in other methods rely on baryonic properties, and hence their selection biases are subject to detailed baryon physics that is not yet fully understood \citep[e.g.][]{Eckert2011}. Thus it is expected that the study of the baryonic property of shear-selected clusters as explored in this paper will help to achieve a better understanding of cluster selection biases.

The structure of the paper is as follows: In section~\ref{sect:galclusamp}, we describe each cluster sample and perform a matching analysis. In section~\ref{sect:method}, we describe the methods used to extract cluster properties as well as the algorithm used for scaling-relation fitting. We present the results of this analysis in section~\ref{sect:results}, discussion in section~\ref{sect:discussion}, and draw conclusions in section~\ref{sect:conclusions}. Throughout this paper, we assume a flat $\Lambda$CDM cosmology with $\Omega_\textrm{m}=0.3$ and $H_0=70~\rm km~s^{-1}~Mpc^{-1}$.


\section{Galaxy cluster samples}
\label{sect:galclusamp}

In this section, the X-ray detected and shear-selected cluster samples are described, as well as their weak-lensing mass calibration. Both samples are cross-matched, and sub-samples are defined in the common sky area.


\subsection{X-ray detected clusters: eFEDS}
\label{sect:eFEDS}

The X-ray extended source sample studied in this work is taken from \cite{Brunner2021}, which presents $\sim 540$~extended X-ray sources detected in the eROSITA Final Equatorial-Depth Survey (eFEDS), which was completed during the Performance Verification (PV) phase of eROSITA \citep{Predehl2021}. eFEDS covers an area of $\sim140$~deg$^{2}$ and has an average, vignetted exposure time of $1.3$~ks in the $0.5-2.0$~keV energy band. 

The X-ray source detection was performed using the eROSITA Standard Analysis Software System (eSASS, {\tt eSASSusers\_201009}) in the $0.2-2.3$~keV energy band. The source detection is based on the sliding-cell algorithm followed by a source characterisation method \citep[see][for further details]{Brunner2021}. The extended X-ray source catalogue was built by setting a minimum detection likelihood of 5 and a minimum extent likelihood of 6.

\cite{LiuA2021} used this X-ray extended source catalogue as a basis for the construction of the eFEDS cluster catalogue and determined several X-ray observables of the detected clusters. Based on X-ray simulations \citep{Comparat2020} and optical follow-up, \cite{LiuA2021} and \cite{Klein2021} determined that this sample has $\sim 20\%$ of contamination.

Using data from the Hyper Suprime-Cam Subaru Strategic Program \citep[HSC-SSP,][]{Aihara2018a, Aihara2018b, Aihara2019} and from the DESI Legacy Imaging Surveys \citep[LS,][]{Dey2019}, photometric redshifts and richness were obtained for all cluster candidates using the Multi-Component Matched Filter (MCMF) cluster confirmation tool \citep{Klein2018}. Approximately $82\%$ of the eFEDS extended X-ray sources have been confirmed to be galaxy clusters by an optical follow- up \citep[see][for further details]{Klein2021}. The redshift range of these clusters is $0<z<1.3$, with the peak of the distribution being around $z\sim0.3$. The confirmed galaxy clusters used in this work are selected using the MCMF cleaning parameter $f_{\rm cont}$ to be $f_{\rm cont}<0.2$, yielding a catalogue purity of $96\%$ \citep{Klein2021}. 

 \begin{figure*}[ht]
     \centering
     \includegraphics[trim=30 20 0 50,clip,width=\textwidth]{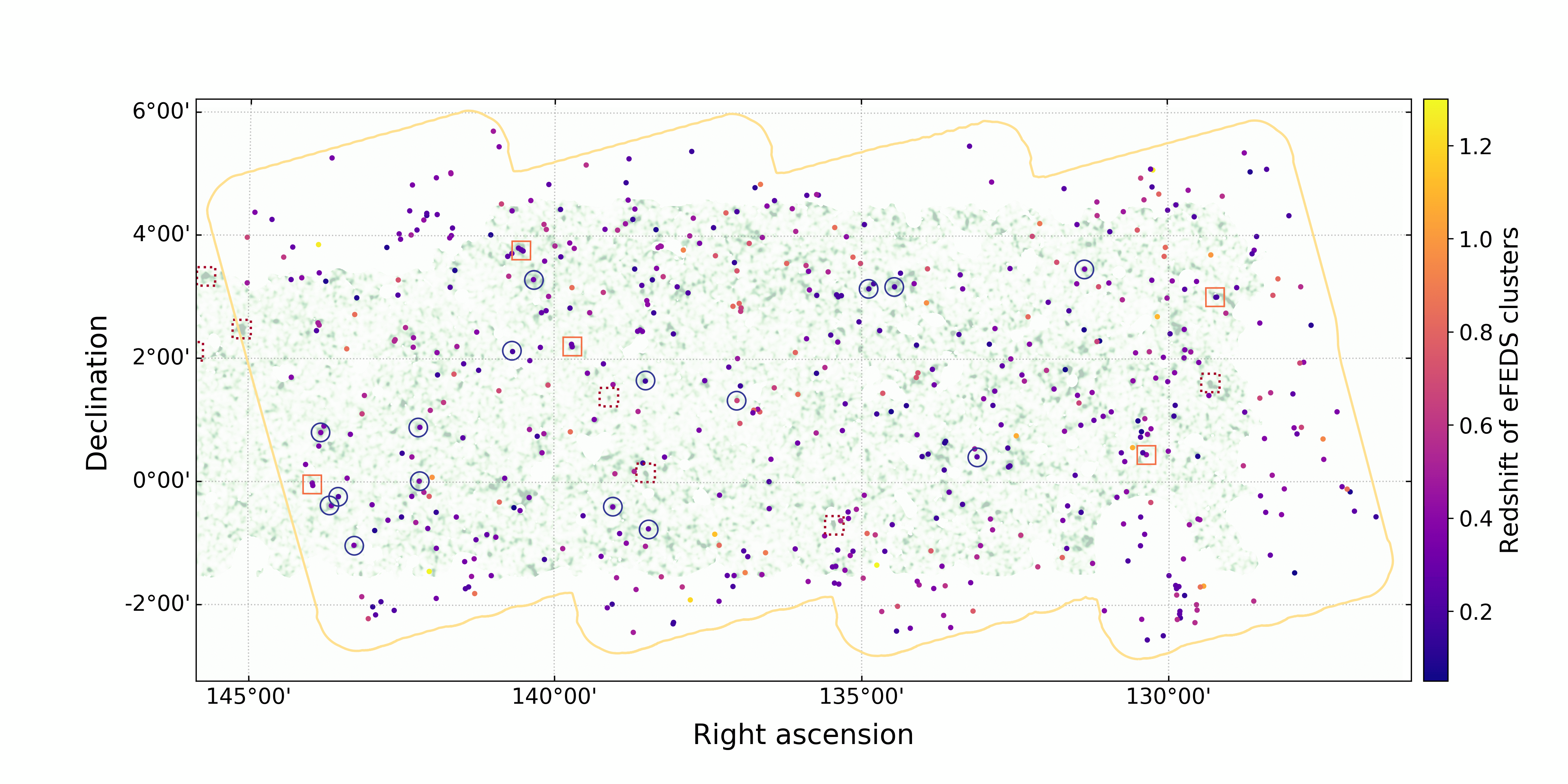}
     \caption{Part of the weak lensing map for the GAMA09H HSC-SSP S19A field (green background). The solid yellow line shows the eFEDS survey footprint. Small filled circles show the location of eFEDS clusters with $z>0.05$, whose colours correspond to their redshift as shown by the colour bar. Open symbols show the position of the shear-selected clusters: blue circles correspond to peaks that have one counterpart in the eFEDS catalogue, orange squares are weak-lensing peaks with two or three eFEDS counterparts, and dotted red squares are peaks with no eFEDS counterparts.}
     \label{fig:hscefedsfootprint}
 \end{figure*}


\subsection{Shear-selected clusters: HSC-SSP S19A}
\label{sect:shearclu}

The shear-selected galaxy cluster sample studied in this paper is taken from \cite{Oguri2021}. Using the HSC-SSP S19A shape catalogue from \cite{Li2021}, \cite{Oguri2021} constructed samples of weak lensing shear-selected clusters with signal-to-noise ratio (S/N) larger than $4.7$ over an area of $\sim 510$~deg$^2$. 

The constructed shear-selected cluster catalogues are the result of two different approaches, which adopt distinct spatial filters. Specifically, \cite{Oguri2021} considered two spatial filters, a truncated Gaussian filter and a truncated isothermal filter. The former is similar to the one used in \cite{Miyazaki2018} and delivers $187$ clusters (the so-called TG15 catalogue), while the latter adopts the functional form proposed in \cite{Schneider1996} and optimises the detection of halos from the mass maps. \cite{Oguri2021} explored the impact of the inner boundary of the truncated isothermal filter when constructing shear-selected cluster catalogues: considering two different values of this parameter \citep[see section~3.3 in][]{Oguri2021}, two shear-selected catalogues of $418$ and $200$ clusters were delivered\footnote{In \cite{Oguri2021} they are referred as TI05 and TI20 catalogues.}. The analysis presented in this work, is focused only on the former shear-selected cluster catalogue because a mass bias correction, derived in \cite{Chen2020}, can be applied to this catalogue (see section~\ref{sect:weaklensmass}). A cross-matching of the other two shear-selected cluster catalogues with eFEDS is discussed in Appendix~\ref{app:A}.

The redshifts of the shear-selected clusters were assigned by cross-matching them with different optically-selected clusters, which include: an optically-selected cluster catalogue obtained from the HSC-SSP using the CAMIRA algorithm \citep{Oguri2014,Oguri2018} as well as several optically-selected clusters from the Sloan Digital Sky Survey \citep[SDSS,][]{York2000} including the redMaPPer catalogue \citep{Rykoff2014} and WHL15 \citep{Wen2012, Wen2015}. In addition, the CODEX cluster catalogue \citep{Finoguenov2020}, which contains galaxy clusters from the ROSAT All-Sky Survey \citep[RASS,][]{Voges1999} with optical confirmations from SDSS, was used. The combination of these catalogues allows covering a redshift range between $0.05<z<1.38$. In short, more than $\sim 97\%$ of the shear-selected clusters have an optical counterpart. The purity of the shear-selected cluster catalogue is estimated to be higher than $95\%$.

\vspace{6pt}
The cluster number in both samples are different due to the distinct selection functions employed in the X-ray and shear detection techniques. The shear-selection requires high cluster masses, that is the reason why the number of the shear-selected clusters is in general smaller.

\subsection{Cluster mass determination}
\label{sect:weaklensmass}

The true masses, that is calibrated weak-lensing masses, in the eFEDS and HSC shear-selected clusters were derived independently, following different approaches. In the following, both methods are briefly described.

\vspace{-10pt}
\paragraph{eFEDS clusters.} \cite{Chiu2021} obtained the weak-lensing mass calibration of the eFEDS clusters using the S19A weak-lensing data from the HSC. They studied $313$ clusters in a redshift range of $0.1<z<1.3$, which are fully covered by HSC data. Using a Bayesian population modelling, a blind analysis for the weak-lensing mass calibration was performed: for each cluster, the observed X-ray count-rate (used as a mass proxy) and the shear-profile were simultaneously modelled using the count-rate-to-true-mass-redshift and weak-lensing mass-to-true-mass-redshift relations. All biases of the different observables were taken into account using simulation-based calibrations. Especially, the weak-lensing mass was extensively calibrated using cosmological hydrodynamical simulations, which take into account, among other things, the cluster member contamination, the miscentring of the X-ray centre and the redshift distribution of the source. It was found that the covered mass range of these eFEDS clusters is $10^{13}<M_{500}~[h^{-1}~{\rm M}_{\odot}]<10^{15}$ with a mass uncertainty of $\sim 30\%$. \cite{Chiu2021} also studied different X-ray observable-to-mass-redshift relations, for example, the bolometric luminosity $-$ mass relation.

\vspace{-10pt}
\paragraph{HSC shear-selected clusters.} \cite{Oguri2021} derived weak-lensing masses following \cite{Miyazaki2018} for all shear-selected clusters with estimated redshift. Briefly, \cite{Oguri2021} derived differential surface density profiles from securely selected background galaxies for each shear-selected cluster. These profiles were fitted with a \cite{Navarro1997} profile, which is parameterised by the mass, $M_{500{\rm c}}$, and concentration, $c_{500{\rm c}}$, parameter for a critical overdensity of $500$. For the shear-selected sample, cluster masses were derived in such a way that the Eddington bias correction for the weak-lensing mass measurements derived in \cite{Chen2020} can be applied. Using large sets of mock cluster samples \cite{Chen2020} found that derived weak-lensing masses for shear-selected clusters are biased high by $\sim 55\%$ with respect to the cluster true mass, on average, due to the up-scatter of weak lensing signals by the shape noise as well as cosmic shear. This mass bias strongly correlates with cluster redshifts, true halo masses, and selection S/N thresholds. Furthermore, an additional scatter due to triaxiality is included in the mass bias correction. \cite{Chen2020} showed that, once this Eddington bias is properly taken into account, the discrepancy of the X-ray luminosity--mass relations for X-ray selected and shear-selected clusters  \citep{Giles2015, Miyazaki2018} is mitigated.

 \begin{figure}[t]
     \centering
     \includegraphics[trim=30 20 40 60,clip,width=\columnwidth]{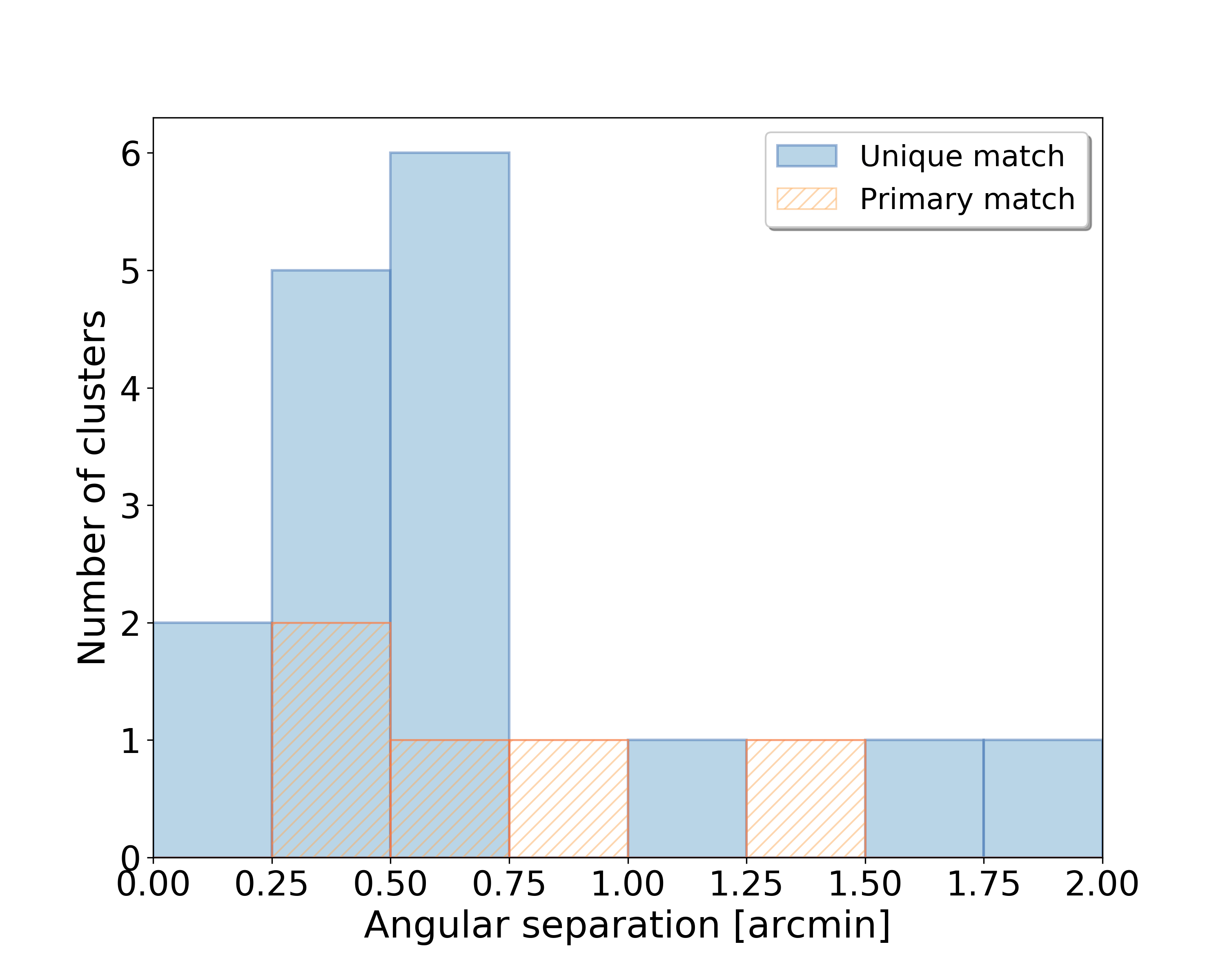}
     \caption{Distribution of the angular separation, in arcmin, between the shear-selected clusters and the closest X-ray detected (eFEDS) clusters. In blue, clusters are shown which have one unique match between the catalogues within $5$~arcmin and with a redshift difference $|\Delta z|<0.1$. The orange histogram displays primary matches for clusters with multiple matches. Primary matches correspond to the most X-ray luminous (and in most cases the closest) eFEDS counterpart.}
     \label{fig:angseparation}
 \end{figure}

The masses of the shear-selected clusters used in this work are bias-corrected. The applied bias is the one calculated for a S/N threshold of $4.7$ and an average surface density of source galaxies of $30$~arcmin$^{-2}$ \citep[see section 2 of][]{Chen2020}. We take into account the cluster redshift dependence of the mass bias and adopt different mass bias values for clusters at different redshifts. Most weak-lensing masses are corrected by $\sim20\%$. Table~\ref{tab:xray_wl_prop} shows the calibrated weak-lensing mass, $M_{500}$, of the shear-selected clusters in the common area between the HSC GAMA09H and eFEDS fields. The masses cover the range $ 8\times10^{13}\lesssim M_{500}~[$M$_\odot]\lesssim 8\times10^{14}$ with a mass uncertainty of $\sim30\%$. This bias-corrected weak-lensing mass is used to determine the radius within which different X-ray observables of the HSC shear-selected clusters are measured (see section~\ref{sect:xray_obs}).

\vspace{10pt}
While true masses are estimated for the eFEDS clusters and HSC shear-selected clusters using a different methodology, we confirm that those of matched clusters of both samples (see section~\ref{sect:catsmatchs}) are on average consistent given their mass uncertainties: $\sim 42\%$ of them have a mass difference of less than $20\%$, $33\%$ have a mass difference between $20-50\%$, and the rest do not exceed a mass difference of $\sim 60\%$. The mean systematic mass difference between these two samples is $\sim 0.1\pm 0.1$~dex. The difference of estimated true masses for individual clusters mainly comes from the distinct assumed centre, the range of radii for the shear profile fitting, and the use of priors.

  \begin{figure}[t]
     \centering
     \includegraphics[trim=0 45 40 80,clip,width=0.9\columnwidth]{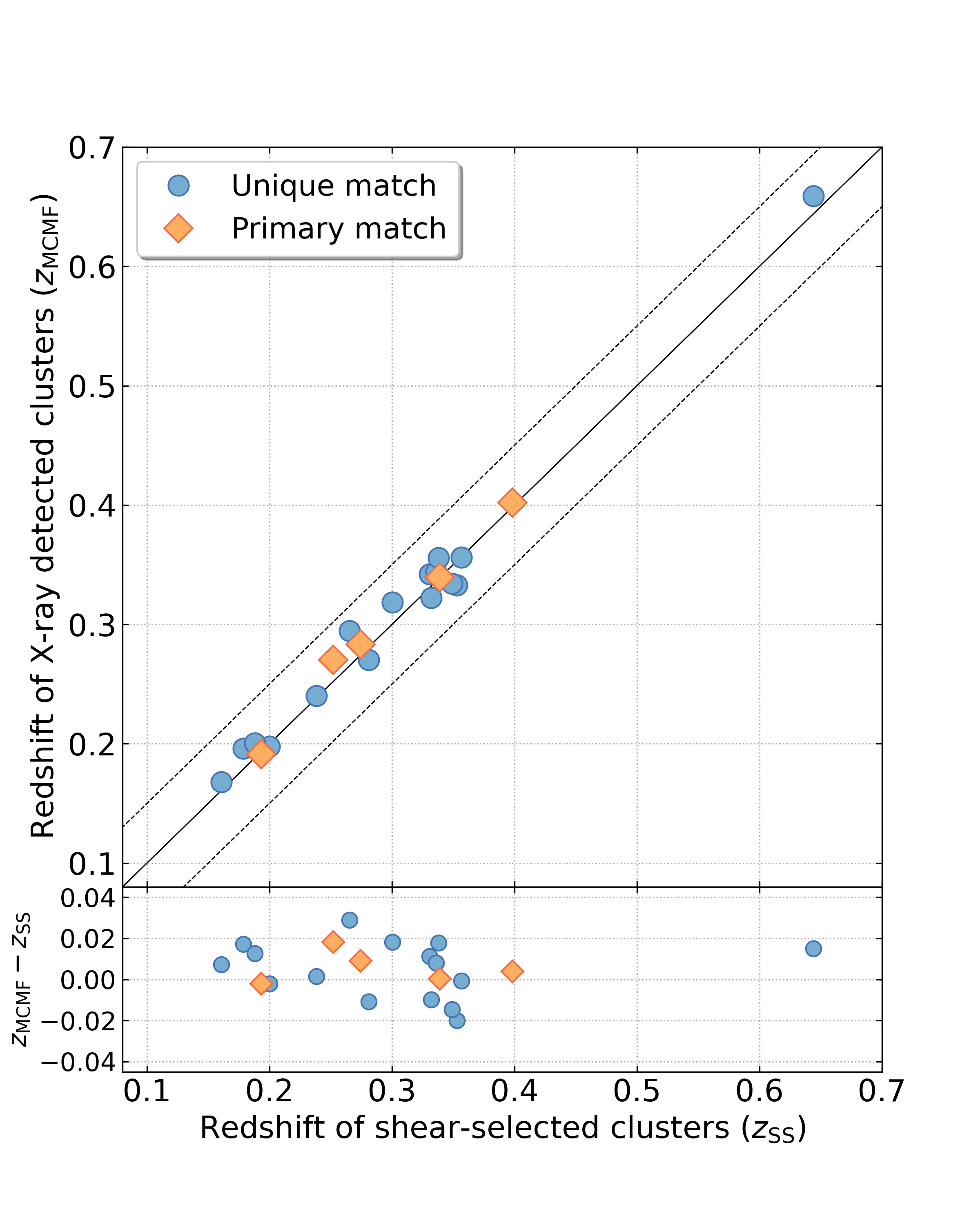}
     \caption{{\it Top}: Comparison between the assigned redshift of the shear-selected and the X-ray detected clusters. In blue, clusters are shown that have one unique match between the catalogues within $5$~arcmin and with a redshift difference $|\Delta z|<0.1$. The orange points display primary matches for clusters with multiple matches. Primary matches correspond to the most X-ray luminous (and in most cases the closest) eFEDS counterpart. The solid black line shows the 1:1 relationship, and the dotted black lines, a $\pm0.05$ offset from this relation. {\it Bottom}: Difference between redshifts.}
     \label{fig:zcomparison}
 \end{figure}


\subsection{Matching between shear-selected clusters and X-ray selected eFEDS clusters}
\label{sect:catsmatchs}

The HSC-SSP S19A shape catalogue consists of six disjoint patches. The patch that overlaps with the eFEDS survey is the so-called GAMA09H. The common sky area between the eFEDS and GAMA09H is $\sim90$~deg$^2$ (see Fig.~\ref{fig:hscefedsfootprint}). There are $313$ eFEDS clusters (with $z>0.05$ and $f_{\rm cont}<0.2$) and $25$ shear-selected clusters lying within the common footprint, that is the area that has been uniformly covered by the HSC survey to enable the extraction of the weak-lensing shear signal. The redshift limit of $z=0.05$ imposed to the eFEDS sample is motivated by the completeness of the weak-lensing peaks, which is small at very low redshift values \citep[see Fig.~9 of][]{Miyazaki2018}, and the incompleteness at such redshifts of the optically-selected cluster samples used for redshift confirmation of the shear-selected clusters. All these shear-selected clusters have an assigned redshift (see section~\ref{sect:shearclu}). Figure~\ref{fig:hscefedsfootprint} displays the location in the sky of the weak-lensing peaks and the eFEDS clusters.

Both catalogues are matched according to a positional and redshift offset in order to assess the number of common clusters to both samples. Galaxy clusters are considered to be matched if, first, they are located within $5$~arcmin and, second, if the redshift difference between the eFEDS and the shear-selected cluster redshifts is $|\Delta z|<0.1$\footnote{Although a redshift difference of $0.1$ seems large, all matched clusters have a redshift difference smaller than 0.04, as shown in Figure~\ref{fig:zcomparison}.}. The shear-selected cluster catalogue has been generated by using a smoothing scale of $1.5$~arcmin, therefore the tolerance radius must be large enough ($5$~arcmin) to identify X-ray counterparts.

The results of the matching procedure show that out of the $25$ shear-selected clusters, only $21$ of them have eFEDS counterparts. Out of these $21$ matches, five of them have multiple eFEDS matches within $5$~arcmin: three shear-selected clusters have two associated eFEDS clusters of similar redshift, one has three eFEDS matches with similar redshift, and one has two eFEDS matches of very different redshift. Clusters with a one to one match will be referred to as unique matches. For the shear-selected clusters with multiple eFEDS counterparts, the most X-ray luminous eFEDS cluster is chosen (only one case is not the closest, in position and redshift); they will be referred to as primary matches.

Figures~\ref{fig:angseparation}~and~\ref{fig:zcomparison} show the distribution of the angular separation and the redshift difference, respectively, between the matched clusters. Both plots show the good accordance between the common clusters: the positional offset between them is less than $2$~arcmin and the redshift difference is less then $0.03$. Using the redshift of the matched shear-selected clusters, the rest-frame transverse physical offset values are between $44$ and $531$~kpc.

The unmatched shear-selected clusters are not the ones with the lowest weak-lensing S/N, but all four have an S/N\ $<5.0$. A visual inspection of their X-ray and optical data reveals that one of the shear-selected clusters without an associated eFEDS cluster is the result of projection effects. For the other three clusters, there is no evident extended X-ray emission. Moreover, there is not any extended X-ray candidate nearby of these clusters, optically confirmed or not. These non-matched clusters are individually discussed in section~\ref{sect:nomatches}.

 \begin{figure}[t]
     \centering
     \includegraphics[trim=10 20 50 60,clip,width=\columnwidth]{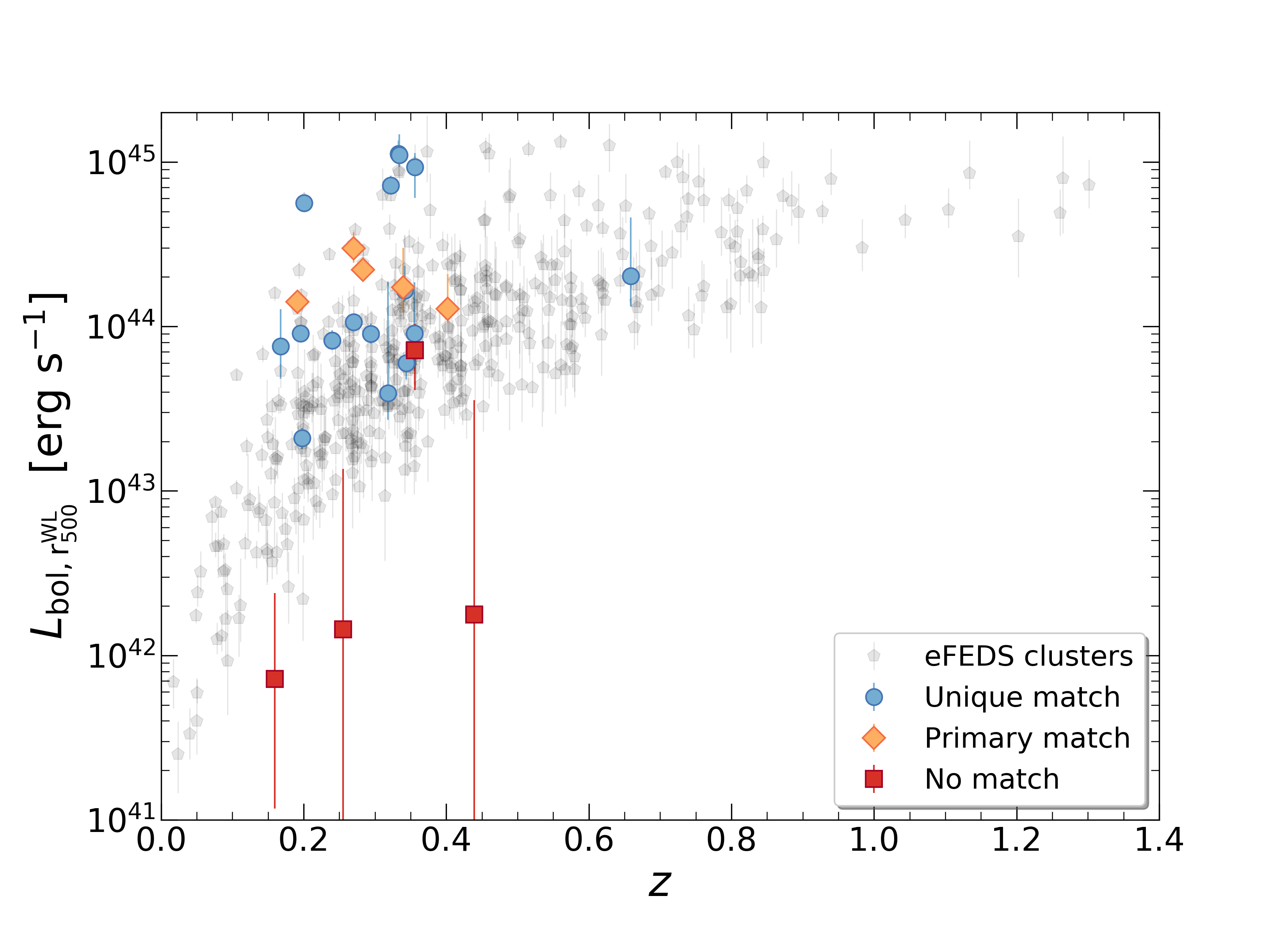}
     \caption{Bolometric luminosities within $r_{500}^{{\rm WL}}$ as a function of redshift of the eFEDS clusters (grey pentagons). Blue circles display the shear-selected clusters with a unique eFEDS match; orange diamonds show weak-lensing peaks with a primary eFEDS counterpart; and red squares show weak-lensing clusters with no eFEDS match.}
     \label{fig:luminvsz}
 \end{figure}


\section{Methodology}
\label{sect:method}

Here, the X-ray cluster analysis used in this work is described. First a brief review of the X-ray analysis is presented, followed by a description of the different morphological parameters that were determined for the clusters.


\subsection{Determination of X-ray observables}
\label{sect:xray_obs}

The X-ray spectral and imaging analysis for the eFEDS clusters is fully described in \cite{Ghirardini2021} and \cite{LiuA2021}. In the following, a short description is presented. 

The main X-ray property of interest for this work is the bolometric X-ray luminosity of each cluster, which is derived from the soft-band luminosity profile. The soft-band luminosity, $L_{\rm X}$, is obtained by integrating the surface brightness along the line of sight in the $0.5-2.0$~keV energy band together with the determined temperature from X-ray spectral analysis. The X-ray surface brightness is calculated by fitting the X-ray cluster image with a projected cluster analytic model \citep{Vikhlinin2006} using a Monte Carlo Markov Chain code \citep[MCMC,][]{Foreman2013}. The faint point sources within the cluster images are excised, while the bright ones are modelled as delta functions convolved with the point spread function (PSF) to eliminate residual emission due to the PSF wings. The temperature was determined by fitting jointly the extracted spectra from the seven eROSITA telescope modules with the X{\tiny SPEC} software \citep{Arnaud1996}. The cluster emission was modelled with an optically thin plasma model {\tt apec} \citep{Smith2001}. Due to the shallow depth of the eFEDS survey, the abundance in the thermal emission model is fixed to $0.3$~Z$_{\odot}$. The model also contains the usual X-ray emission background components: Local Hot Bubble, Galactic Halo and unresolved X-ray sources \cite[see][for further details]{Ghirardini2021}. The temperature determined within a given radius is used in the calculation of the cluster luminosity. Therefore, the luminosity profile is obtained by multiplying the surface-brightness profile by a conversion factor using the determined temperature, thus taking into account uncertainties from the spatial and spectral analysis. In the same manner, the bolometric luminosity (in the $0.01-100$~keV energy range), $L_{\rm bol}$, is obtained.

For the shear-selected clusters, the bolometric luminosity and temperature are measured within $r_{500}^{\rm WL}$, that is $r_{500}$\footnote{$r_{500}$ is the radius within which the mean over-density of the galaxy cluster is $500$ times the critical density at the cluster redshift.} from the calibrated weak-lensing mass, that is, bias corrected by the \cite{Chen2020} correction. As described in section~\ref{sect:weaklensmass}, the difference in the true mass between the \cite{Oguri2021} and \cite{Chiu2021} estimations can reach up to $\sim 60\%$, however, this is translated as a maximum of $30\%$ difference in the bolometric luminosity using the two estimations of $r_{500}^{\rm WL}$. In fact, $\sim76\%$ of the matched clusters have a bolometric luminosity difference of less than $20\%$ (i.e. a mean systematic luminosity difference of $0.02\pm0.04$~dex) using one or the other $r_{500}^{\rm WL}$ estimations. The apertures are centred on the X-ray position of the corresponding eFEDS counterpart for shear-selected clusters with a unique or primary matches. Both X-ray observables were measured including the core of the cluster. In this work, core excised derived quantities are omitted since the X-ray observations are relatively shallow and cluster counts are low. The same approach is also applied at the positions of the non-matched shear-selected clusters to obtain X-ray luminosity measurements for them.

The bolometric luminosities and masses of the shear-selected clusters are shown in Table~\ref{tab:xray_wl_prop}. The bolometric luminosity of all eFEDS clusters and weak-lensing peaks as function of redshift is shown in Figure~\ref{fig:luminvsz}. The matched shear-selected clusters are among the most luminous eFEDS clusters, while the non-matched shear-selected clusters lie in the low luminosity regime.


\subsection{X-ray morphological properties}
\label{sect:morphoparam}

\cite{Ghirardini2021b} have determined different X-ray morphological parameters of the eFEDS sample, which help us to understand the dynamical state of the cluster sample. Using the procedure briefly described in section~\ref{sect:xray_obs}, to obtain surface brightness and density profiles, \cite{Ghirardini2021b} calculated eleven X-ray morphological parameters for each cluster in the eFEDS sample, but used only a subset of this sample\footnote{\cite{Ghirardini2021b} applied selection criteria of extent likelihood and detection likelihood values larger than 12 on the eFEDS sample to obtain a cleaner sample. This selection reduces the fraction of spurious clusters to $\sim 14\%$ in the eFEDS sample, decreasing the sample size to 325 clusters.} to study the overall trend of the morphological analysis. The morphological parameters were calculated within $r_{500}^{{\rm WL}}$ \citep{Chiu2021}. In the following, a brief description of each X-ray morphological parameter is presented.

\vspace{-10pt}
\paragraph{Concentration ($c_{\rm SB}$).} This parameter indicates how concentrated the X-ray emission is and it correlates with the presence of a cool core (CC) in the cluster \citep{Santos2008}. It is defined as the ratio between the integrated surface brightness in two different circular apertures. \cite{Ghirardini2021b} presented two concentration measurements: $c_{{\rm SB,}r_{500}}$, whose apertures are $0.1r_{500}$ and $r_{500}$, and $c_{{\rm SB,}40-400~{\rm kpc}}$, whose apertures are $40$ and $400$~kpc.

\vspace{-10pt}
\paragraph{Central gas density ($n_{0}$).} This value also indicates the state of relaxation of the clusters. It is based on the findings of several studies, which have shown that relaxed systems tend to have a higher gas density in the core \citep[e.g.][]{Hudson2010}. \cite{Ghirardini2021b} used a representative value of the density computed at $0.02r_{500}$ (at radius $r=0$ the density profile might diverge).

\vspace{-10pt}
\paragraph{Cuspiness ($\alpha$).} This parameter measures the slope of the cluster density profile at a fixed radius \citep[][]{Vikhlinin2007}. As in \cite{Lovisari2017}, \cite{Ghirardini2021b} fixed the radius at $r=0.04r_{500}$, avoiding the cluster core where cooling affects the measurements the most, but also staying close to the cluster centre to avoid the flattening of the profile caused by AGN outflows.

\vspace{-10pt}
\paragraph{Ellipticity ($\epsilon$).} This quantity is defined as the ratio between the semi-minor and the semi-major axis. \cite{Ghirardini2021b} obtained these values from fitting an elliptical and rotated density profile. It is expected that relaxed clusters have a rounder shape ($\epsilon\sim 1$) than disturbed ones ($\epsilon \ll 1$).

\vspace{-10pt}
\paragraph{Power ratios.} These parameters are obtained through a two-dimensional multipole expansion of the cluster surface brightness distribution within a given aperture. The radial fluctuations are detected by the higher order moments, which are sensitive to smaller scales. This concept is based on the idea that by using the projected mass profile (in this case using the surface brightness) the power ratios are related to the cluster potential \citep[][]{Buote1995}. \cite{Ghirardini2021b} considered only the power ratios up to order 4 ($P_{10},~P_{20},~P_{30},$ and $P_{40}$).

\vspace{-10pt}
\paragraph{Gini coefficient.} This parameter measures the X-ray flux distribution in galaxy clusters. If the total flux is equally distributed among the considered pixels in a given aperture, then the Gini coefficient is equal to $0$, on the contrary, if the total flux is concentrated in a single pixel, then the value is $1$. This parameter is used in economics, but it was used in astronomy by \cite{Abraham2003} and many other optical galaxy studies.

\vspace{-10pt}
\paragraph{Photon asymmetry ($A_{\rm phot}$).} This parameter quantifies the degree of rotational symmetry of the light coming from the galaxy clusters \citep[][]{Nurgaliev2013}. 

\vspace{6pt}
Thanks to the large luminosity and redshift coverage of the eFEDS sample, \cite{Ghirardini2021b} successfully constrained the evolution of morphological parameters with both redshift and luminosity. These allow us to obtain evolution-independent morphological estimators. Furthermore, \cite{Ghirardini2021b} introduced a new morphological parameter, the relaxation score, $R_{\rm score}$. This parameter is derived taking into account all the information of the different redshift and luminosity independent morphological parameters, their correlations with respect to the concentration parameter and the cluster selection function.


\begin{table}[t]
\centering
\begin{threeparttable}
\renewcommand{\arraystretch}{1.3}
\caption{Properties of the shear-selected clusters within the eFEDS footprint.}
{\tiny
\begin{tabular}{cccccc}
\toprule
eFEDS ID & $z_{\rm  MCMF}$ & $L_{{\rm bol},r_{500}^{\rm WL}}$ & ID & $z$ & $M_{500}$ \\
 (eFEDSJ+) &  &  [$10^{43}$~erg~s$^{-1}$] & & & [$10^{14}$~M$_{\odot}$] \\
\midrule
\midrule
091403.3+013846   & 0.168 & $~~7.56_{- 2.75}^{+ 5.14}$ & 114 & 0.161 & $1.95_{-0.51}^{+0.54}$ \\
085931.9+030839   & 0.196 & $~~9.05_{- 0.59}^{+ 0.72}$ &  62 & 0.179 & $2.02_{-0.52}^{+0.56}$ \\
092241.9+020719   & 0.198 & $~~2.10_{- 0.30}^{+ 0.32}$ & 117 & 0.200 & $1.76_{-0.48}^{+0.52}$ \\
085751.6+031039   & 0.201 & $~56.30_{- 4.20}^{+ 5.20}$ &  16 & 0.188 & $4.38_{-0.92}^{+0.91}$ \\
093403.5$-$001422 & 0.240 & $~~8.21_{- 0.99}^{+ 1.28}$ & 133 & 0.238 & $2.19_{-0.54}^{+0.57}$ \\
085230.6+002457   & 0.270 & $~10.60_{- 1.00}^{+ 1.30}$ &  49 & 0.281 & $2.17_{-0.55}^{+0.58}$ \\
091351.1$-$004507 & 0.294 & $~~9.00_{- 1.14}^{+ 1.40}$ &  53 & 0.265 & $3.10_{-0.75}^{+0.71}$ \\
092844.0+005318   & 0.318 & $~~3.94_{- 1.23}^{+14.86}$ & 158 & 0.300 & $1.64_{-0.70}^{+0.69}$ \\
091610.1$-$002348 & 0.322 & $~71.80_{- 5.20}^{+10.50}$ &  29 & 0.332 & $4.35_{-0.90}^{+0.93}$ \\
092121.2+031726   & 0.333 & $112.00_{-13.00}^{+23.00}$ &  24 & 0.353 & $5.35_{-0.90}^{+0.94}$ \\
084528.6+032739   & 0.334 & $110.00_{-16.30}^{+37.00}$ &   6 & 0.349 & $5.23_{-0.86}^{+0.91}$ \\
093431.3$-$002309 & 0.342 & $~16.50_{- 2.30}^{+ 6.90}$ &  20 & 0.331 & $4.33_{-1.06}^{+1.05}$ \\
092846.5+000056   & 0.344 & $~~5.98_{- 1.21}^{+ 2.08}$ & 162 & 0.336 & $3.06_{-1.05}^{+0.92}$ \\
093302.7$-$010145 & 0.356 & $~~9.08_{- 2.39}^{+ 9.52}$ &  12 & 0.338 & $2.78_{-0.67}^{+0.74}$ \\
093513.0+004757   & 0.356 & $~93.00_{-32.50}^{+20.00}$ &   8 & 0.357 & $8.08_{-1.14}^{+1.20}$ \\
090805.9+011952   & 0.659 & $~20.30_{- 7.10}^{+25.80}$ & 159 & 0.644 & $3.66_{-1.44}^{+1.58}$ \\
\midrule
083654.6+025954   & 0.191 & $~14.10_{- 0.70}^{+ 1.30}$ &  94 & 0.193 & $2.94_{-0.64}^{+0.68}$ \\
091849.0+021204   & 0.283 & $~22.10_{- 2.20}^{+ 3.50}$ &  18 & 0.274 & $2.65_{-0.57}^{+0.63}$ \\
092209.3+034628   & 0.270 & $~29.80_{- 5.40}^{+ 7.50}$ &  35 & 0.252 & $4.20_{-0.80}^{+0.80}$ \\
093546.3$-$000115 & 0.339 & $~17.30_{- 5.10}^{+12.70}$ &  58 & 0.339 & $2.68_{-0.67}^{+0.69}$ \\
084129.0+002645   & 0.402 & $~12.80_{- 2.20}^{+ 8.00}$ & 104 & 0.398 & $2.38_{-0.93}^{+0.89}$ \\
\midrule
                      &   & $~~0.07_{- 0.06}^{+ 0.17}$ & 141 & 0.160 & $0.85_{-0.59}^{+0.51}$ \\
                      &   & $~~0.14_{- 0.14}^{+ 1.22}$ & 169 & 0.255 & $2.11_{-0.83}^{+0.83}$ \\
                      &   & $~~7.15_{- 3.04}^{+ 3.25}$ & 146 & 0.356 & $2.68_{-0.76}^{+0.85}$ \\
                      &   & $~~0.18_{- 0.17}^{+ 3.38}$ & 144 & 0.439 & $1.94_{-0.92}^{+0.90}$ \\
\bottomrule
\end{tabular}
\tablefoot{The horizontal lines divide the unique, primary and non-matches categories (from top to bottom). Weak-lensing peaks not matched to an eFEDS cluster have no corresponding eFEDS ID and redshift.}
}
\label{tab:xray_wl_prop}
\end{threeparttable}
\end{table}


\section{Results}
\label{sect:results}

In this section the measurements of the $L_{\rm bol}-M$ relation of the shear-selected clusters are presented, as well as their dynamical state analysis.

 \begin{figure}[t]
     \centering
     \includegraphics[trim=20 30 80 85,clip,width=\columnwidth]{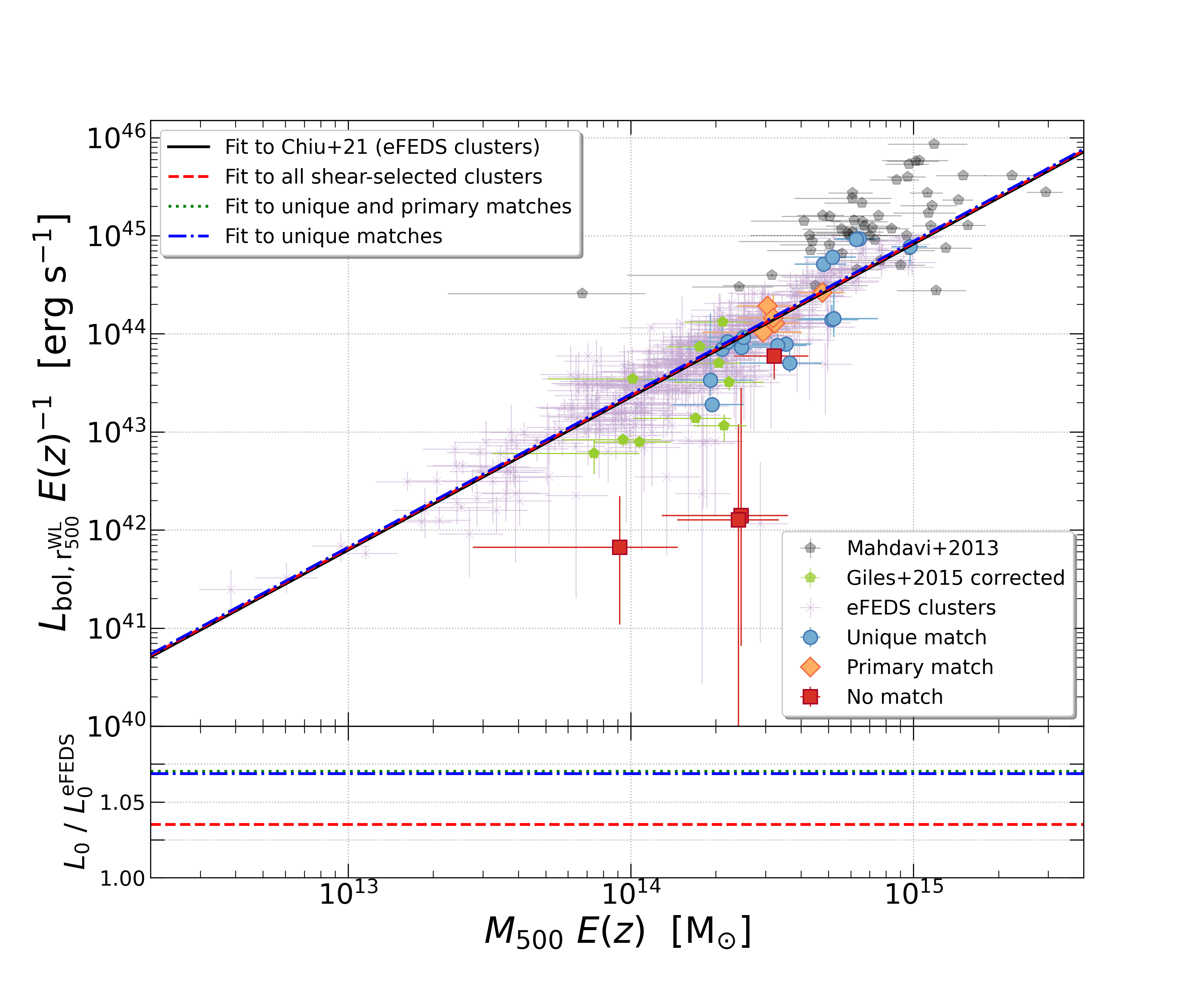}
     \caption{{\it Top:} X-ray bolometric luminosity--mass relation for the weak-lensing peaks in the eFEDS footprint. Blue circles show the shear-selected clusters with a unique eFEDS match; orange diamonds display weak-lensing peaks with a primary eFEDS counterpart; and red squares show weak-lensing clusters with no eFEDS match. eFEDS clusters are displayed as light purple asterisks and its corresponding fit is given by the black solid line. For comparison, the data of the X-ray selected clusters presented in \citet[][grey pentagons]{Mahdavi2013}, and the shear-selected clusters studied in \citet[][green pentagons]{Giles2015} are plotted. The fit to the shear-selected cluster is given by the dashed-red line, and for two sub-samples of them the fits are shown by the dotted-green and dashed-dotted blue lines. {\it Bottom:} Ratio of the different normalisations of the distinct shear-selected cluster to normalisation of the eFEDS clusters fit.}
     \label{fig:lbolrwL_Mobs}
 \end{figure}


\subsection{Bolometric luminosity--mass relation}
\label{sect:lmrelation}

The X-ray bolometric luminosity--mass, $L_{\rm bol}-M$, relation for the shear-selected clusters is derived, with the masses estimated from the weak-lensing analysis. The masses, in units of $h^{-1}M_\odot$ in \cite{Oguri2021}, are converted to those in units of $M_\odot$ adopting $H_{0}=70$~km~s$^{-1}$~Mpc$^{-1}$ and are bias corrected according to \cite{Chen2020}, obtaining their true masses (see section~\ref{sect:weaklensmass}). Figure~\ref{fig:lbolrwL_Mobs} shows the $L_{\rm bol}-M$ relation of the shear-selected clusters, which  are colour coded according to the eFEDS cross-matching. The error correlation between $L_{\rm bol}$ and $M$ is ignored in this analysis since the correlation is very small.

The bolometric luminosity measurements of the shear-selected clusters are compared with the ones from the eFEDS X-ray selected cluster sample, since both samples cover approximately the same cluster mass range. As in \cite{Chiu2021}, only eFEDS clusters with $f_{\rm cont}<0.2$ are used (see section~\ref{sect:eFEDS}). The bolometric luminosities and calibrated weak-lensing masses in all data sets were corrected for the expected self-similar evolution, that is $\times\,E(z)^{-1}$ and $\times\,E(z)$, respectively, where $E(z)$ is the dimensionless Hubble parameter.

The data are fitted with a power law of the form
\begin{equation}
 E(z)^{-1}\frac{L_{\rm bol}}{L_0}=\Bigg[E(z)\frac{M}{M_0}\Bigg]^{B_{LM}}   
\end{equation}
\label{eq:lmrelation}
using the BCES orthogonal regression in logarithmic space \citep{Akritas1996,Nemmen2012}. The pivot mass $M_0$ is taken to be $2\times10^{14}~$M$_\odot$. The slope, $B_{LM}$, and the normalisation, $L_0$, are determined from the data fit. First, the eFEDS sample is fitted, yielding $B_{LM}^{\rm eFEDS}=1.56\pm0.05$ and $L_0^{\rm eFEDS}=(6.67\pm0.21)\times10^{43}$~erg~s$^{-1}$ (black solid line in Fig.~\ref{fig:lbolrwL_Mobs}). Since there is a small number of shear-selected clusters, the slope $B_{LM}$ is fixed to the best-fit value $B_{LM}^{\rm eFEDS}$ when fitting the shear-selected cluster sample. It is found that ${L_0}=(6.91\pm0.87)\times10^{43}$~erg~s$^{-1}$ (red dashed line in Fig.~\ref{fig:lbolrwL_Mobs}) when fitting the $21$ weak-lensing peaks. There is a slight offset in the normalisation of the fits between the eFEDS and shear-selected samples, but they are consistent within $1\sigma$. As mentioned in section~\ref{sect:weaklensmass}, there exists an additional, but small systematic difference in the cluster mass between the eFEDS and the shear-selected cluster samples. As shown in Table~\ref{tab:Lbol-Mrelation}, this consistency remains even if the fit is performed using only unique or unique and primary weak-lensing peaks. This confirms that the mass-bias correction introduced in \cite{Chen2020} reduces the tension in the $L_{\rm bol}-M$ normalisation between X-ray selected and shear-selected cluster samples. 

\cite{Chiu2021} calculated the parameter constraints of several X-ray scaling relations for the eFEDS clusters taking into account the eFEDS X-ray selection function and the flexibility of possible redshift-dependent deviations from the self-similar prediction. For the $L_{\rm bol}-M$ relation, they found a slope of $B_{LM}^{\rm Chiu+21}=1.55^{+0.16}_{-0.14}$, which is consistent with the slope obtained here using the BCES orthogonal regression method. The normalisation obtained by \cite{Chiu2021} is $L_{0}^{\rm Chiu+21}=9.2^{+1.6}_{-1.3}\times10^{43}$~erg~s$^{-1}$ at their pivotal redshift $z_{\rm piv}=0.35$, and it is slightly higher than the one obtained here.



\begin{table}[t]
\centering
\begin{threeparttable}
\renewcommand{\arraystretch}{1.1}
\caption{Fit results of the $L_{\rm bol}-M$ relation.}
{\small
\begin{tabular}{lcc}
\toprule
Fitted sample & Slope ($B_{LM}$) & normalisation ($L_0$) \\
 &  & [$\times10^{43}$~erg~s$^{-1}$] \\
\midrule
\midrule
eFEDS & $1.56\pm0.05$ & $6.67\pm0.21$ \\
\midrule
All shear-selected clusters & $1.56$ & $6.91\pm0.87$ \\
Unique and primary matches & $1.56$ & $7.14\pm0.91$ \\
Unique matches & $1.56$ & $7.13\pm1.16$ \\
\bottomrule
\end{tabular}
\tablefoot{The scaling relation is fitted with a power-law of the form $E(z)^{-1}(L/L_0)=[E(z)(M/M_0)]^{B_{LM}}$, where $M_0=2\times10^{14}~$M$_\odot$. The slope is fixed for the shear-selected clusters.}
}
\label{tab:Lbol-Mrelation}
\end{threeparttable}
\end{table}

\begin{table}[ht]
\centering
\begin{threeparttable}
\renewcommand{\arraystretch}{1.1}
\caption{Linear regression parameters (slope and normalisation) for the $L_{\textrm X}-M$ relation.}
{\small
\begin{tabular}{lcc}
\toprule
Fitted sample & Slope & normalisation \\
 &  &  [$\times 10^{43}$~erg~s$^{-1}$] \\
\midrule
\midrule
All shear-selected clusters (fixed slope) & $1.52$ & $0.70_{-0.11}^{+0.12}$\\  
All shear-selected clusters (slope prior) & $1.52_{-0.03}^{+0.03}$ & $0.69_{-0.11}^{+0.12}$ \\
\midrule
eFEDS clusters & $1.52_{-0.03}^{+0.03}$ & $0.87_{-0.02}^{+0.02}$ \\
\midrule
XXL clusters by \citet{Akino2021} & $1.38^{+0.27}_{-0.18}$ & $1.34^{+0.19}_{-0.16}$ \\
\bottomrule
\end{tabular}
}
\label{tab:Lx-Mrelation}
\end{threeparttable}
\end{table}

\begin{table*}[t]
\centering
\begin{threeparttable}
\renewcommand{\arraystretch}{1.0}
\caption{Median values of the luminosity and redshift corrected and non-corrected X-ray morphological parameters of the eFEDS and shear-selected samples.}
{\footnotesize
\begin{tabular}{lcccc}
\toprule
Morphological & \multicolumn{2}{c}{Not corrected} & \multicolumn{2}{c}{$L$ and $z$ corrected} \\
 parameter & eFEDS sample & Shear-selected sample & eFEDS sample & Shear-selected sample  \\
\midrule
\midrule
                   $n_0$ & $(0.57 \pm 0.60)\times10^{-2}$ & $(0.65 \pm 0.41)\times10^{-2}$ & $(0.59 \pm 0.41)\times10^{-2}$ & $(0.60 \pm 0.32)\times10^{-2}$ \\
  $c_{{\rm SB,}r_{500}}$ & $0.16 \pm 0.12$                & $0.17 \pm 0.06$                & $0.15 \pm 0.10$                & $0.18 \pm 0.06$ \\
 $c_{\rm SB,40-400~kpc}$ & $(0.90 \pm 0.85)\times10^{-1}$ & $(0.83 \pm 0.42)\times10^{-1}$ & $(0.90 \pm 0.70)\times10^{-1}$ & $(0.96 \pm 0.49)\times10^{-1}$ \\
                $\alpha$ & $0.72 \pm 0.36$                & $0.69 \pm 0.21$                & $0.70 \pm 0.22$                & $0.72 \pm 0.23$ \\
              $\epsilon$ & $0.73 \pm 0.22$                & $0.79 \pm 0.11$                & $0.73 \pm 0.11$                & $0.80 \pm 0.12$ \\
                $P_{10}$ & $(0.52 \pm 1.19)\times10^{-3}$ & $(0.22 \pm 0.35)\times10^{-3}$ & $(0.45 \pm 0.64)\times10^{-3}$ & $(0.40 \pm 0.29)\times10^{-3}$ \\
                $P_{20}$ & $(0.71 \pm 1.57)\times10^{-4}$ & $(0.22 \pm 0.34)\times10^{-4}$ & $(0.57 \pm 0.69)\times10^{-4}$ & $(0.39 \pm 0.41)\times10^{-4}$ \\
                $P_{30}$ & $(0.18 \pm 0.47)\times10^{-4}$ & $(0.05 \pm 0.11)\times10^{-4}$ & $(0.16 \pm 0.20)\times10^{-4}$ & $(0.13 \pm 0.08)\times10^{-4}$ \\
                $P_{40}$ & $(0.80 \pm 1.80)\times10^{-5}$ & $(0.29 \pm 0.50)\times10^{-5}$ & $(0.60 \pm 0.70)\times10^{-5}$ & $(0.50 \pm 0.60)\times10^{-5}$ \\
                    Gini & $0.65 \pm 0.07$                & $0.70 \pm 0.06$                & $0.66 \pm 0.03$                & $0.66 \pm 0.02$ \\
          $A_{\rm phot}$ & $0.47 \pm 0.62$                & $0.21 \pm 0.27$                & $0.44 \pm 0.37$                & $0.39 \pm 0.24$ \\
\bottomrule
\end{tabular}
\tablefoot{The number after the median shows the $16^{\rm th}$ and $84^{\rm th}$ percentiles of the distributions.}
}
\label{tab:xray_morpho}
\end{threeparttable}
\end{table*}

\subsection{Soft-band luminosity--mass relation}
\label{app:D}

As in the previous section, the soft-band ($0.5-2$~keV) luminosity-mass, $L_{\textrm X}-M$, relation for the shear-selected clusters is derived. However, the fitting methodology is slightly changed in order to compare the results with the ones obtained by the XXL collaboration \citep[][]{Pierre2016}. In the following, the methodology and results of this comparison are presented.

The XXL survey is the one of the largest {\it XMM-Newton} surveys, covering $50$~deg$^2$ with nearly $7$~Ms of exposure. In the same manner, as eFEDS, the XXL survey was designed to provide a well-characterised sample of X-ray detected galaxy clusters. In 2018, \cite{Adami2018} published a sample of $365$ clusters down to a flux of $10^{-15}$~erg~s$^{-1}$~cm$^{-2}$ in the $0.5-2.0$~keV energy band. The clusters span a redshift range between $0<z<1.2$. \cite{Umetsu2020}, determined weak-lensing masses for $136$~of these XXL clusters using HSC-SSP data, covering a mass range of $10^{13}<M_{500}^{\rm WL}$~[M$_\odot$]$<6\times10^{14}$. The median redshift of this sub-sample is $z=0.31$, very similar to the one of eFEDS clusters. Using this sub-sample, \citet[][]{Akino2021} investigated X-ray observable-to-mass scaling relations, among them the $L_{\rm X}-M$ relation. The redshift, mass range and weak-lensing mass calibration covered by the XXL clusters, make them the ideal external X-ray sample to compare the HSC shear-selected and the eFEDS samples presented in this work.

\begin{figure}[t]
     \centering
     \includegraphics[trim=20 10 80 50,clip,width=\columnwidth]{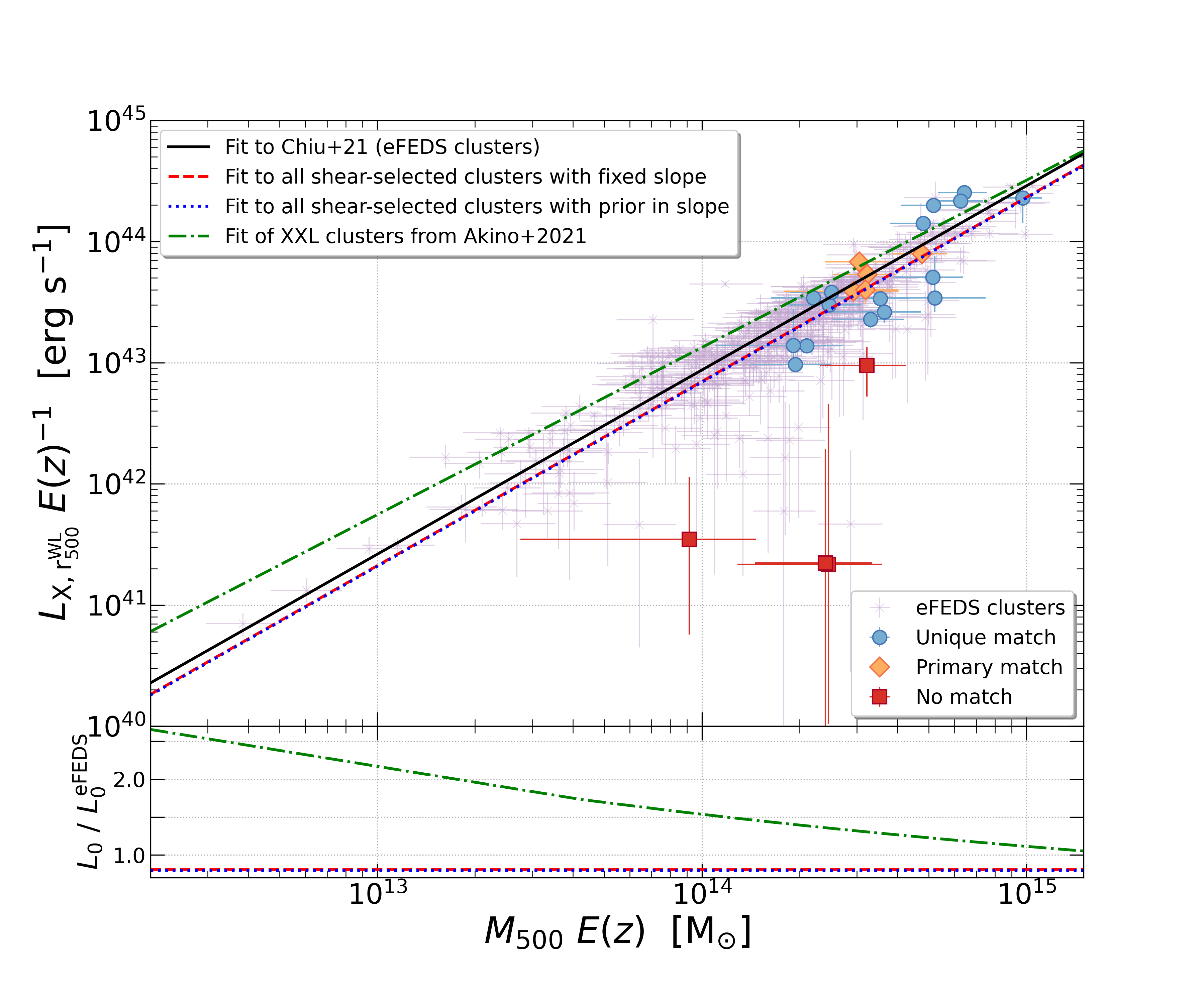}
     \caption{X-ray soft band luminosity--mass relation for the weak-lensing peaks in the eFEDS footprint. Blue circles show the shear-selected clusters with a unique eFEDS match; orange diamonds display weak-lensing peaks with a primary eFEDS counterpart; and red squares show weak-lensing clusters with no eFEDS match. eFEDS clusters are displayed as purple asterisks and its corresponding fit is given by the black solid line. Two fits, with fixed slope and with a prior in the slope, for the shear-selected clusters are shown by the red dashed and blue dotted lines, respectively. For comparison, the fit obtained by \citet{Akino2021} on the XXL clusters is shown by the green dotted-dashed line.}
     \label{fig:lXrwL_Mobs}
 \end{figure}

To compare the HSC shear-selected sample studied in this work with the XXL sample, the $L_{\rm X}-M$ relation is investigated. Figure~\ref{fig:lXrwL_Mobs} shows the $L_{\textrm X}-M$ relation of the HSC shear-selected sample (colour coded accordingly to the eFEDS cross-matching) and the eFEDS clusters (purple points).

\citet{Akino2021} carried out the scaling relation fitting using a linear regression method, obtaining a slope of $1.38^{+0.27}_{-0.18}$ and a normalisation of $1.34^{+0.19}_{-0.16}~\times 10^{43}~{\rm erg~s^{-1}}$ for the XXL clusters (green dotted-dashed line in Fig.~\ref{fig:lXrwL_Mobs}). In order to compare with the XXL results, we use the same fitting method for the $L_{\rm X}-M$. For the eFEDS sample, we obtain a slope of $1.52_{-0.03}^{+0.03}$ (black solid line), which is in good agreement with the XXL sample; however, the normalisation is $0.87_{-0.02}^{+0.02}~\times 10^{43}~{\rm erg~s^{-1}}$. By keeping the slope fixed to the eFEDS value and with a prior around this value (red dashed and blue dotted lines, respectively), we performed a fit of the HSC shear-selected sample. While there is consistency within $\sim 1\sigma$ between the normalisations of the eFEDS and HSC shear-selected samples, the disagreement increases between the HSC shear-selected sample and the XXL sample (XXL luminosities are larger by a factor of $4\sigma$ at $10^{14}$~M$_\odot$ but there is $\sim1.5\sigma$ agreement at $10^{13}$~M$_\odot$ and $10^{15}$~M$_\odot$ because of the uncertainty of slope). The results are summarised in Table~\ref{tab:Lx-Mrelation}. A possible source of this discrepancy is the used radius to extract the X-ray luminosity: the XXL X-ray luminosity is measured not at $r_{500}^{\rm WL}$ but at the maximum detection radius of X-ray source count. Moreover, the XXL does not cover the low mass group regime as eFEDS ($M_{500}^{\rm WL}<10^{13}$~M$_\odot$), which can cause an overestimation of the slope in their XXL fits.

For the $L_{\rm X}-M$ relation, \citet{Chiu2021} found a slope of $1.50^{+0.15}_{-0.14}$, which is consistent with the slope obtained here using the linear regression method. It is also consistent with the XXL slope within the uncertainties. The normalisation obtained by \cite{Chiu2021} is $3.36^{+0.53}_{-0.49}\times10^{43}$~erg~s$^{-1}$, at their pivotal redshift $z_{\rm piv}=0.35$, is higher than the one obtained with the linear regression method.

\subsection{Dynamical state of the shear-selected clusters}
\label{sect:dynamicalstate}

One of our goals is to investigate the dynamical state of the shear-selected clusters and compare it to the one of X-ray selected cluster samples, here with an emphasis on the eFEDS sample. In the following X-ray and optical morphological parameters are discussed for the shear-selected clusters and the eFEDS sample.


\subsubsection{X-ray morphology of the shear-selected clusters}

The X-ray morphological parameters described in section~\ref{sect:morphoparam} and obtained in \cite{Ghirardini2021b}, are compared for the eFEDS sample and the shear-selected sample. We note that for the shear-selected samples, the values of their corresponding eFEDS counterparts are used, that is no new morphological estimators are calculated within $r_{500}^{\rm WL}$ from \cite{Oguri2021}. Moreover, not all the shear-selected clusters fall into the cluster selection applied by \cite{Ghirardini2021b} (see section~\ref{sect:morphoparam}), $4$ shear-selected clusters have an extension likelihood lower than $12$ (between $\sim 9-11$ values). This clusters can be still be corrected by the luminosity and redshift factors described in \cite{Ghirardini2021b}, but there is no relaxation score parameter for them.

The median values (together with the difference with the $16^{\rm th}$ and $84^{\rm th}$ percentiles of their distribution) of the eleven morphological parameters for the eFEDS sample\footnote{The $17$ shear-selected clusters are a sub-sample of the eFEDS clusters, therefore they are removed from the main eFEDS sample in this X-ray morphological analysis.} and for the shear-selected sample are shown in Table~\ref{tab:xray_morpho}. These values are presented with and without luminosity and redshift correction (see section~\ref{sect:morphoparam} and \cite{Ghirardini2021b} for further details). The luminosity and redshift corrected morphological parameter distributions are shown in Fig.~\ref{fig:morphoparam} (not-corrected morphological estimators show similar distributions).

The results in Table~\ref{tab:xray_morpho} (and Fig.~\ref{fig:morphoparam}) show that the distribution of the corrected (luminosity and evolution-independent) and non-corrected morphological parameters between eFEDS and the shear-selected clusters are consistent with each other. Especially, the median values of the central density ($n_{0}$), cuspiness ($\alpha$) and Gini coefficient remain very similar before and after the luminosity and redshift correction between both samples. Both concentrations ($c_{{\rm SB,}r_{500}}$ and $c_{{\rm SB,}40-400~{\rm kpc}}$) and ellipticity ($\epsilon$) values for the shear-selected clusters remain slightly higher after the luminosity and redshift correction, although the difference is not significant. In fact, larger values of concentration and ellipticity are likely associated with relaxed clusters. This might be a direct consequence of the Gaussian filter used to select this sample (see section~\ref{sect:shearclu}). Finally, although consistent, after the luminosity and redshift correction, the median values of the power-ratios ($P_{10}, P_{20}, P_{30}, P_{40},$) and photon asymmetry ($A_{\rm phot}$) of the shear-selected sample approach more to the median values of the eFEDs sample but still they are smaller. Lower values of power-ratios are usually associated with more relaxed clusters.

\begin{figure}[t]
    \centering
    \includegraphics[trim=25 20 60 60,clip,width=\columnwidth]{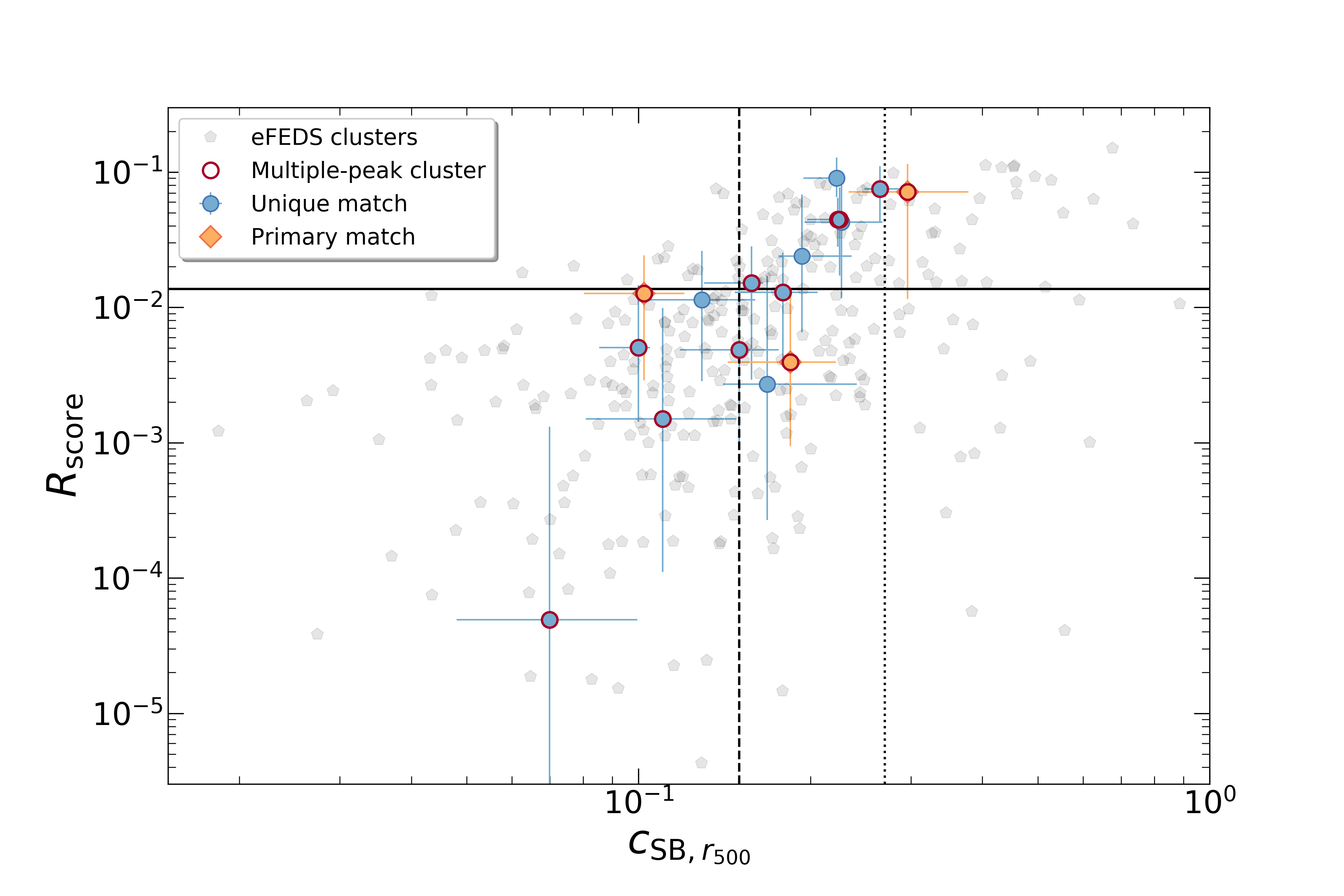}
    \caption{Relaxation parameter, $R_{\rm score}$, as function of the concentration parameter, $c_{\rm SB, r_{500}}$. Filled-blue circles show the shear-selected clusters with a unique eFEDS match; filled-orange diamonds display weak-lensing peaks with a primary eFEDS counterpart. Points with a red ring around are clusters with multiple-peaks, which were found by peak-finding method described in section~\ref{sect:peakfinding}. The solid black line indicates the threshold to distinguish relaxed clusters (above) from disturbed (below) clusters as suggested in \cite{Ghirardini2021b}, while the dashed and dotted vertical lines display a similar criteria suggested by \cite{Lovisari2017}. Errors in the eFEDS clusters are omitted for clarity purposes.}
    \label{fig:rscore_concentation}
\end{figure}

As described in section~\ref{sect:morphoparam}, \cite{Ghirardini2021b} determined the relaxation score parameter for the eFEDS clusters. The correlation between this parameter and the concentration is shown in Fig.~\ref{fig:rscore_concentation}. In this figure, the shear-selected clusters for which $R_{\rm score}$ is measured are highlighted by the blue-filled circles and orange-filled diamonds. \cite{Ghirardini2021b} suggested a threshold of $R_{\rm score}=0.0137$ to distinguish between relaxed (above) and disturbed (below) clusters. \cite{Lovisari2017} proposed a similar criteria using the concentration parameter: relaxed clusters have $c_{{\rm SB,} r_{500}}>0.27$ and disturbed clusters $c_{{\rm SB,} r_{500}}<0.15$. The shear-selected clusters do not show a preferred dynamical state according to the $R_{\rm score}-c_{\rm SB, r_{500}}$ plane. Following the $R_{\rm score}$ criterion, half of them seem to be in a relaxed state, and a similar conclusion can be drawn from the $c_{\rm SB, r_{500}}$ threshold.

\subsubsection{Optically defined merging state of the cluster samples}
\label{sect:peakfinding}

We search for merging clusters by a peak-finding method of galaxy distributions following \citet{2019PASJ...71...79O}. The lifetime of galaxy subhalos whose distribution is similar to the dark matter distribution is much longer than that of gas subhalos \citep{2008PASJ...60..345O}. The number of luminous galaxies is almost conserved during mergers, while the X-ray luminosity is significantly affected by mergers because of the collisional nature of the ICM \citep{2001ApJ...561..621R}. The angular resolution of the galaxy distribution, which is higher than that of the weak-lensing mass reconstruction, resolves subhalos inside clusters. Therefore, the galaxy distribution enables to search for merger candidates unbiasedly. However, since it is difficult to distinguish between pre- and post-mergers only from the galaxy distribution, X-ray morphological parameters are essential.

We select red-sequence galaxies in the colour-magnitude plane in a similar way as in \citet{2018PASJ...70S..24N}, and then make galaxy maps with a Gaussian kernel of ${\rm FWHM}=200$~kpc. Since both the number of red galaxies and the angular size of the smoothing scale depends on cluster redshifts, we adopt a redshift-dependent threshold corresponding to the peak height of CAMIRA richness $N=15$ that roughly correspond to $M_{500}\sim 5\times10^{13}M_\odot$. We also subtract the contamination of the extended distribution of a galaxy peak from other peaks, where we assume the average extended distribution of the CAMIRA clusters for the distribution of the highest peak and a Gaussian distribution for other peaks. The galaxy maps are shown in Appendix~\ref{app:B} and Fig.~\ref{fig:nomatchesxrayoptical}. We search for peaks within $500$~kpc from the centres of the shear-selected clusters. The number of clusters with multiple peaks is $15$ out of $25$ clusters, that is the merger fraction is $60 \%$. For comparison, we repeat the same analysis for $425$ from the $444$ eFEDS clusters in the HSC-SSP footprint and find $108$ multiple-peak clusters ($24\%$). We note that the number of eFEDS clusters is larger than that of weak-lensing analysis because the colour analysis does not requires the strict full-colour and full-depth conditions. 

To further check the above result, we first construct eFEDS sub-samples that are more comparable to the shear-selected cluster sample by selecting eFEDS clusters with $M_{500}>1.5\times10^{14}M_\odot$ in the redshift ranges $0.15<z<0.45$ and $0.15<z<0.65$. These sub-samples give $22\%$ ($59$ out of $264$) and $19\%$ ($70$ out of $364$), respectively, of multiple-peak clusters. These percentages are still lower than the one from the shear-selected sample. The same analysis was performed for the entire sample of shear-selected clusters with redshift information  available ($182$). The final percentages are: $76\%$ of the shear-selected clusters ($138$) show a single peak, while $24\%$ ($44$) have are multiple peaks. Therefore, no significant differences in the merger fraction between X-ray and shear-selected samples are found. The result of a high-merger fraction present in the shear-selected clusters located in the eFEDS footprint can be explained by a statistical fluctuation due to the small number of the shear-selected clusters in the eFEDS region. The sample of X-ray detected clusters will significantly increase in the near future with the analysis of the first eROSITA all-sky survey, in the same fashion the number of shear-selected clusters will be higher with further HSC observations. Therefore larger cluster samples will help us to identify if there might be a higher incidence of mergers in the shear-selected cluster population.

The multiple-peak shear-selected clusters that have an eFEDS counterpart, and therefore an X-ray morphological parameter estimation, are shown in Fig.~\ref{fig:rscore_concentation} (and also Fig.~\ref{fig:morphoparam}) as red rings enclosing the corresponding clusters. In both figures, the distributions of these multiple-peak clusters are not different from those of the shear-selected and eFEDS samples.


\subsection{Weak-lensing peaks in superclusters}

\cite{LiuA2021} searched for superclusters in the eFEDS area. In short, employing a friends-of-friends (FoF) algorithm with an redshift-dependent linking length, $19$~superclusters candidates in the redshift range $z=0.1-0.8$ were found. Each supercluster has at least four galaxy cluster members. Of the $542$ eFEDS clusters, $\sim 18\%$ of them lie in superclusters. Out of these $19$ superclusters, $2$ lie outside the common GAMA09/eFEDS area (no. $5$ and $16$ in \cite{LiuA2021}).

Out of the $21$ shear-selected clusters with an eFEDS counterpart (see section~\ref{sect:catsmatchs}), $9$ are located in an eFEDS supercluster. This represents $\sim 43\%$ of the weak-lensing peaks with at least one eFEDS counterpart. Of these $9$, $5$ are unique matches and $4$ primary matches. The fact that a higher percentage of shear-selected clusters lie in superclusters is not surprising since weak-lensing searches identify the integrated shear signal using a broad lensing kernel along the line-of-sight direction. Therefore the lensing signal is affected, not only by foreground and background structures but also by nearby objects whose angular separation is smaller than the kernel size, which may cause the enhancement of the shear signals. We leave the quantitative estimate of the impact of superclusters on weak lensing shear-selected clusters for future work.

 \begin{figure*}[ht]
    \centering
     \includegraphics[width=0.24\textwidth]{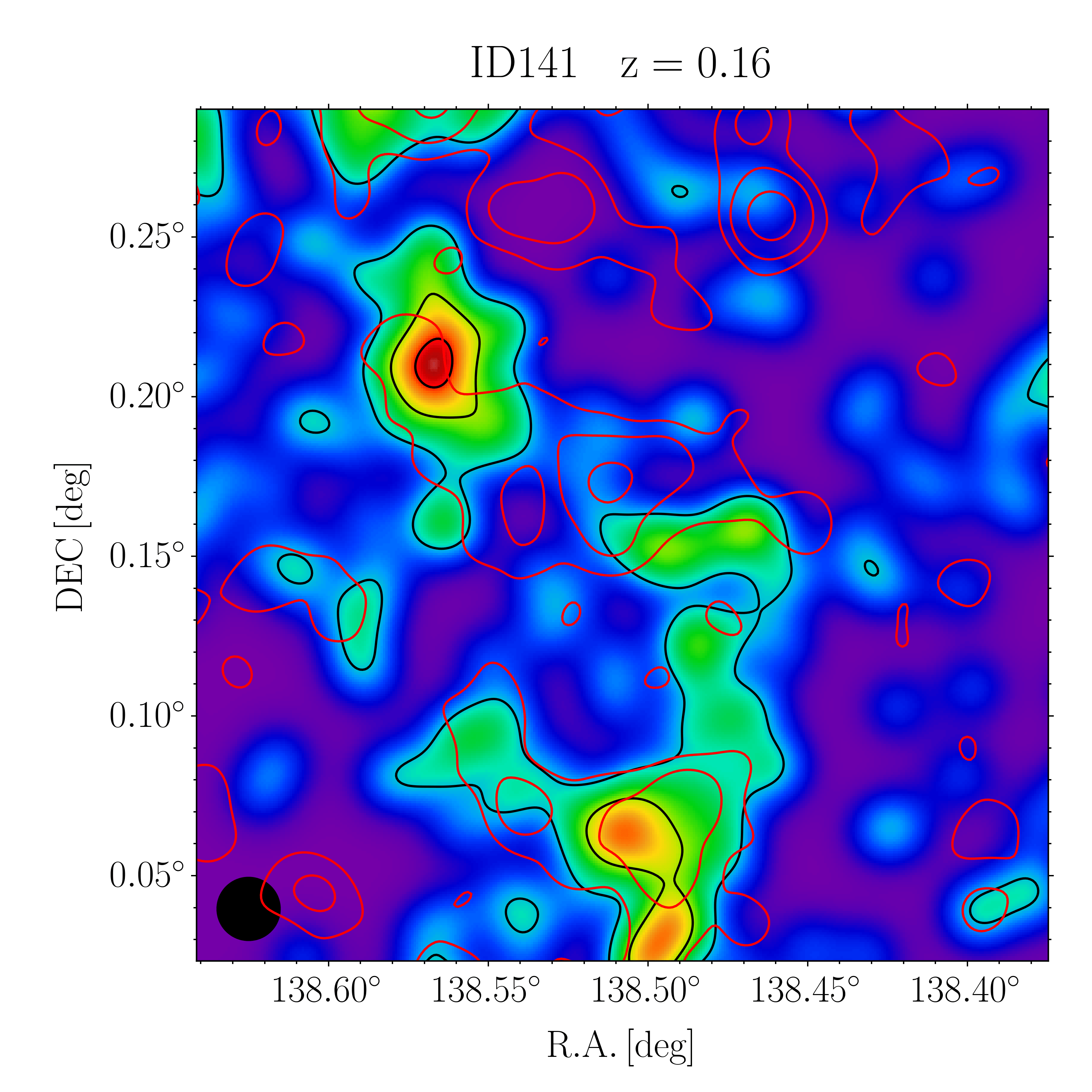}
     \includegraphics[width=0.24\textwidth]{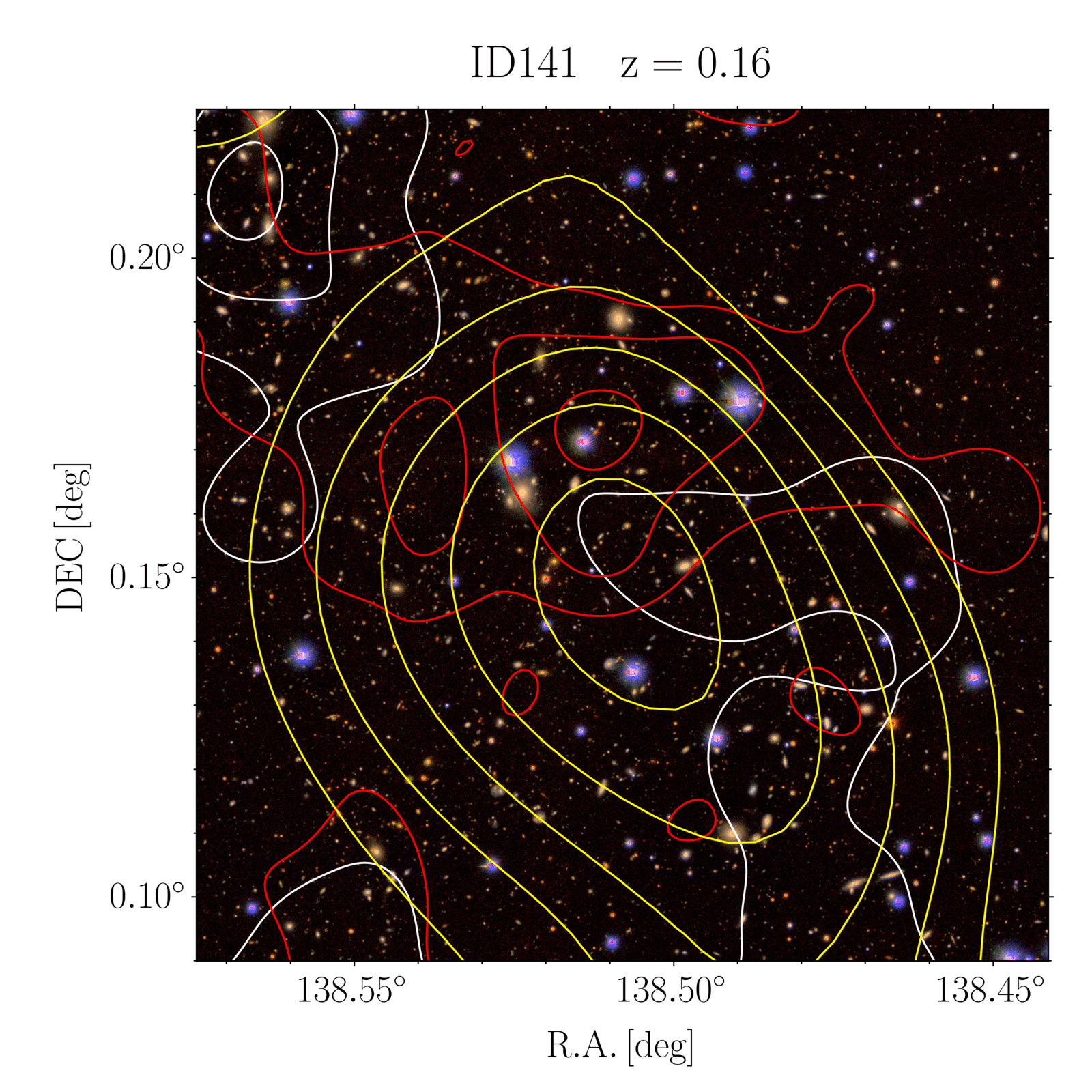}
     \hspace{0.4cm}            
     \includegraphics[width=0.24\textwidth]{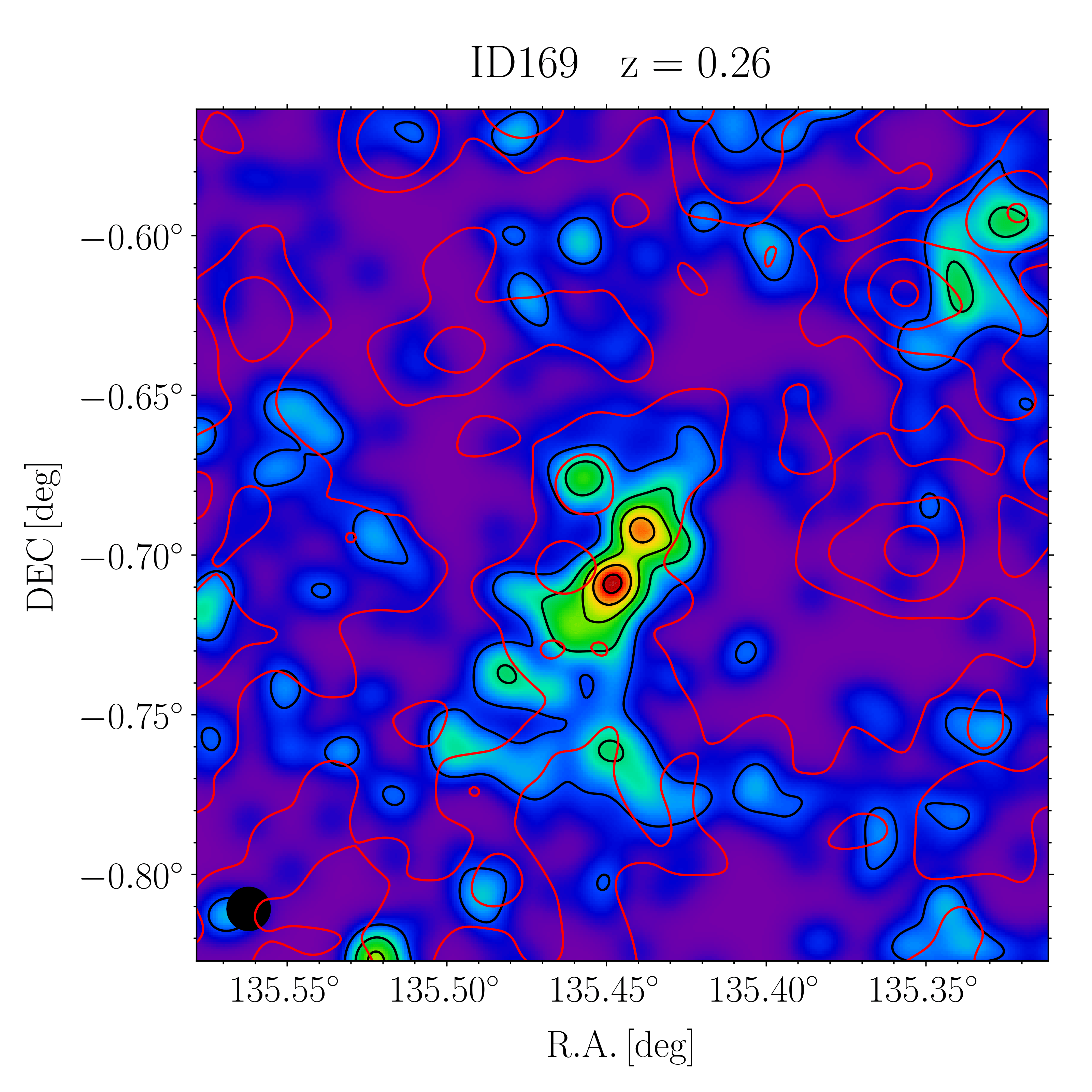}
     \includegraphics[width=0.24\textwidth]{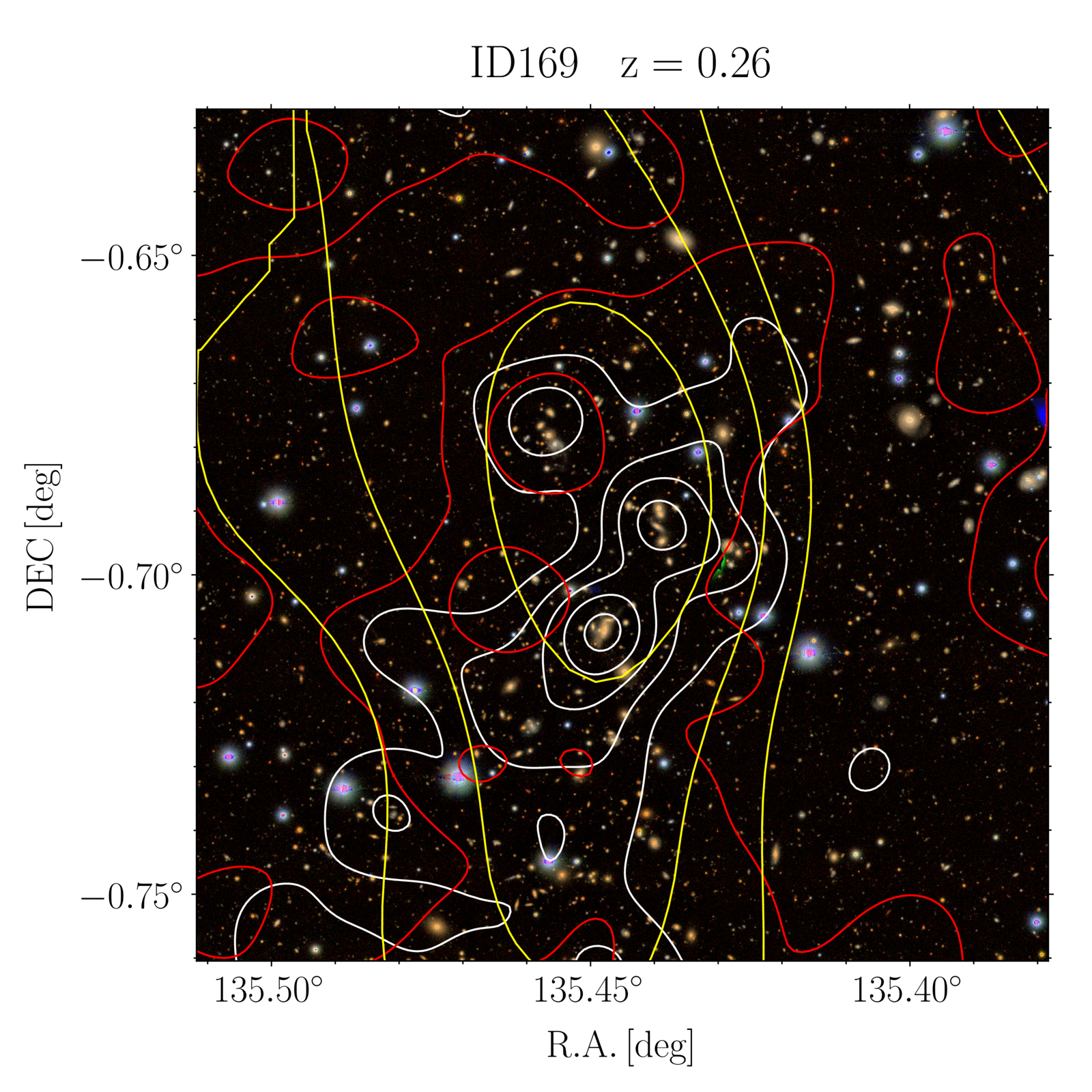}
                 
     \includegraphics[width=0.24\textwidth]{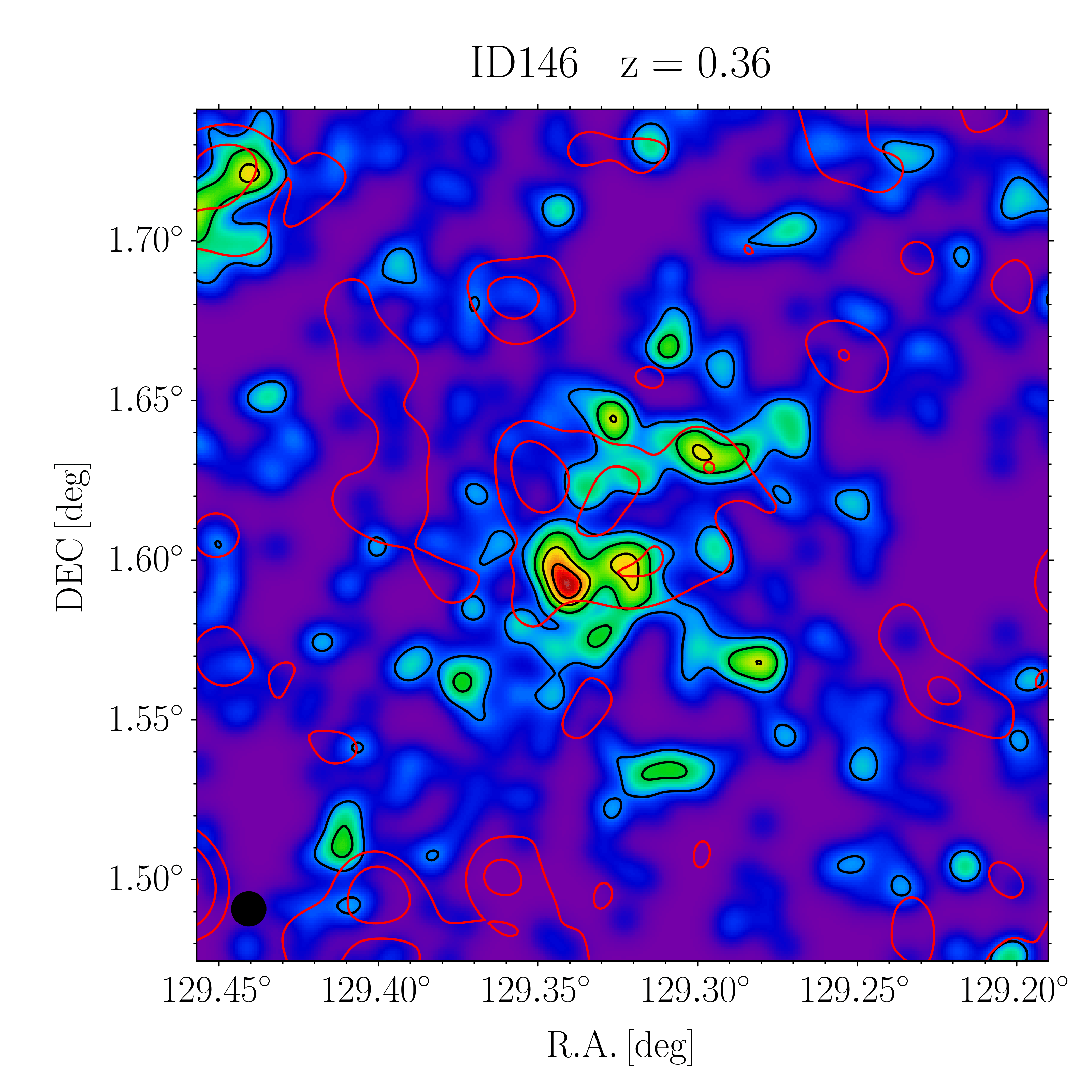}
     \includegraphics[width=0.24\textwidth]{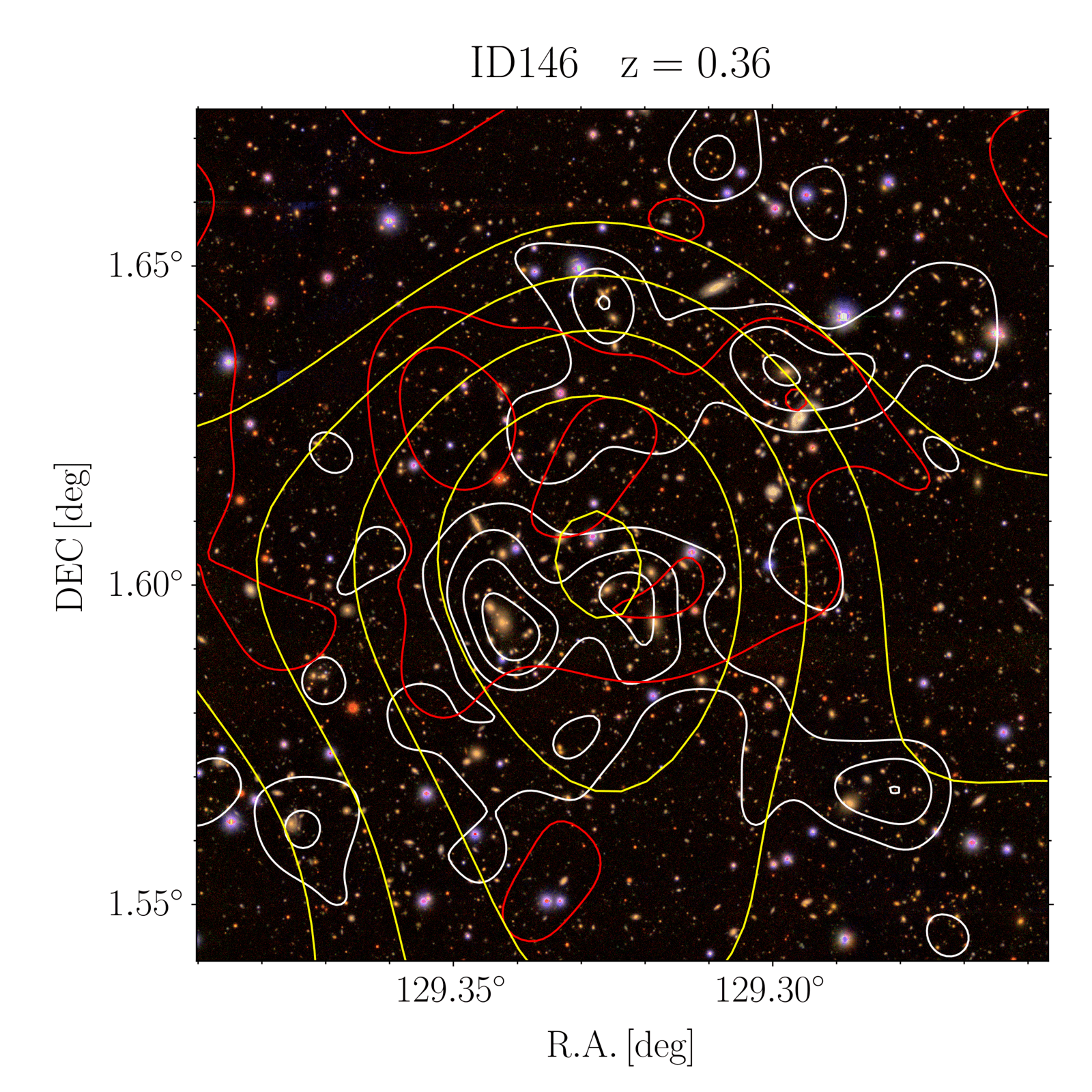}
     \hspace{0.4cm}             
     \includegraphics[width=0.24\textwidth]{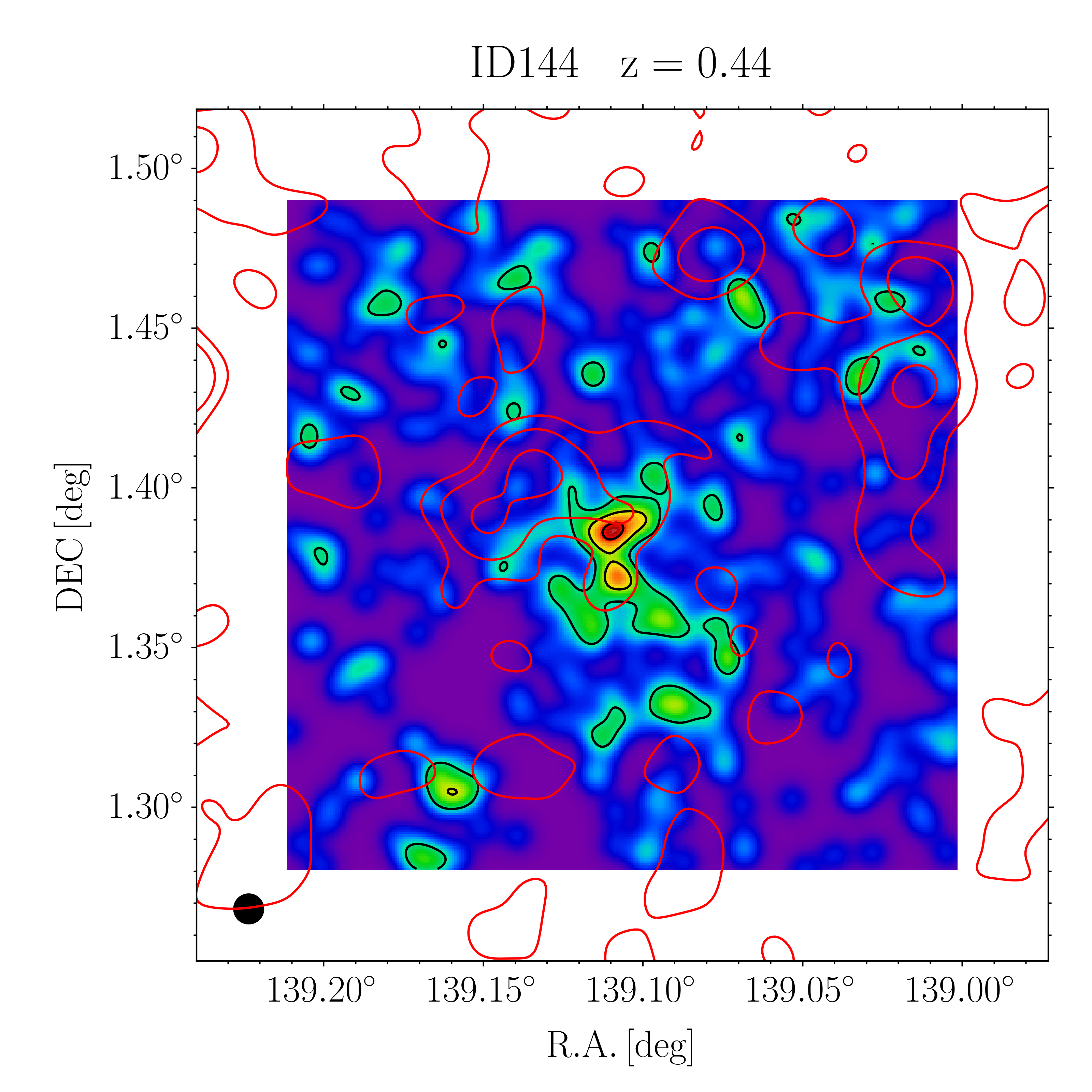}
     \includegraphics[width=0.24\textwidth]{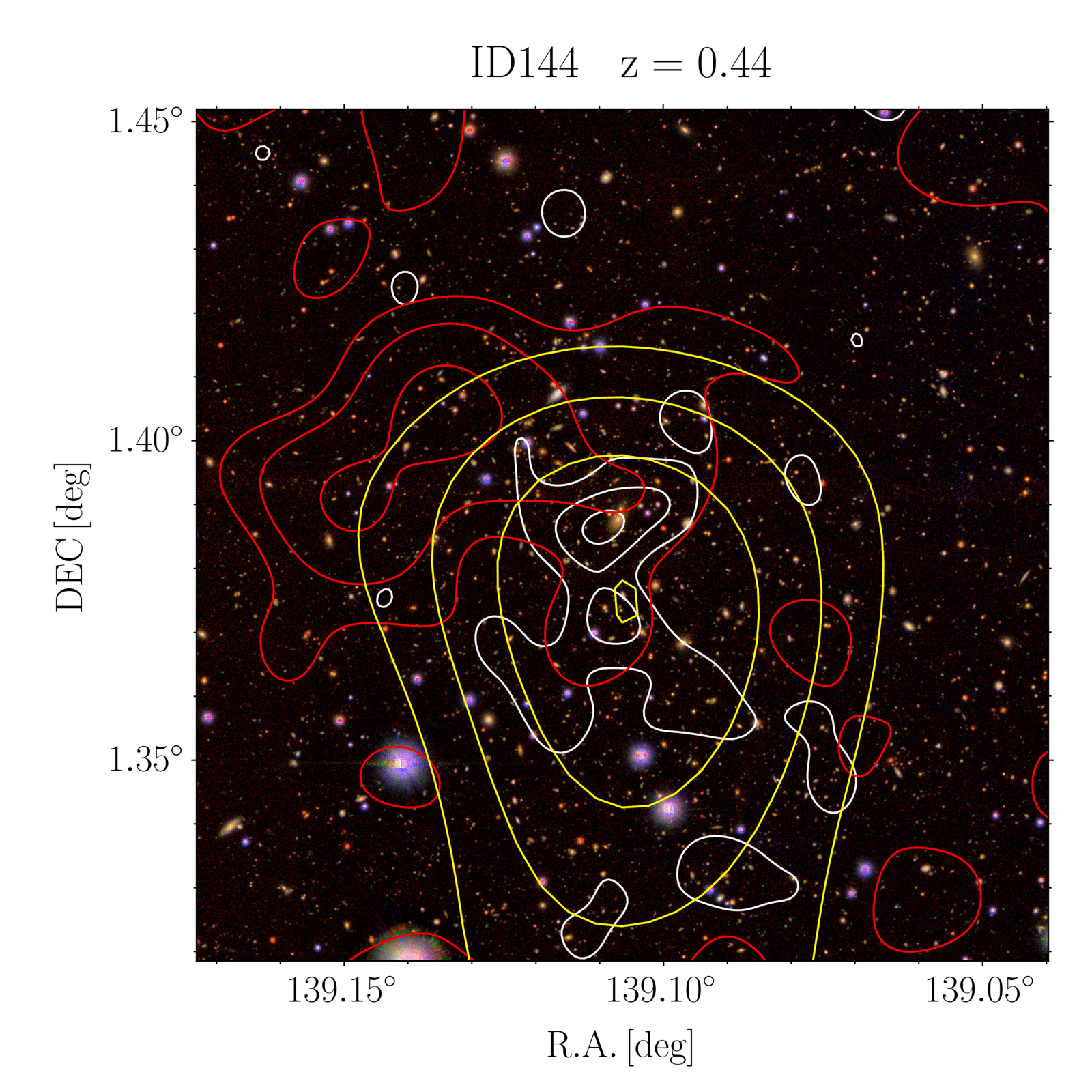}
     \caption{{\it Left}: Galaxy density maps ($16\times16$ arcmin) centred on the HSC shear-selected cluster positions without an eFEDS match. Overlaid in red are X-ray contours, which were obtained by smoothing the raw X-ray image in the $0.5-2.0$~keV energy band with a Gaussian of $24$~arcsec. The black circles in the lower-left corners show the smoothing scale, FWHM\ =\ $200$~kpc. {\it Right}: HSC-SSP optical images centred on the HSC shear-selected cluster positions. The $8\times8$ arcmin optical images of the central region are created using the $z$, $i$, and $r$ bands. X-ray contours are shown in red, galaxy density contours in white (they are the same contours as the black ones in the corresponding galaxy density maps), and weak-lensing mass contours in yellow.}
     \label{fig:nomatchesxrayoptical}
 \end{figure*}


\subsection{Notes on clusters without eFEDS counterpart}
\label{sect:nomatches}

As found in section~\ref{sect:catsmatchs}, there are four HSC shear-selected clusters without a counterpart in the eFEDS cluster catalogue. Figure~\ref{fig:nomatchesxrayoptical} shows the galaxy map distribution and optical images of these systems. In each case, the galaxy maps are smoothed by a Gaussian of FWHM$=200$~kpc. The red contours are obtained from the smoothed (Gaussian of $24$~arcsec) raw X-ray image ($0.5-2.0$~keV energy band). The HSC-SSP optical images ($z$, $i$, and $r$ bands) are also centred on the HSC shear-selected cluster positions and have a size of $8\times8$ arcmin. These images have overlaid the following: X-ray contours are shown in red, galaxy density contours in white and weak-lensing mass contours in yellow. The X-ray images show that none of these clusters have significant extended X-ray emission. Here, we discuss each of these clusters in detail. In each case, the cluster is identified by its ID as in \cite{Oguri2021}.

\subsubsection{ID 141}

\citet{Oguri2021} found that the closest optical cluster to this weak-lensing peak is a CAMIRA cluster located $0.93$~Mpc ($\sim 5.5$~arcmin) away at $z=0.160$ with richness $N=22.9$. The lack of a clear BCG in the optical image in Fig.~\ref{fig:nomatchesxrayoptical} of this cluster (top panel) confirms that the corresponding optical counterpart is far away from the peak position. Furthermore, the galaxy distribution analysis (see section~\ref{sect:peakfinding}) reveals that there are two prominent peaks located more than $\sim 6$~arcmin away from the shear-selected cluster position, but none around the shear-selected cluster. Most likely, this is a spurious shear-selected cluster.

\subsubsection{ID 144}

This shear-selected cluster has a WHL15 counterpart $\sim 50$~kpc away at $z=0.439$. The concentration of galaxies shown in Fig.~\ref{fig:nomatchesxrayoptical} is consistent with the small spatial offset. The cluster is also detected by CAMIRA with richness $N=16.7$ and redshift $z=0.453$. There is another CAMIRA cluster with richness $N=31.5$ and redshift $z=0.80$ at $\sim 1$~arcmin away from the mass map peak. This is likely to be a real cluster at $z=0.44$ but possibly with an enhancement of the weak lensing signal due to the line-of-sight structure. The faint X-ray signal may be explained by the high redshift of the cluster as well as the projection effect.

\subsubsection{ID 146}

This shear-selected cluster has a CAMIRA counterpart $\sim 260$~kpc away at $z=0.356$.
Fig.~\ref{fig:nomatchesxrayoptical} shows a complex but significant concentration of galaxies. The high richness of this cluster, $N=38.1$, implies that this is a sufficiently massive cluster that is capable of producing a mass map peak. Indeed the X-ray flux is detected by our forced measurement at more than $2\sigma$ level at the shear cluster position, even though this cluster is not included in the eFEDS X-ray cluster catalogue. 

\subsubsection{ID 169}

This shear-selected cluster has a WHL15 counterpart $218$~kpc away at $z=0.255$ with relatively high WHL15 richness of $R_{L_{\ast},500}=35.2$. Fig.~\ref{fig:nomatchesxrayoptical} shows some concentration of galaxies that is consistent with the mass map peak position.


\section{Discussion}
\label{sect:discussion}


\subsection{Comparison with previous studies}

As mentioned in section~\ref{sect:intro}, \cite{Giles2015} investigated the $L_{\rm bol}-M$ relation of a small sample of shear-selected clusters. They obtained X-ray properties for $10$ low S/N shear-selected clusters in the redshift range of $0.13<z<0.27$ and compared them with the X-ray selected sampled of \cite{Mahdavi2013}. \cite{Mahdavi2013} analysed X-ray and optical data of $50$ X-ray selected clusters in the redshift range of $0.15<z<0.55$. Both samples can be compared with our work because they both extracted X-ray properties within $r_{500}^{\rm WL}$. Of special interest for the analysis presented here is the sample of \cite{Giles2015}, since their weak-lensing masses can also be bias corrected\footnote{The \cite{Chen2020} bias correction for this sample is obtained by using a S/N threshold of $3.69$ and an average surface density of source galaxies of $30$~arcmin$^{-2}$.} using the adjustment presented in \cite{Chen2020}. For reference, Fig.~\ref{fig:lbolrwL_Mobs} shows the location of the \cite{Giles2015} and \cite{Mahdavi2013} samples (green and grey pentagons, respectively) in the $L_{\rm bol}-M$ plane in comparison with the eFEDS and the HSC shear-selected samples. Since the cosmological parameters adopted in this work and the \cite{Mahdavi2013} and \cite{Giles2015} studies are the same, we can compare them directly.

First of all, one can see that the \cite{Chen2020} bias correction on the \cite{Giles2015} sample brings this sample in agreement with eFEDS and with the HSC weak-lensing peaks. We do not perform a fit on this sample because we do not know how much the luminosity is effected by the mass bias correction, that is the \cite{Chen2020} correction reduces the weak-lensing masses of the clusters, leading to a smaller radius, and therefore smaller luminosities. In this sense, the \cite{Giles2015} luminosities in Fig.~\ref{fig:lbolrwL_Mobs} can be considered as upper limits. Second, the masses in \cite{Mahdavi2013} are larger than the ones in eFEDS and HSC shear-selected cluster samples. As in section~\ref{sect:lmrelation}, we fitted the \cite{Mahdavi2013} sample using Eq.~\ref{eq:lmrelation}. We obtain a slope of $B_{LM}^{\rm M13}=2.35\pm0.89$, which is steeper than that of the eFEDS clusters.

\cite{Giles2015} also studied the dynamical state of their weak-lensing peaks. They found that the majority of their clusters appear unrelaxed, and one third of them seem to host a cool-core. As shown in section ~\ref{sect:dynamicalstate}, the HSC shear-selected clusters studied in this work have X-ray morphological properties that do not prefer a relaxed or unrelaxed state. This was also found for the eFEDS clusters. Using the peak finding method of galaxy distribution, $60\%$ of the HSC shear-selected clusters appear to be in a merging state, although we show that this might be a statistical fluctuation due to the small number of clusters in the sample.

\subsection{No X-ray underluminous shear-selected clusters}

In section~\ref{sect:catsmatchs}, we have shown that four shear-selected clusters do not have an eFEDS counterpart. Their galaxy distribution map (see section~\ref{sect:nomatches}) revealed that one of them (ID $141$) is the result of some galaxy conglomerates being close in projection and therefore giving rise to a false weak-lensing peak. As shown in Fig.~\ref{fig:nomatchesxrayoptical}, the shear-selected clusters without an eFEDS counterparts do not show any sign of extended X-ray emission. The nearest, in projection, point-like sources to these weak-lensing peak positions are located between $20$ and $80$~arcsec, and have fluxes lower than $9\times10^{-15}$~erg~s$^{-1}$~cm$^{-2}$ in the $0.5-2$~keV energy band \citep{LiuT2021a,Salvato2021}. These point-like sources are removed from the X-ray analysis as explained in section~\ref{sect:xray_obs}.

\cite{LiuA2021} described the eFEDS selection function from extensive simulations, which took into account the instrument response and a realistic emission of galaxy clusters and Active Galactic Nuclei (AGN). Using training classifiers, \cite{LiuA2021} found that the cluster (true) $0.5-2.0$~keV energy band X-ray counts, its flux in this soft band, the redshift, mass and luminosity are the most discriminating cluster features when constructing the selection function of eFEDS. We have a look at the detection probability of the non-matched shear-selected clusters based on their soft-band luminosity and fluxes.

We found that based on the soft-band luminosity, clusters ID $141$, $144$ and $169$ have a probability $\ll 1$\footnote{Probability of $1$ means that a cluster has a $100\%$ probability of being detected. See Fig.~5 in \cite{LiuA2021}.} of being detected by the pipeline used in eFEDS. Their fluxes, which are in the range of $0.4-1.5\times10^{-15}$~erg~s$^{-1}$~cm$^{-2}$ in the $0.5-2.0$~keV energy band, are also below the flux limit of $1.5\times10^{-15}$~erg~s$^{-1}$~cm$^{-2}$ where eFEDS has $\sim 40\%$ percent completeness \citep[see][]{LiuA2021}. Therefore it is not surprising that such clusters are not detected in the eFEDS given the shallow observations.

Cluster ID $146$ has a probability of $\sim0.1$ of being detected in eFEDS given its soft-band luminosity. It has a flux of $\sim 3.2\times10^{-14}$~erg~s$^{-1}$~cm$^{-2}$ in the $0.5-2.0$~keV energy band and within $r_{500}^{\textrm WL}$. At this flux, the eFEDS sample has a completeness of $65\%$. Therefore, the absence of this cluster in the eFEDS cluster catalogue is again not surprising.

We calculated the significance of these four non-matched clusters to check if they are consistent with the tail of the $L_{\rm bol}-M$ scaling relation. We found that these clusters are not statistically different ($\lesssim2.5\sigma$) to the eFEDS sample, that is their presence is consistent with the log-normal distribution around the mean eFEDS scaling relation taking into account its scatter. These clusters can be explained as a stochastic fluctuation at the low luminosity end of the $L_{\rm bol}-M$ relation at a given mass. With deeper eROSITA (pointed) observations, we would be able to detect them.

X-ray underluminous clusters have been a subject of study in the last years \citep[e.g.][]{Dietrich2009,Castellano2011,Andreon2011,Trejo2014,Andreon2019}. Some of these works focus only on one or a few clusters that have not been detected in previous X-ray observations. The main conclusions of such works are that X-ray underluminous clusters have not yet formed, which translate in a low X-ray luminosity, or they are located in filaments, which, when seen along the line of sight, would look like massive clusters. Our results indicate that only one of the non-matched shear-selected clusters is a spurious weak-lensing peak, while the other three, given the available X-ray observation and X-ray derived properties, have a low probability of being detected. Therefore we do not find any clear examples of X-ray underluminous clusters in our shear-selected cluster sub-sample.

In addition, if the X-ray underluminous clusters constitute a significant fraction of clusters in such a way that they are not well captured by the standard unimodal, log-normal X-ray luminosity-mass relation, they would cause a discrepancy of scaling relations derived from X-ray- and shear-selected cluster samples even after correcting for the selection biases assuming the unimodal, log-normal distribution. The consistency of the bias-corrected scaling relations between these cluster samples found in this paper indicates that such assumption on the shape of the X-ray luminosity-mass relation is valid and there is no significant population of such X-ray underluminous clusters.


\section{Summary and conclusions}
\label{sect:conclusions}

We have presented a complete census of X-ray properties of shear-selected clusters. We have achieved this by comparing the HSC weak-lensing shear-selected clusters with the eFEDS X-ray selected cluster sample. Both samples share $\sim90$~deg$^2$ of a common area.  This study is one of the first of its kind comparing X-ray and shear-selected cluster samples over the same region in the sky.

We have found that $21$ out of $25$ shear-selected clusters have an eFEDS counterpart. The physical separation between them is between $44$ and $531$~kpc, which correspond to $\sim0.05r_{500}^{\rm WL}$ and $\sim0.5r_{500}^{\rm WL}$, respectively. Their redshifts are consistent at the $5$\% level. We have found that the scaling relation between X-ray bolometric luminosity and true mass of shear-selected clusters is consistent with the eFEDS X-ray selected clusters. This is achieved once the weak-lensing mass of the shear-selected clusters is corrected by the Eddington mass bias as quantified in \citet{Chen2020} (by approximately $20\%$). Similarly, the X-ray soft-band luminosity and true mass scaling relation is consistent between the shear-selected clusters and the eFEDS sample. Furthermore, the results are also comparable to the results of the XXL survey.

Different X-ray morphological indicators (e.g. cuspiness, ellipticity, power ratios, etc.) show that the shear-selected clusters do not have a preferable dynamical state compared with X-ray selected clusters. The peak-finding method of galaxy distribution shows that the $60\%$ of shear-selected cluster located in the eFEDS footprint appear to be in a merging stage. However, once we compare the full HSC shear-selected cluster sample, we do not find a significant difference in the merger fraction between X-ray and shear-selected samples. We have also found that $43\%$ of the shear-selected clusters lie in super-clusters detected in the eFEDS field.

The galaxy distribution of one of the four shear-selected clusters without an eFEDS counterpart reveals that such a weak-lensing peak is likely a spurious shear-selected cluster. Two other shear-selected clusters have X-ray fluxes way below the flux limit where the eFEDS sample has $40\%$ completeness. The last shear-selected cluster without an eFEDS counterpart has a flux where the eFEDS sample is $65\%$ complete. 

Overall, we have found a good consistency of scaling relations and the dynamical state between shear-selected and X-ray selected cluster samples. Our results indicate that there is no significant population of X-ray underluminous clusters that were previously advocated. Such X-ray underluminous clusters, if sufficiently abundant, pose a challenge for using the abundance of X-ray clusters as a cosmological probe. The absence of significant population of X-ray underluminous clusters suggests that X-ray cluster samples can be regarded as nearly complete and therefore can be used as an accurate cosmological probe.

\vspace{0.5cm}
\begin{acknowledgements}
The authors thank the referee for the useful comments on the manuscript.\\
This work was supported in part by the World Premier International Research Center Initiative (WPI Initiative), MEXT, Japan, and JSPS KAKENHI Grant Nos. JP19KK0076, JP18K03693, JP20H00181, JP20H05856, JP16KK0101.
\\
This work made use of the Gravity Supercomputer at the Department of Astronomy, Shanghai Jiao Tong University.
\\
TS acknowledges support from the German Federal Ministry for Economic Affairs and Energy (BMWi) provided through DLR via project 50OR1803.
\\
We thank the HSC-XXL collaboration for letting us use their $L_{\rm X}-M$ relation (in Fig.~\ref{fig:lXrwL_Mobs}) in advance of publication.
\\
This work is based on data from eROSITA, the soft X-ray instrument aboard SRG, a joint Russian-German science mission supported by the Russian Space Agency (Roskosmos), in the interests of the Russian Academy of Sciences represented by its Space Research Institute (IKI), and the Deutsches Zentrum f{\"{u}}r Luft und Raumfahrt (DLR). The SRG spacecraft was built by Lavochkin Association (NPOL) and its subcontractors, and is operated by NPOL with support from the Max Planck Institute for Extraterrestrial Physics (MPE).
\\
The development and construction of the eROSITA X-ray instrument was led by MPE, with contributions from the Dr. Karl Remeis Observatory Bamberg \& ECAP (FAU Erlangen-Nuernberg), the University of Hamburg Observatory, the Leibniz Institute for Astrophysics Potsdam (AIP), and the Institute for Astronomy and Astrophysics of the University of T{\"{u}}bingen, with the support of DLR and the Max Planck Society. The Argelander Institute for Astronomy of the University of Bonn and the Ludwig Maximilians Universit{\"{a}}t Munich also participated in the science preparation for eROSITA.
\\
The eROSITA data shown here were processed using the \texttt{eSASS/NRTA} software system developed by the German eROSITA consortium.
\\
The Hyper Suprime-Cam (HSC) collaboration includes the astronomical communities of Japan and Taiwan, and Princeton University. The HSC instrumentation and software were developed by the National Astronomical Observatory of Japan (NAOJ), the Kavli Institute for the Physics and Mathematics of the Universe (Kavli IPMU), the University of Tokyo, the High Energy Accelerator Research Organization (KEK), the Academia Sinica Institute for Astronomy and Astrophysics in Taiwan (ASIAA), and Princeton University.  Funding was contributed by the FIRST program from the Japanese Cabinet Office, the Ministry of Education, Culture, Sports, Science and Technology (MEXT), the Japan Society for the Promotion of Science (JSPS), Japan Science and Technology Agency  (JST), the Toray Science  Foundation, NAOJ, Kavli IPMU, KEK, ASIAA, and Princeton University.
\end{acknowledgements}


\bibliographystyle{aa}
\bibliography{references}


\appendix
\onecolumn


\section{Matching to other shear-selected cluster catalogues}
\label{app:A}

In section~\ref{sect:shearclu} we briefly described the methodology used in \cite{Oguri2021} to construct shear-selected galaxy cluster samples using the HSC-SSP S19A data. \cite{Oguri2021} obtained three cluster catalogues, which were the result of applying different shape filters. In the main part of the paper, the clusters selected using a truncated Gaussian filter (the TG15 catalogue) that lie in the eFEDS footprint were studied since a mass bias correction derived in \cite{Chen2020} can be applied. Since the other two catalogues, TI05 and TI20 were constructed with a different technique, the same mass bias correction cannot be applied to these two catalogues. Nevertheless, here we cross-match these two catalogues with the eFEDS cluster catalogue and show the results in this section. For completeness, we also consider the work of \citet[][hereafter Ha20]{Hamana2020}, who also constructed a weak-lensing cluster catalogue, but using the HSC-SSP first-year data (S16A). They obtained a sample of 124 shear-selected clusters with S/N larger than $5$. The technique used considers the dilution effect of cluster-member and foreground galaxies on weak-lensing signals from galaxy clusters.

\begin{figure*}[h!]
     \centering
     \includegraphics[trim=20 10 20 50,clip,width=0.85\textwidth]{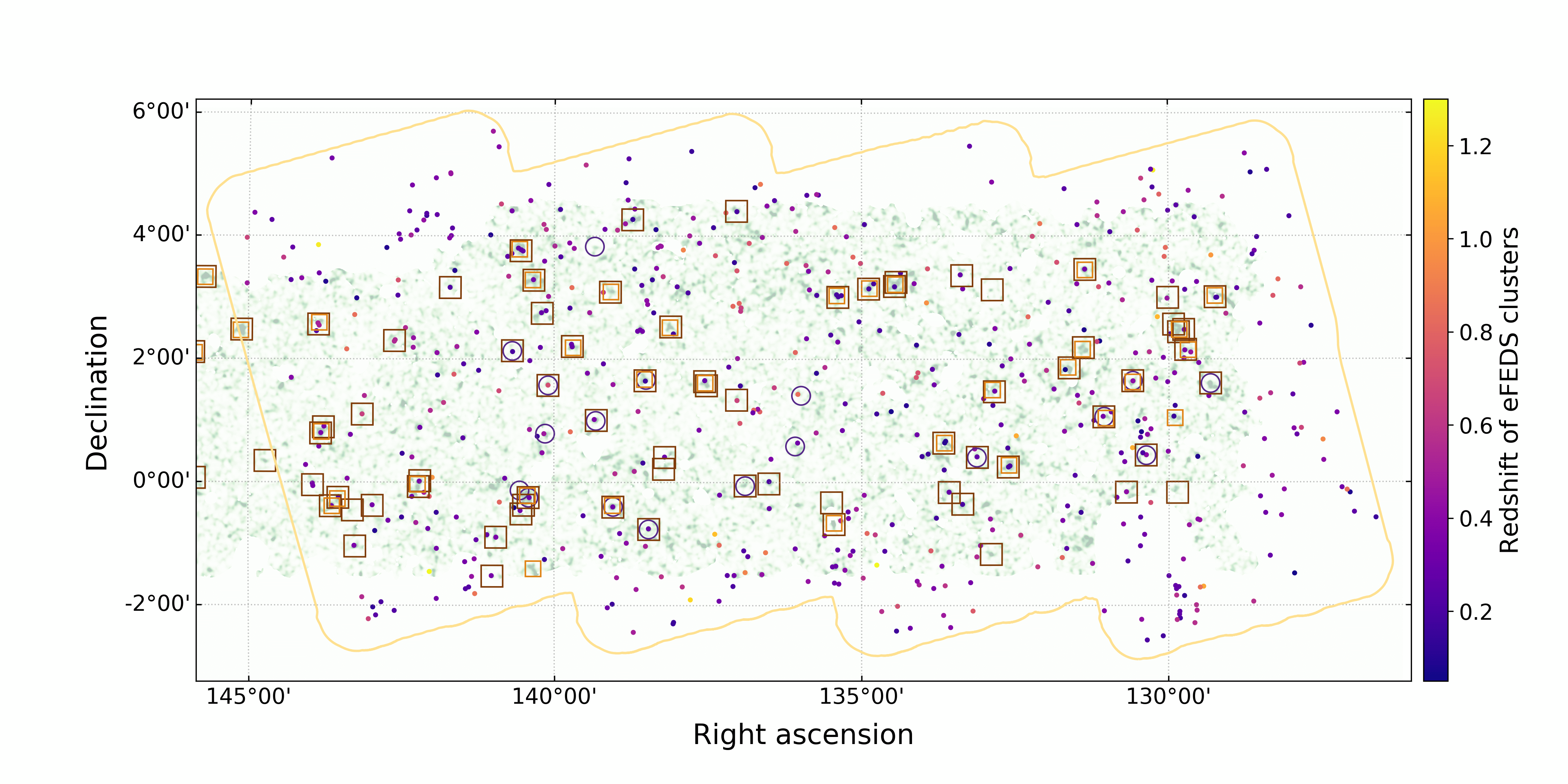}
     \caption{Part of the weak lensing map for the GAMA09H HSC-SSP S19A field (green background). The solid yellow line shows the eFEDS survey footprint. Small filled circles show the location of eFEDS clusters with $z>0.05$, whose colour correspond to their redshift as shown by the colour bar. Open symbols show the position of the shear-selected clusters: purple circles correspond to the \cite{Hamana2020} catalogue, orange and brown squares are weak-lensing peaks from the TI20 and TI05 catalogues, respectively \citep{Oguri2021}.}
     \label{fig:hscefedsfootprintallshear}
 \end{figure*}

Table~\ref{tab:diffcats_matching} shows the total number of shear-selected clusters in the HSC-SSP GAMA09H patch for the TI05, TI20 and Ha20 catalogues, as well as the number of such clusters that lie within the common area with the eFEDS survey. Figure~\ref{fig:hscefedsfootprintallshear} shows the location of the shear-selected clusters in these catalogues along with the position of the X-ray detected clusters in eFEDS. Using the same criteria as in section~\ref{sect:catsmatchs} ($\Delta z<0.1$ within a matching radius of $5$~arcmin), these catalogues were cross-matched with the eFEDS cluster catalogue. These three catalogues obtain between $\sim 48-62\%$ of unique matches with the eFEDS sample, while the TG15 catalogue has a unique match percentage of $\sim 68\%$. In an opposite fashion, the TI05, TI20 and Ha20 catalogues have $\sim30\%$ of their clusters without an eFEDS counterpart, while the TG15 has only $16\%$ its clusters without a match. The three catalogues constructed in \cite{Oguri2021} show that $\sim15-22\%$ of the clusters have multiple eFEDS counterparts, while for the Ha20 catalogue these are only $5\%$. The three shear-selected cluster catalogues in \cite{Oguri2021} are constructed using different background galaxies, for example, there are more clusters at higher redshifts ($z>0.6$) in TI05 than in TG15 and TI20. This might explain why we find more clusters without an eFEDS counterpart in this catalogue.

Finally, the weak-lensing cluster catalogues in \cite{Oguri2021} show that $30-43\%$ of their clusters are located in super-clusters. Only $\sim16\%$ of the Ha20 clusters are located in super-clusters, but this percentage is difficult to compare with those obtained in the other catalogues given that the used data are different. 

\begin{table*}[h!]
\centering
\caption{Number and percentages (in parentheses) of shear-selected clusters in the HSC-SSP GAMA09H area from three catalogues.}
{\small
\begin{tabular}[width=\columnwidth]{l c c c c c c}
\toprule
 Catalogue & HSC-SSP GAMA09H & Unique matches & Primary matches & Non-matched & In superclusters & Outside the eFEDS area \\
\midrule
TI05 & 86 & 37~($43$) & 11~($13$) & 22~($25$) & 21~($24$) & 16~($19$) \\ 
TI20 & 39 & 15~($38$) & ~~7~($18$) &  ~~9~($23$) & 12~($31$) & ~~8~($21$)\\ 
Ha20 & 18 & 11~($61$) &  ~~1~($6$) &  ~~6~($33$) & ~~3~($17$) & ~0~(0) \\ 
\toprule
\end{tabular}
\tablefoot{TI05 and TI20 catalogues are from \cite{Oguri2021} using the S19A data release, and the catalogue from \citet[][Ha20]{Hamana2020} uses the S16A data release. These catalogues are matched with eFEDS clusters.}
\label{tab:diffcats_matching}
}
\end{table*}

\onecolumn

\section{Images of matched shear-selected clusters}
\label{app:B}

As found and discussed in section~\ref{sect:catsmatchs}, there are $21$ clusters in the shear-selected sample in the eFEDS footprint that have a corresponding eFEDS cluster associated with them. Figures~\ref{fig:uniquematchesxrayoptical}~and~\ref{fig:primarymatchesxrayoptical} show the galaxy map distribution and optical images of these systems. In each case, the galaxy maps have been smoothed by a Gaussian of FWHM$=200$~kpc. The red contours are obtained from the smoothed (Gaussian of $24$~arcsec) raw X-ray image ($0.5-2.0$~keV energy band). The HSC-SSP optical images ($z$, $i$ and $r$ bands) are also centred on the HSC shear-selected cluster positions and have a size of $8\times8$ arcmin. These images have overlaid the following: X-ray contours are shown in red, galaxy density contours in white and weak-lensing mass contours in yellow.

 \begin{figure*}[ht]
     \captionsetup[subfigure]{labelformat=empty}
     \centering
     \subfloat[][]{\includegraphics[width=0.24\columnwidth]{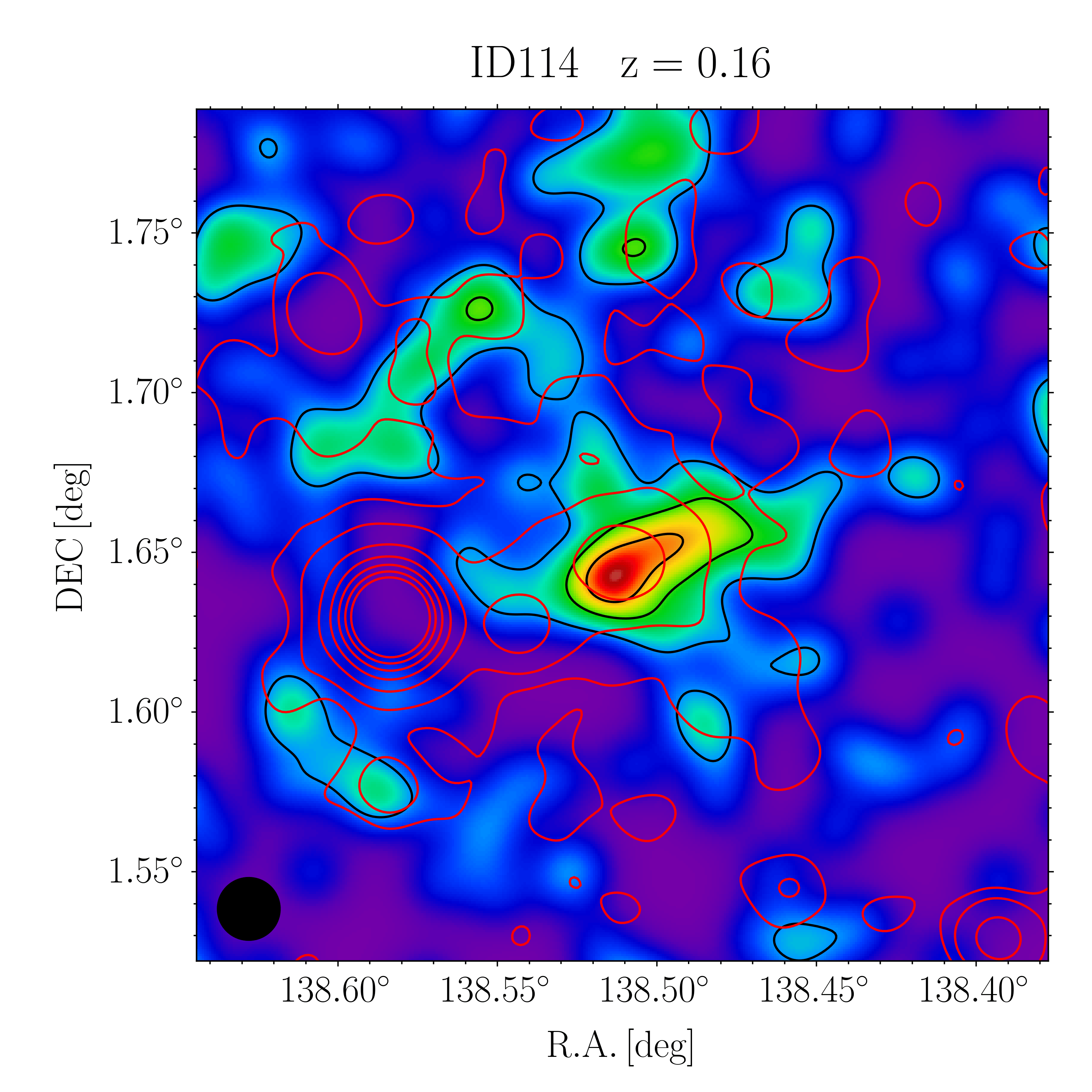}}
     \subfloat[][]{\includegraphics[width=0.24\columnwidth]{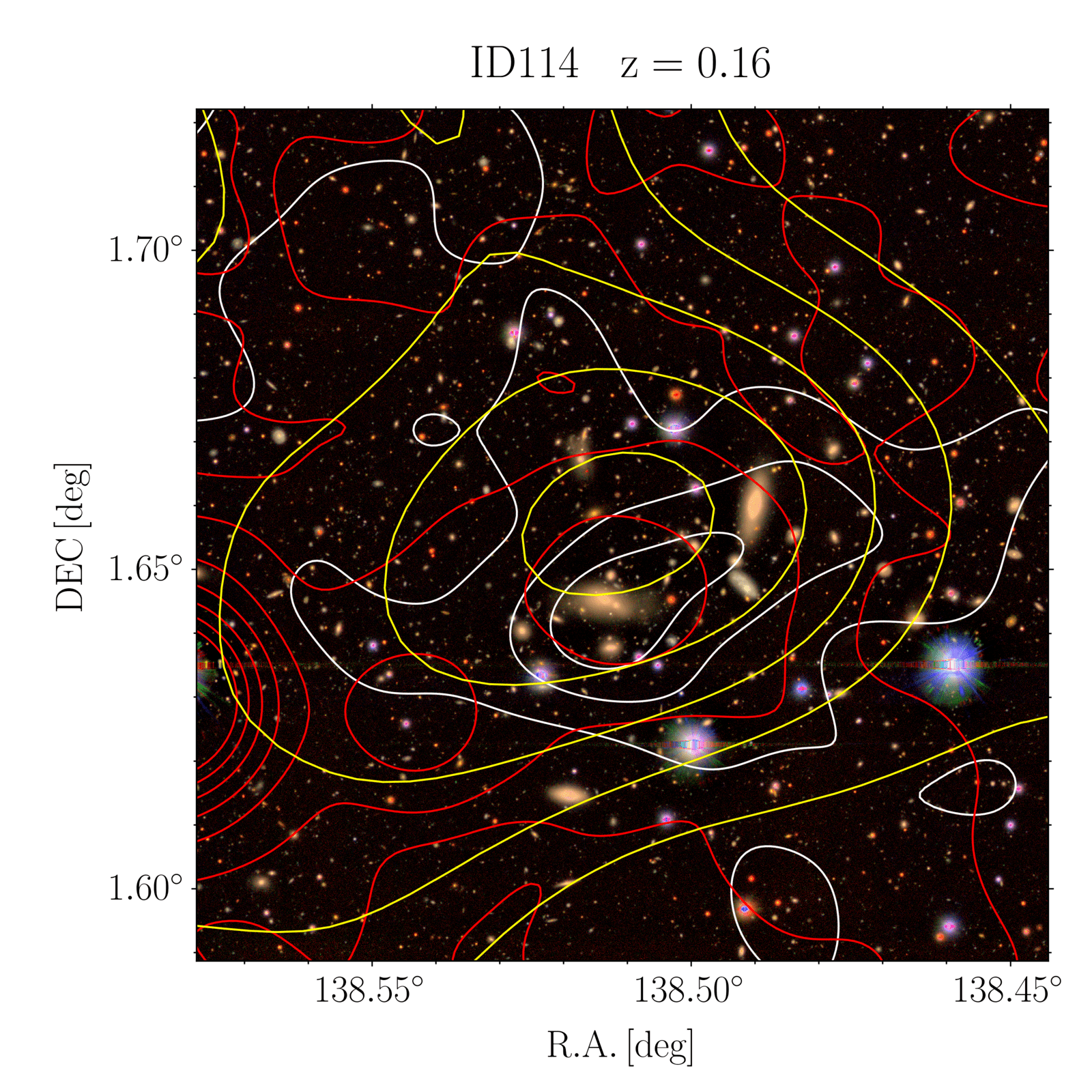}}
     \hspace{0.4cm}
    \subfloat[][]{\includegraphics[width=0.24\columnwidth]{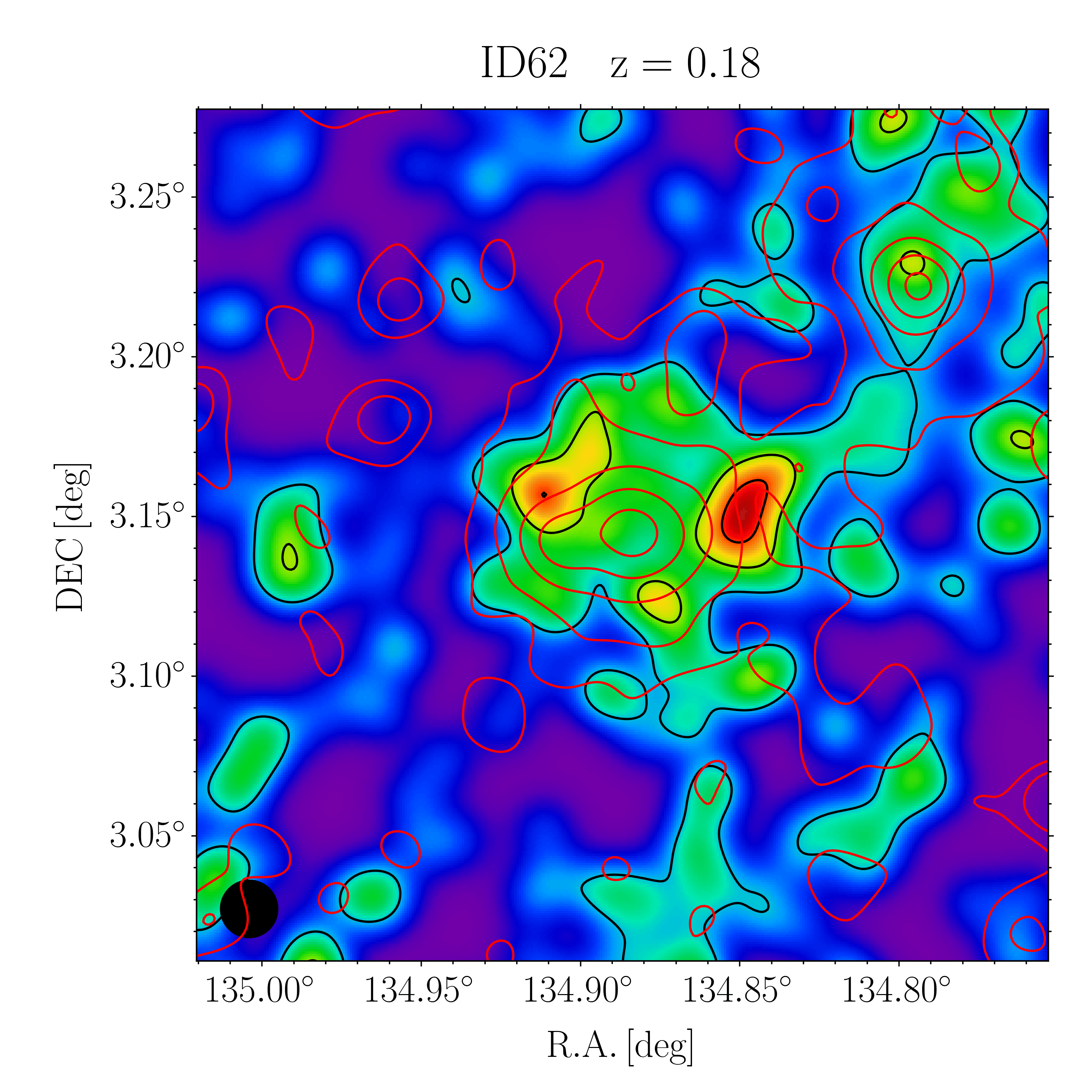}}
     \subfloat[][]{\includegraphics[width=0.24\columnwidth]{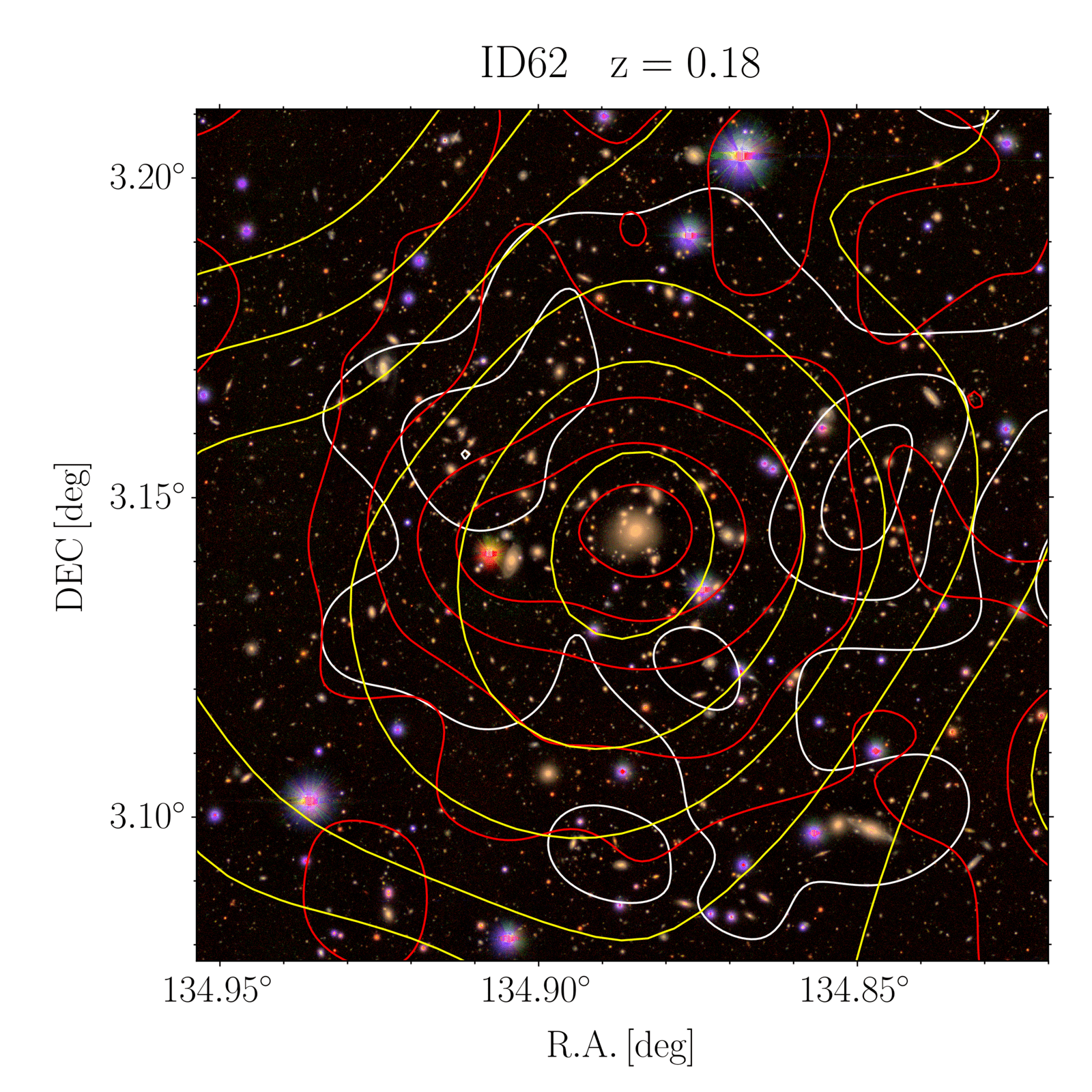}}\\[-5ex]
     
     \subfloat[][]{\includegraphics[width=0.24\columnwidth]{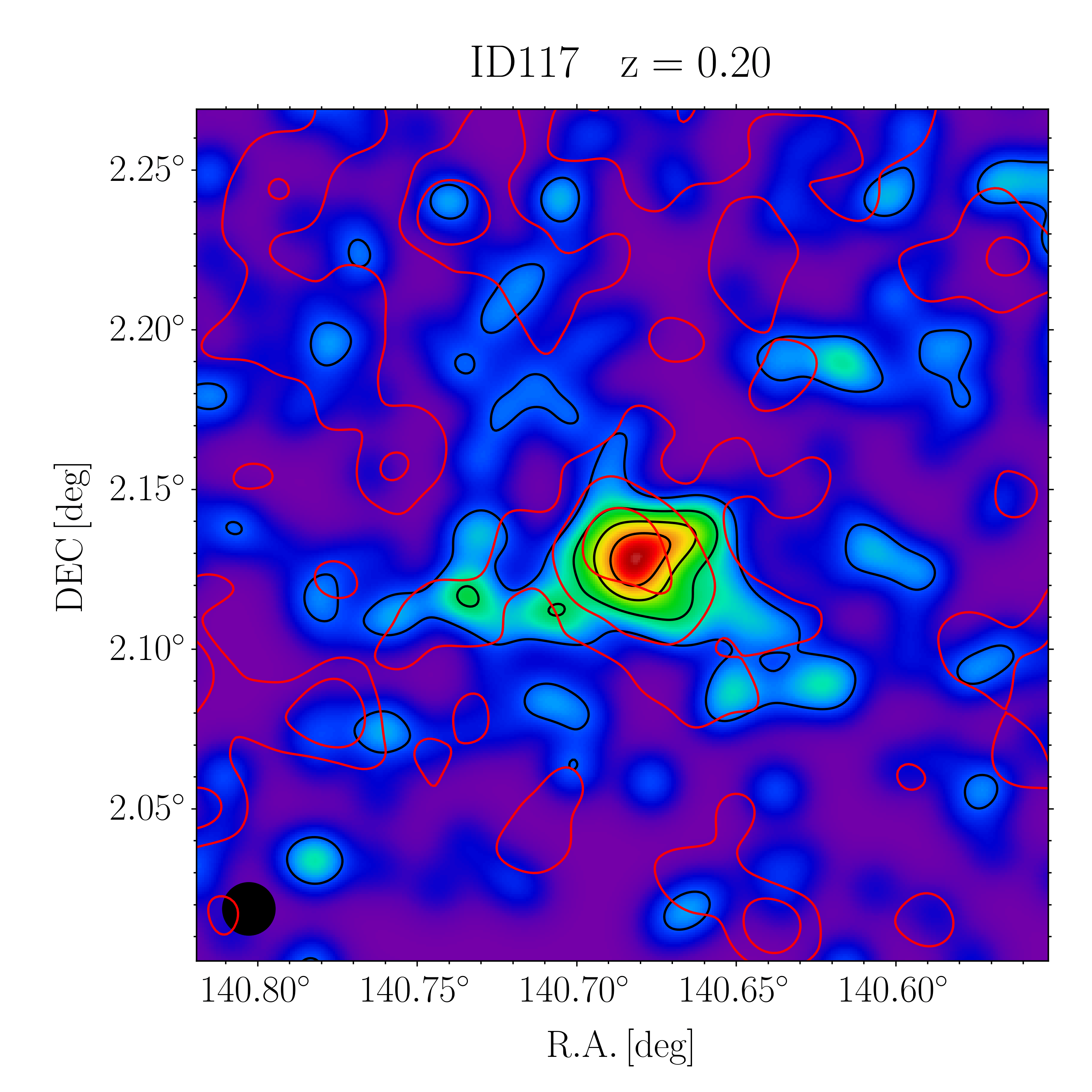}}
     \subfloat[][]{\includegraphics[width=0.24\columnwidth]{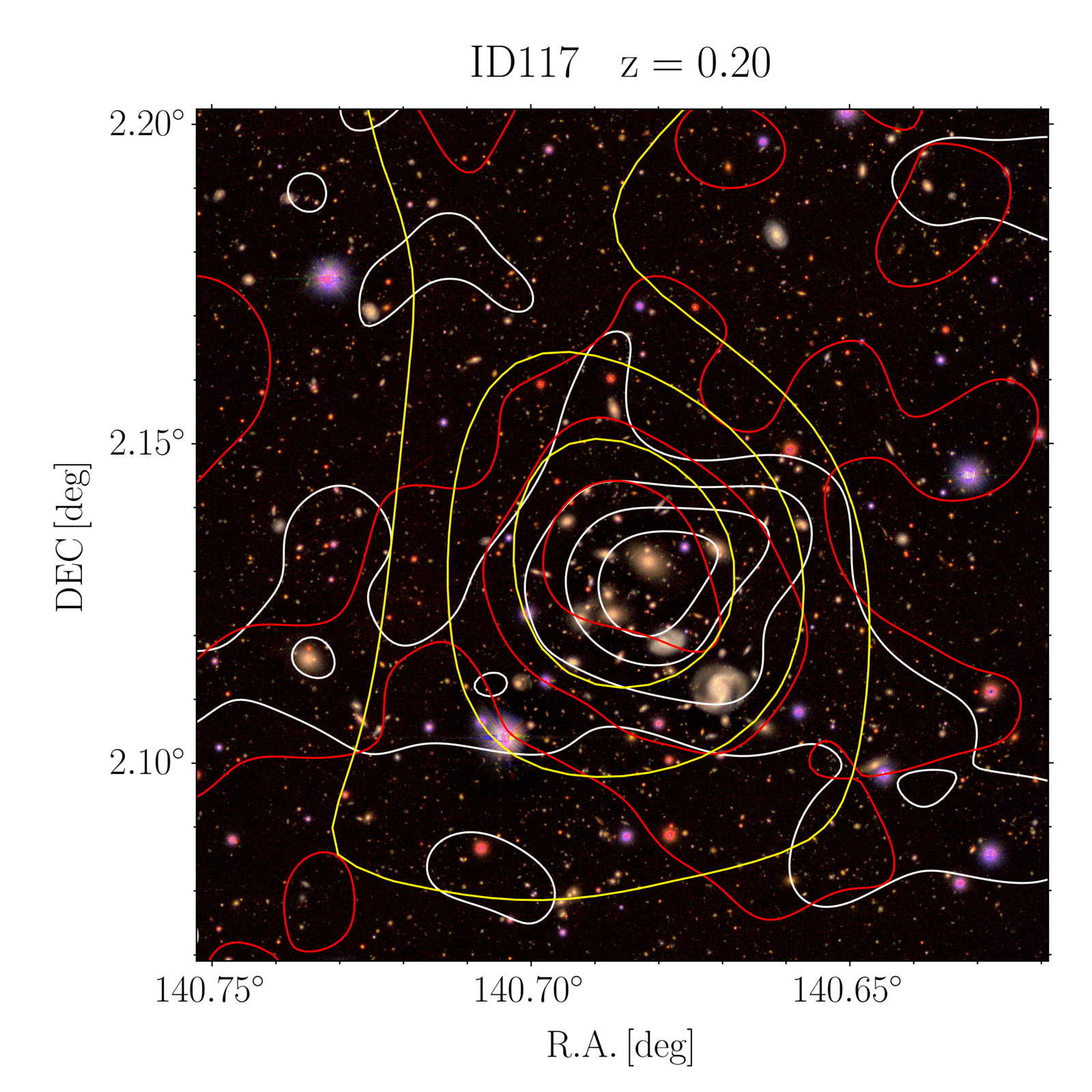}}
     \hspace{0.4cm}     
     \subfloat[][]{\includegraphics[width=0.24\columnwidth]{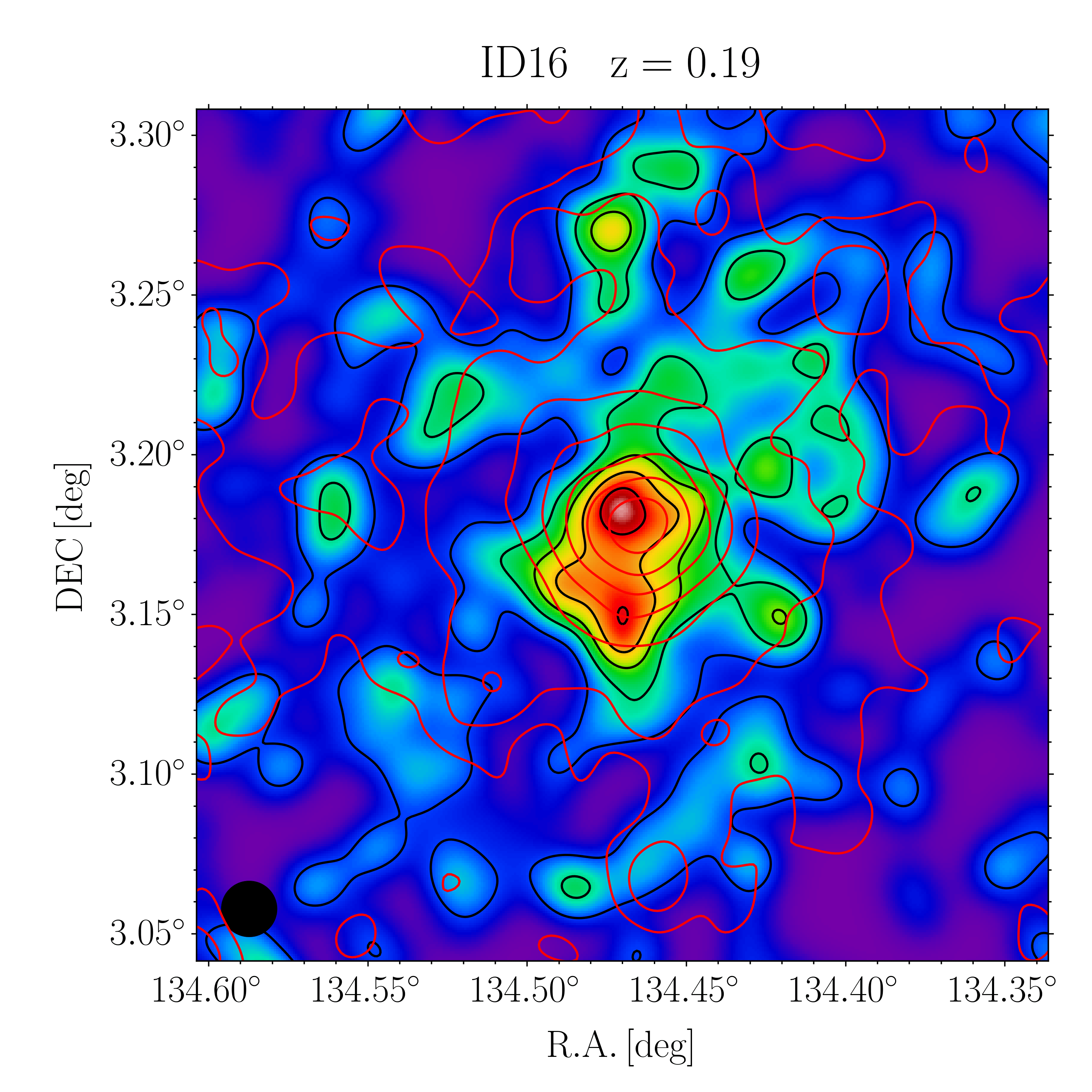}}
     \subfloat[][]{\includegraphics[width=0.24\columnwidth]{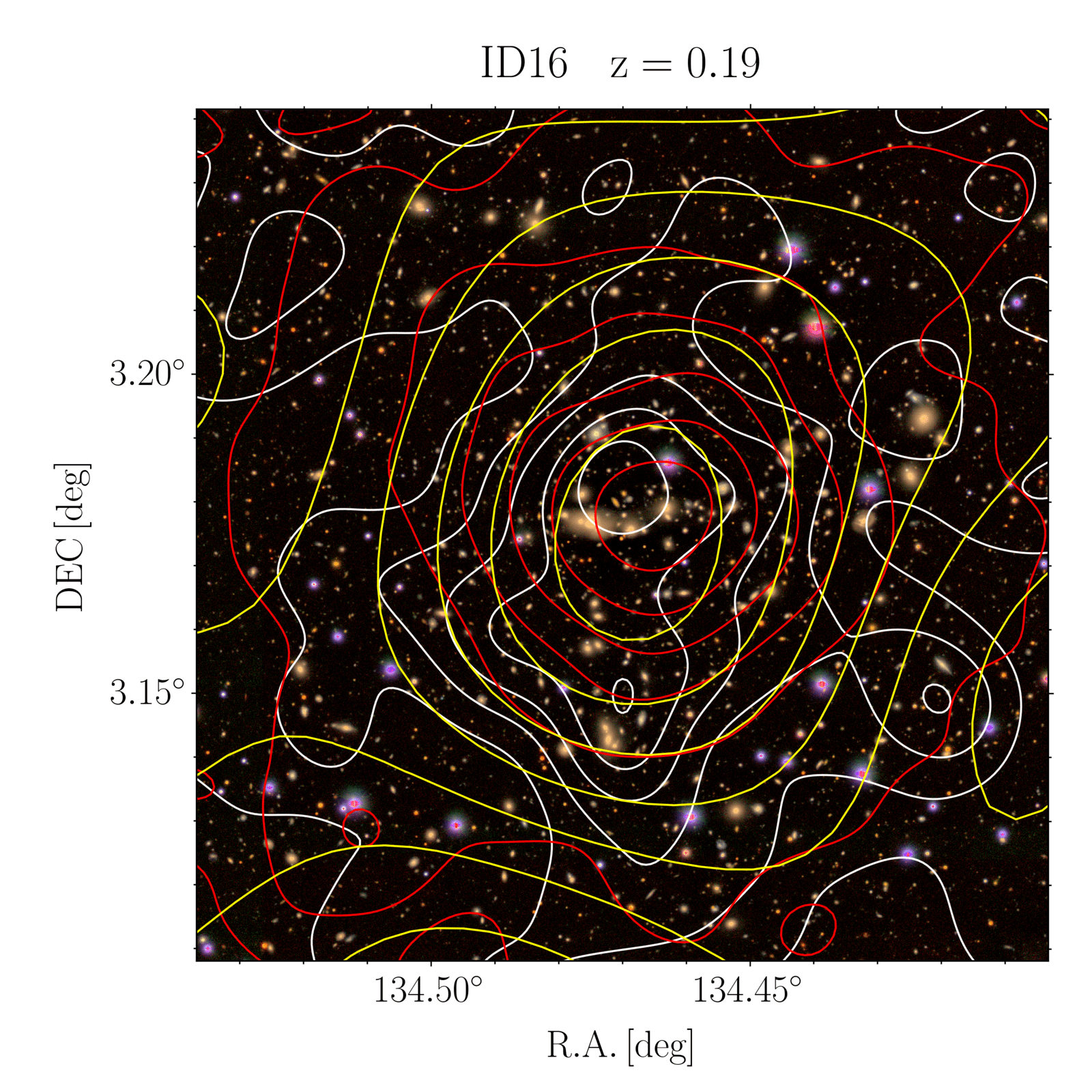}}\\[-5ex]
          
     \subfloat[][]{\includegraphics[width=0.24\columnwidth]{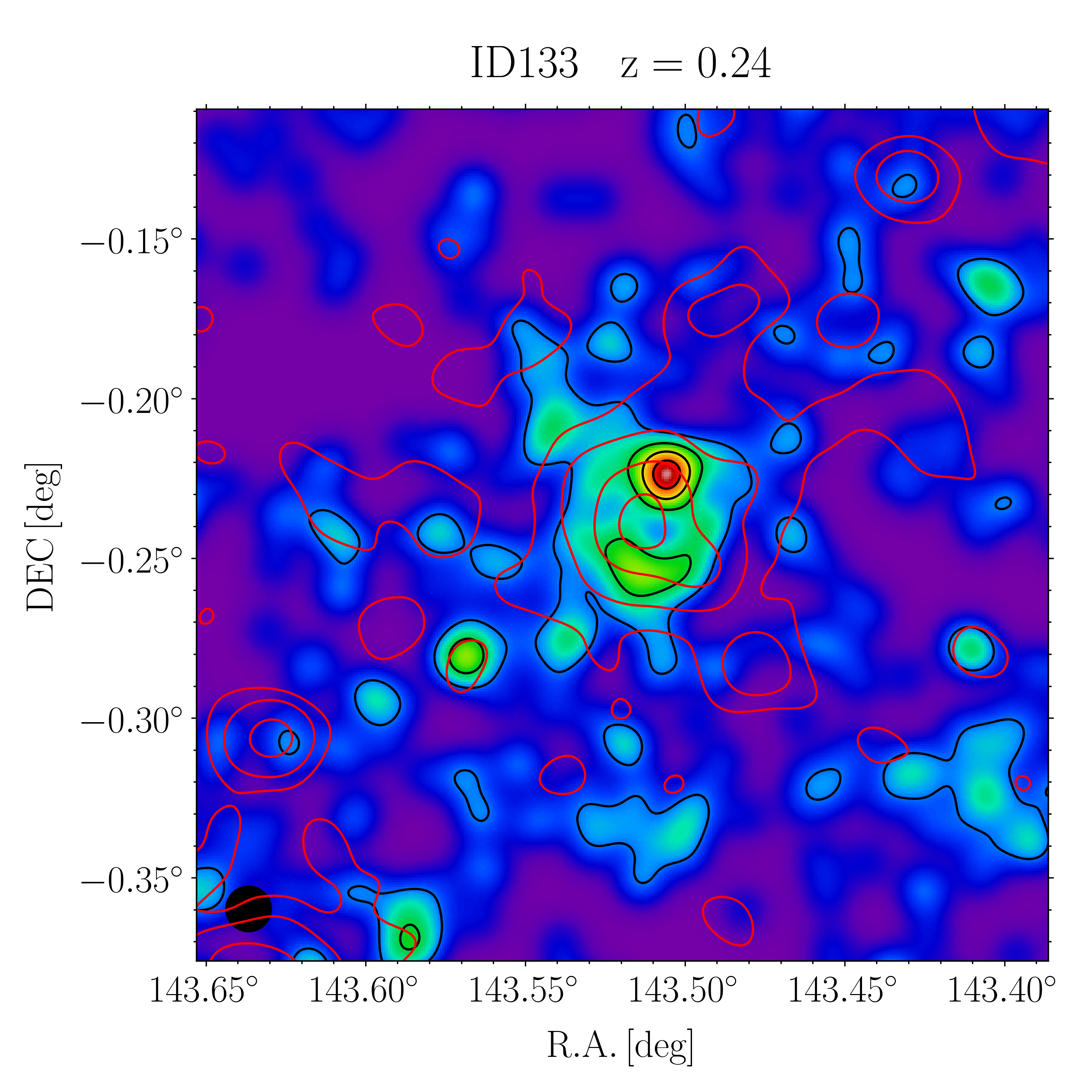}}
     \subfloat[][]{\includegraphics[width=0.24\columnwidth]{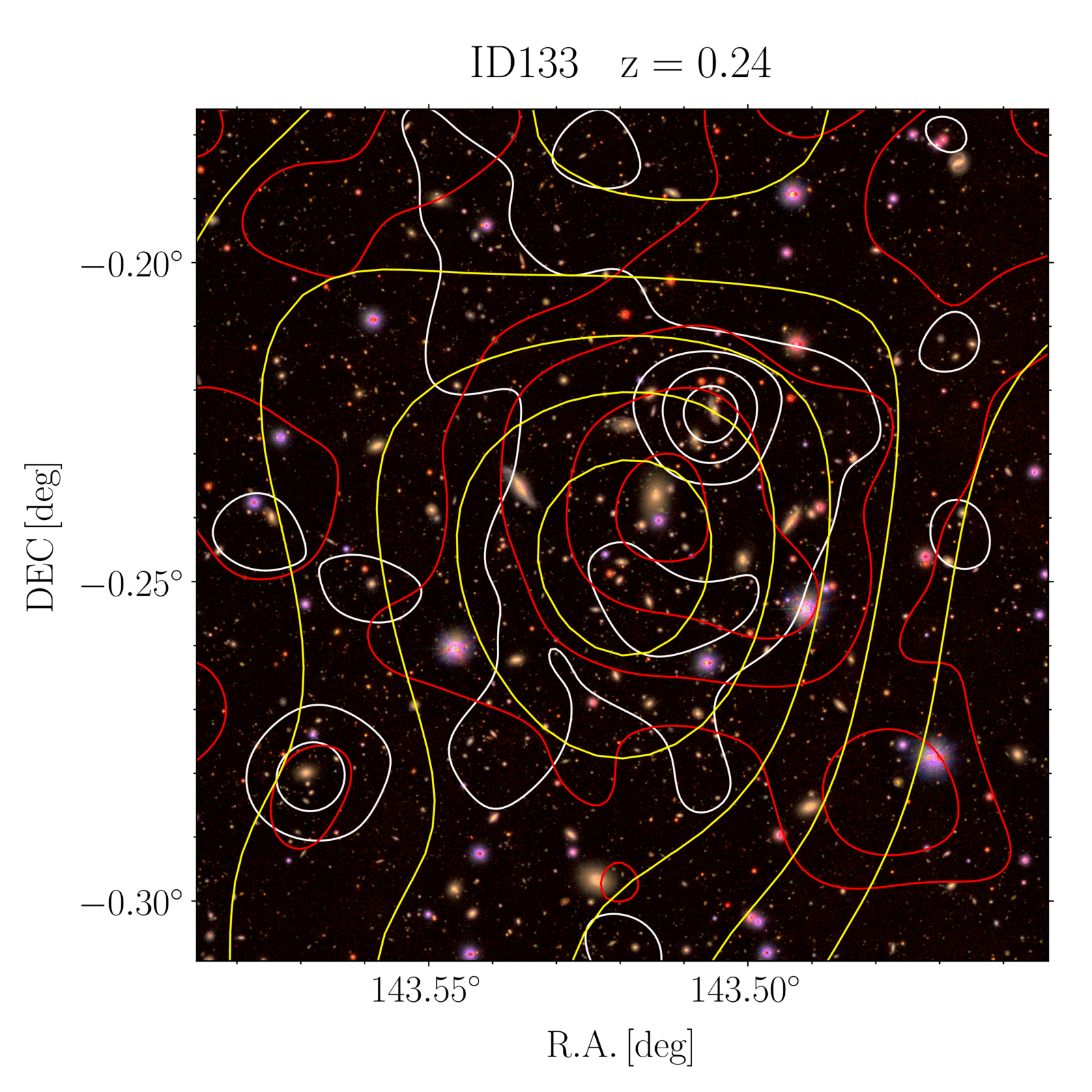}}
     \hspace{0.4cm}          
     \subfloat[][]{\includegraphics[width=0.24\columnwidth]{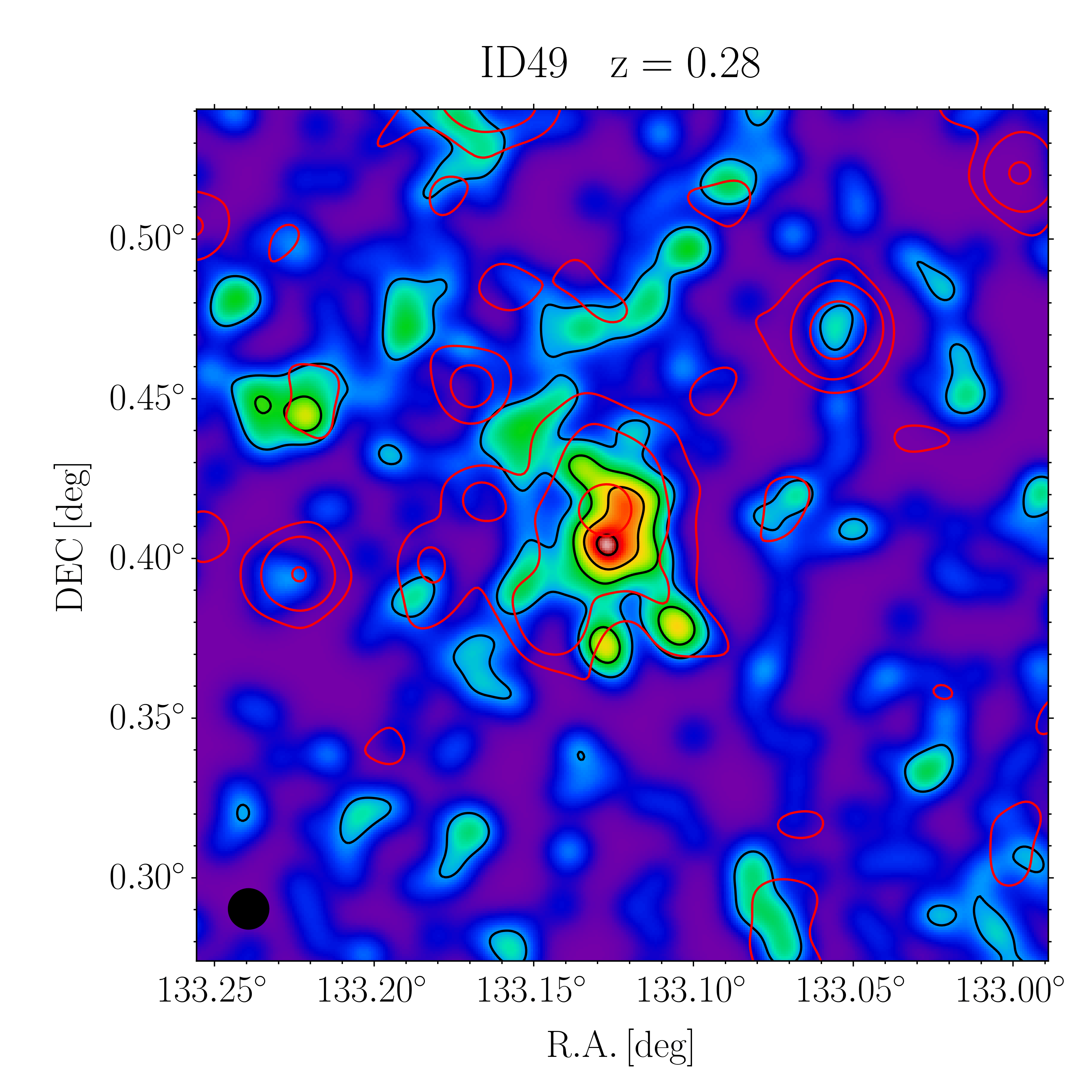}}
     \subfloat[][]{\includegraphics[width=0.24\columnwidth]{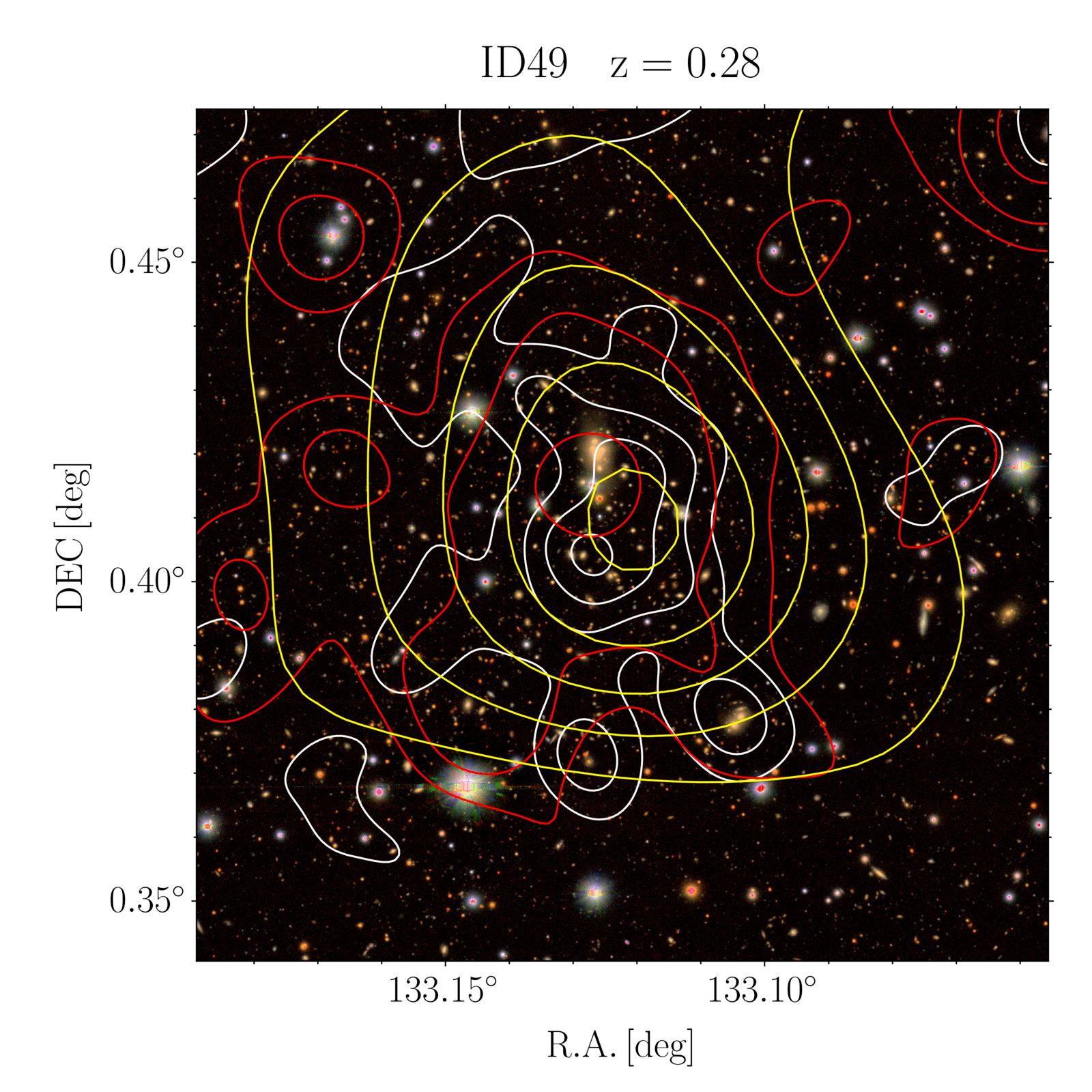}}\\[-5ex]
          
     \subfloat[][]{\includegraphics[width=0.24\columnwidth]{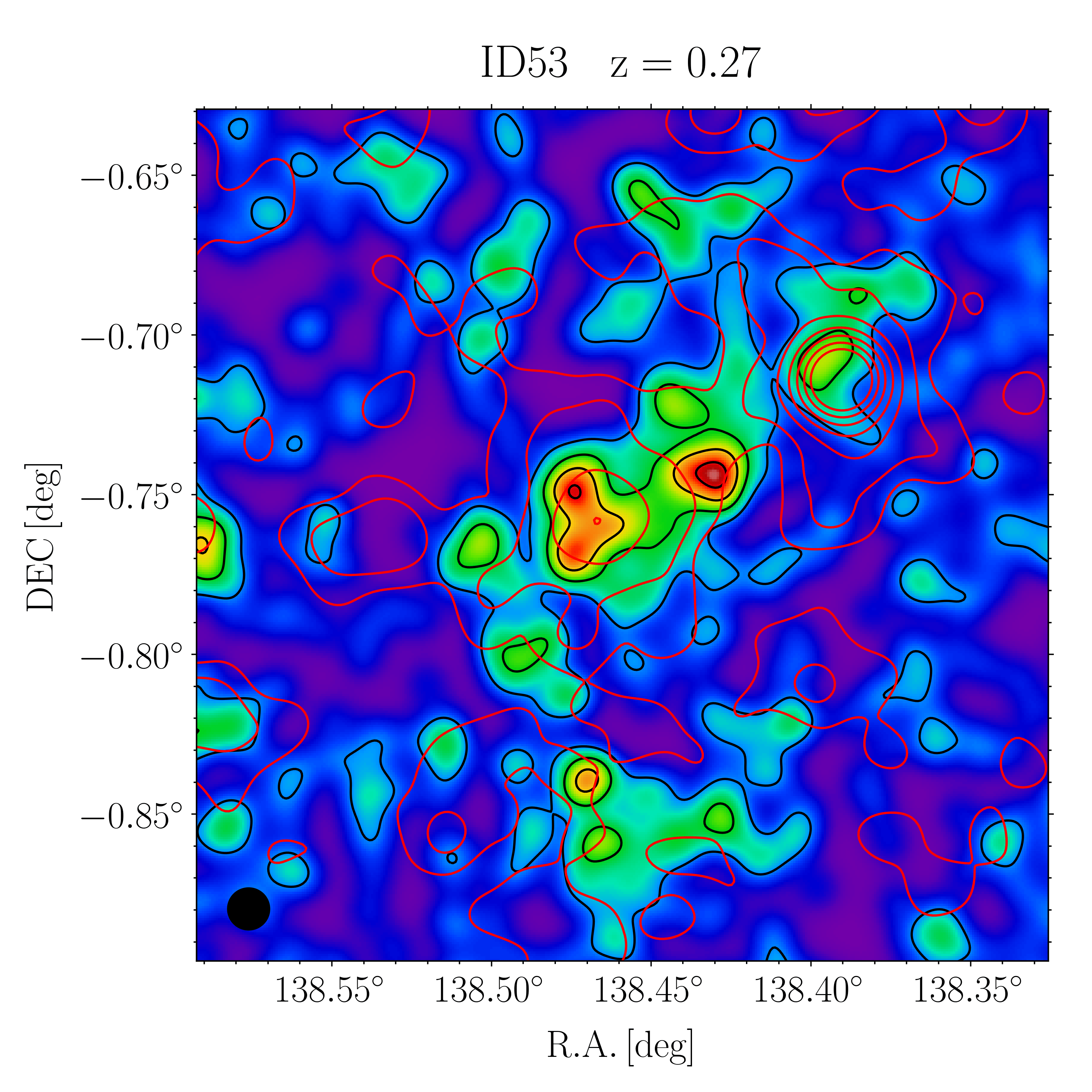}}
     \subfloat[][]{\includegraphics[width=0.24\columnwidth]{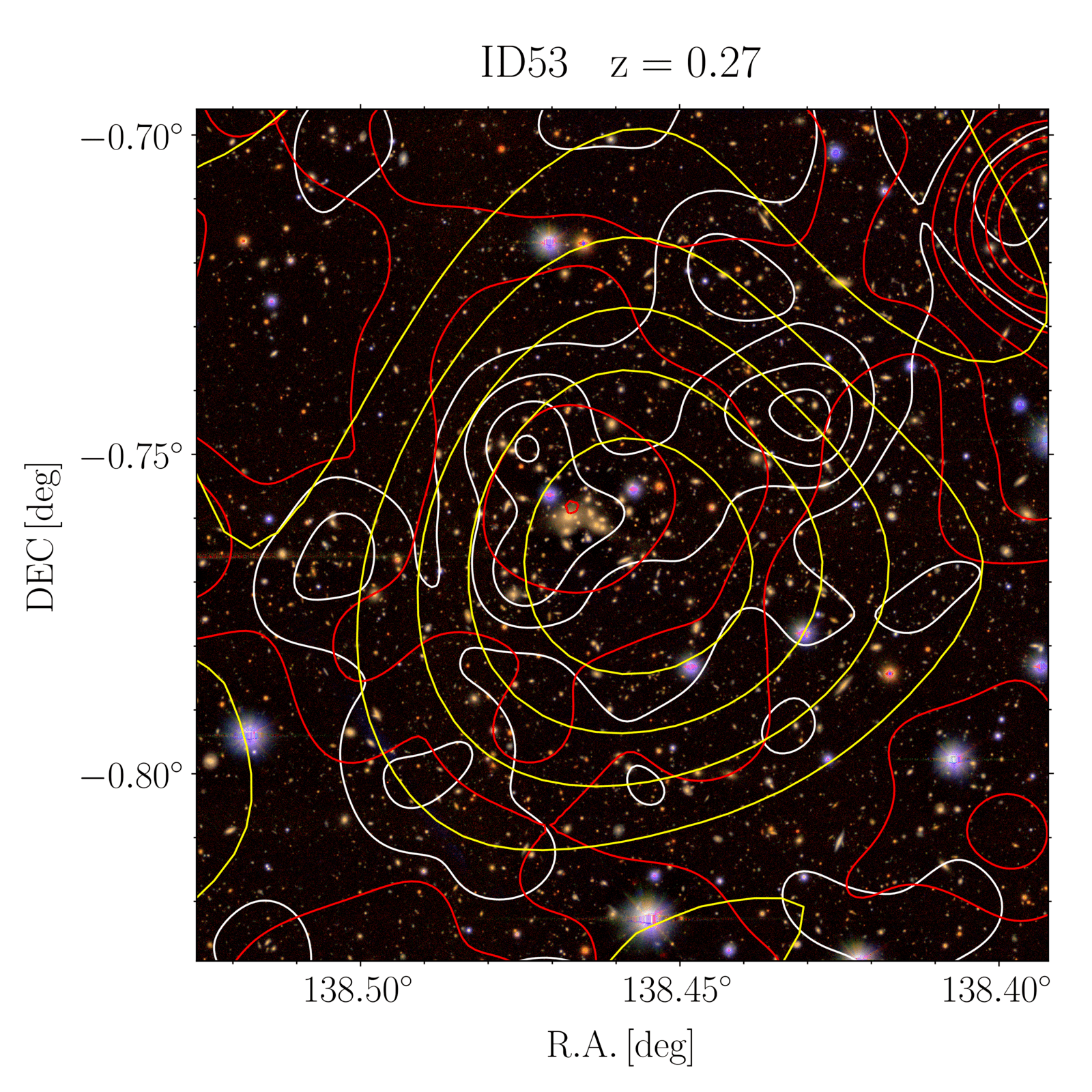}}
     \hspace{0.4cm}          
     \subfloat[][]{\includegraphics[width=0.24\columnwidth]{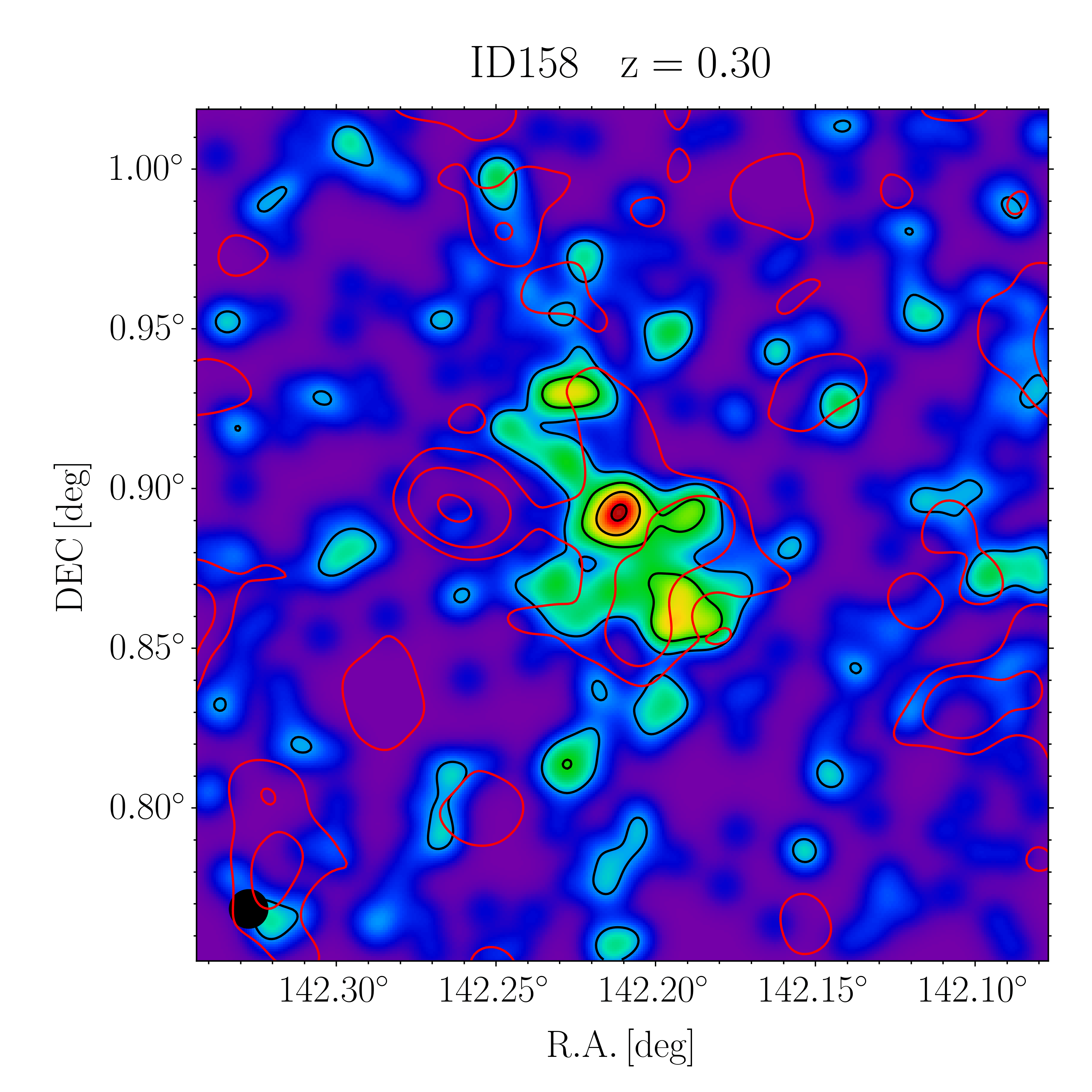}}
     \subfloat[][]{\includegraphics[width=0.24\columnwidth]{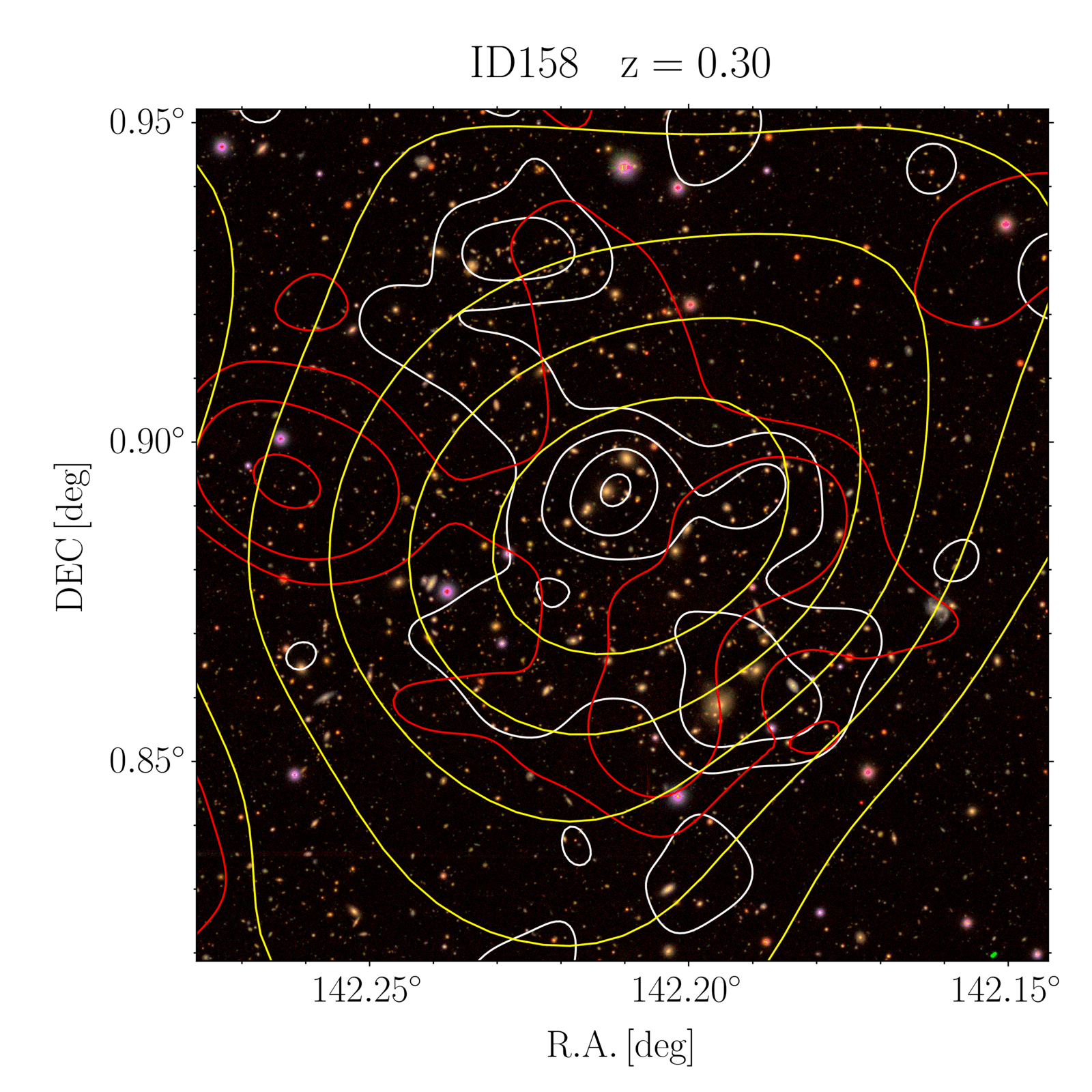}}\\[-5ex]
     \caption{{\it Left}: Galaxy density maps ($16\times16$ arcmin) centred on the HSC shear-selected cluster positions with a unique eFEDS match. Overlaid in red are X-ray contours, which were obtained by smoothing the raw X-ray image in the $0.5-2.0$~keV energy band with a Gaussian of $24$~arcsec. The black circles in the lower-left corners show the smoothing scale, FWHM\ =\ $200$~kpc. {\it Right}: HSC-SSP optical images centred on the HSC shear-selected cluster positions. The $8\times8$ arcmin optical images of the central region are created using the $z$, $i$, and $r$ bands. X-ray contours are shown in red, galaxy density contours in white (they are the same contours as the black ones in the corresponding galaxy density maps), and weak-lensing mass contours in yellow.}
     \label{fig:uniquematchesxrayoptical}
 \end{figure*}
               
 \begin{figure*}[t]
     \ContinuedFloat
     \captionsetup[subfigure]{labelformat=empty}
     \centering
     \subfloat[][]{\includegraphics[width=0.24\columnwidth]{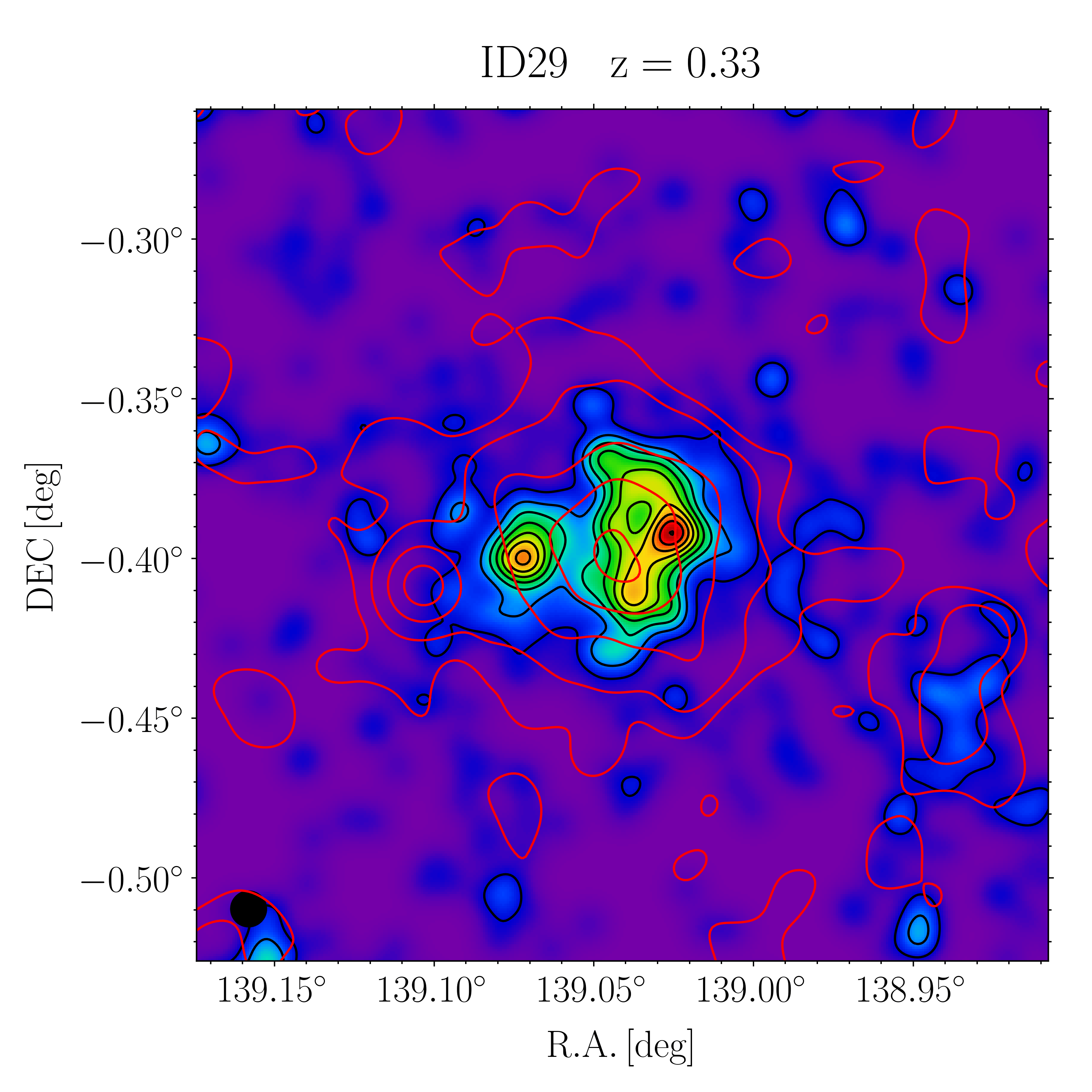}}
     \subfloat[][]{\includegraphics[width=0.24\columnwidth]{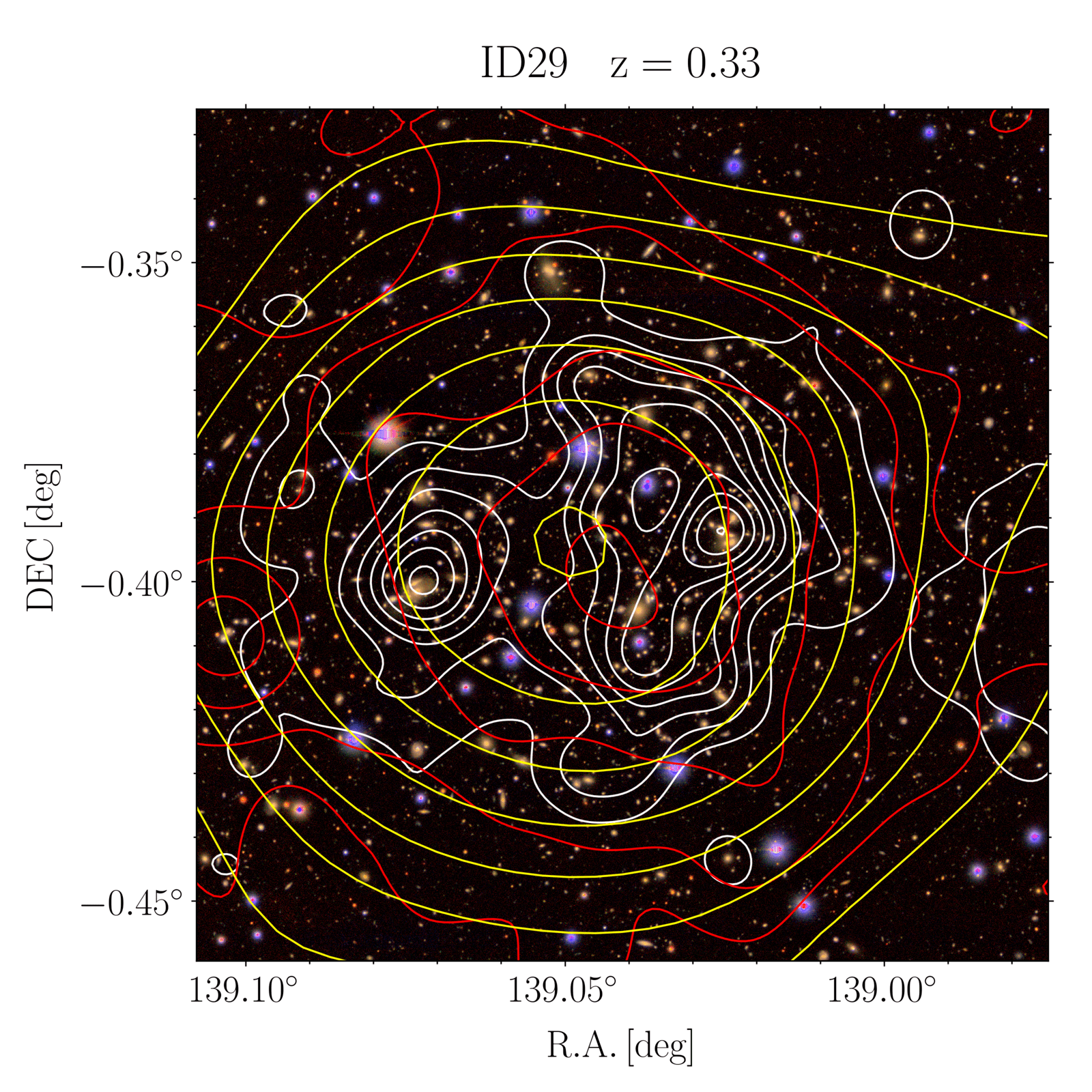}}
     \hspace{0.4cm}          
     \subfloat[][]{\includegraphics[width=0.24\columnwidth]{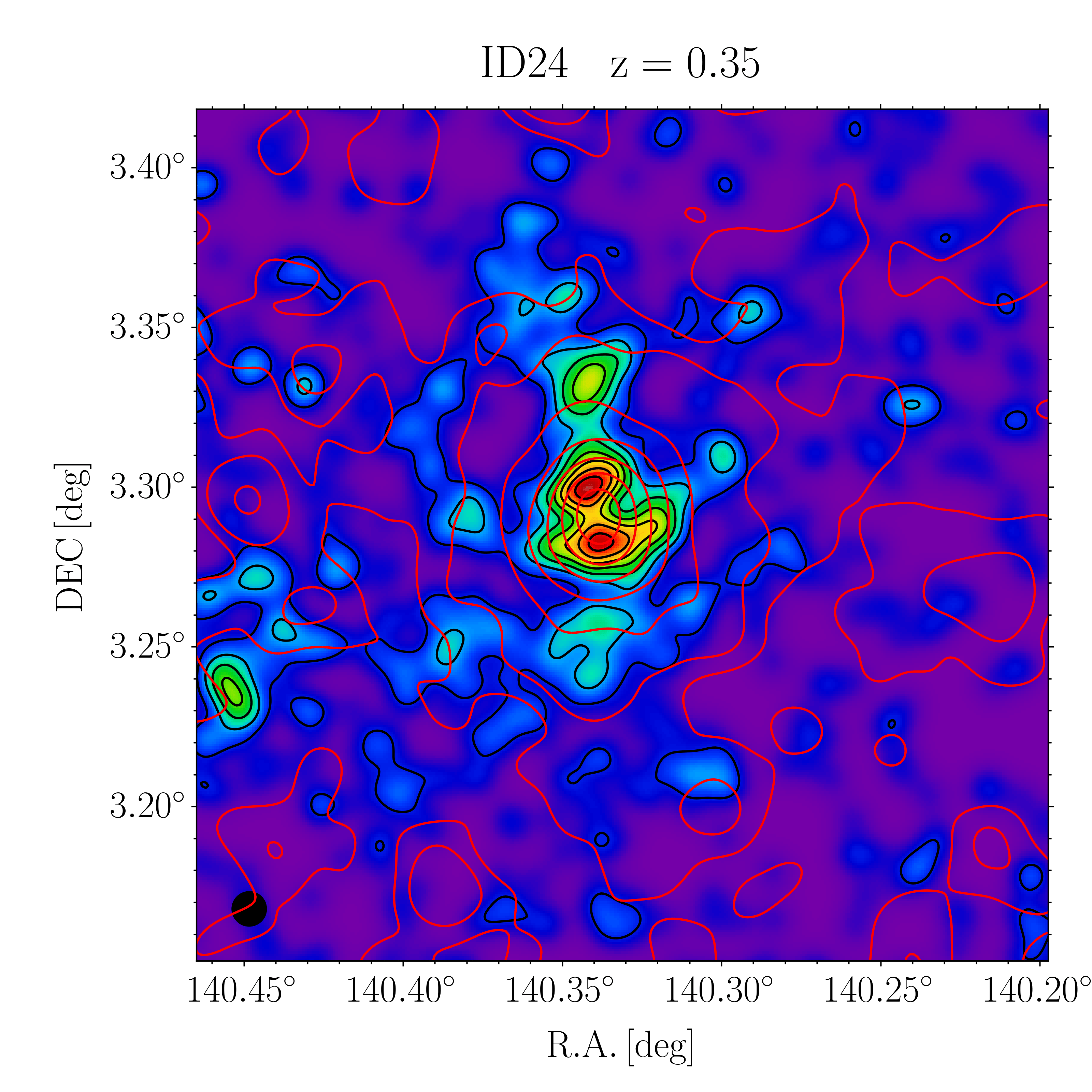}}
     \subfloat[][]{\includegraphics[width=0.24\columnwidth]{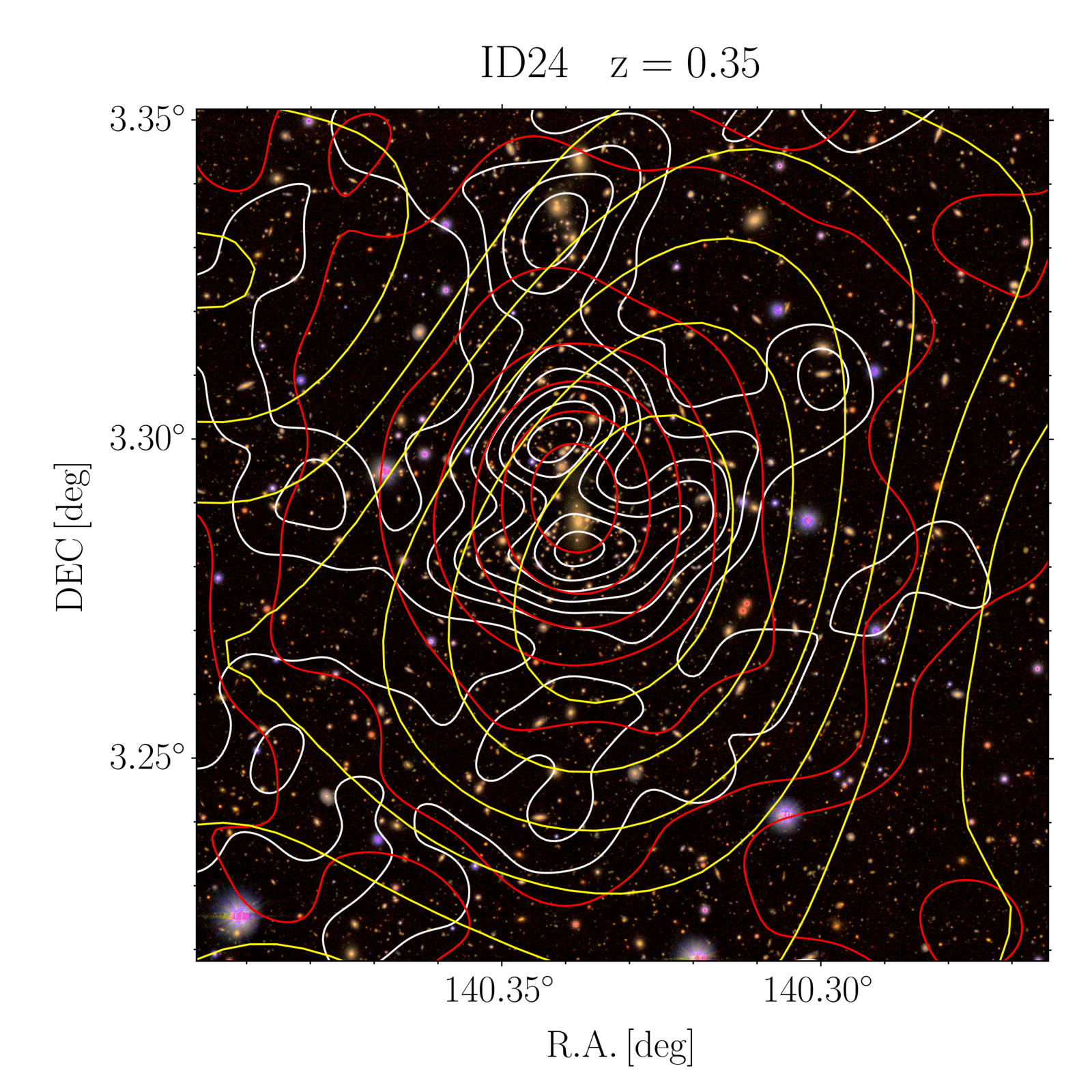}}\\[-5ex]
           
     \subfloat[][]{\includegraphics[width=0.24\columnwidth]{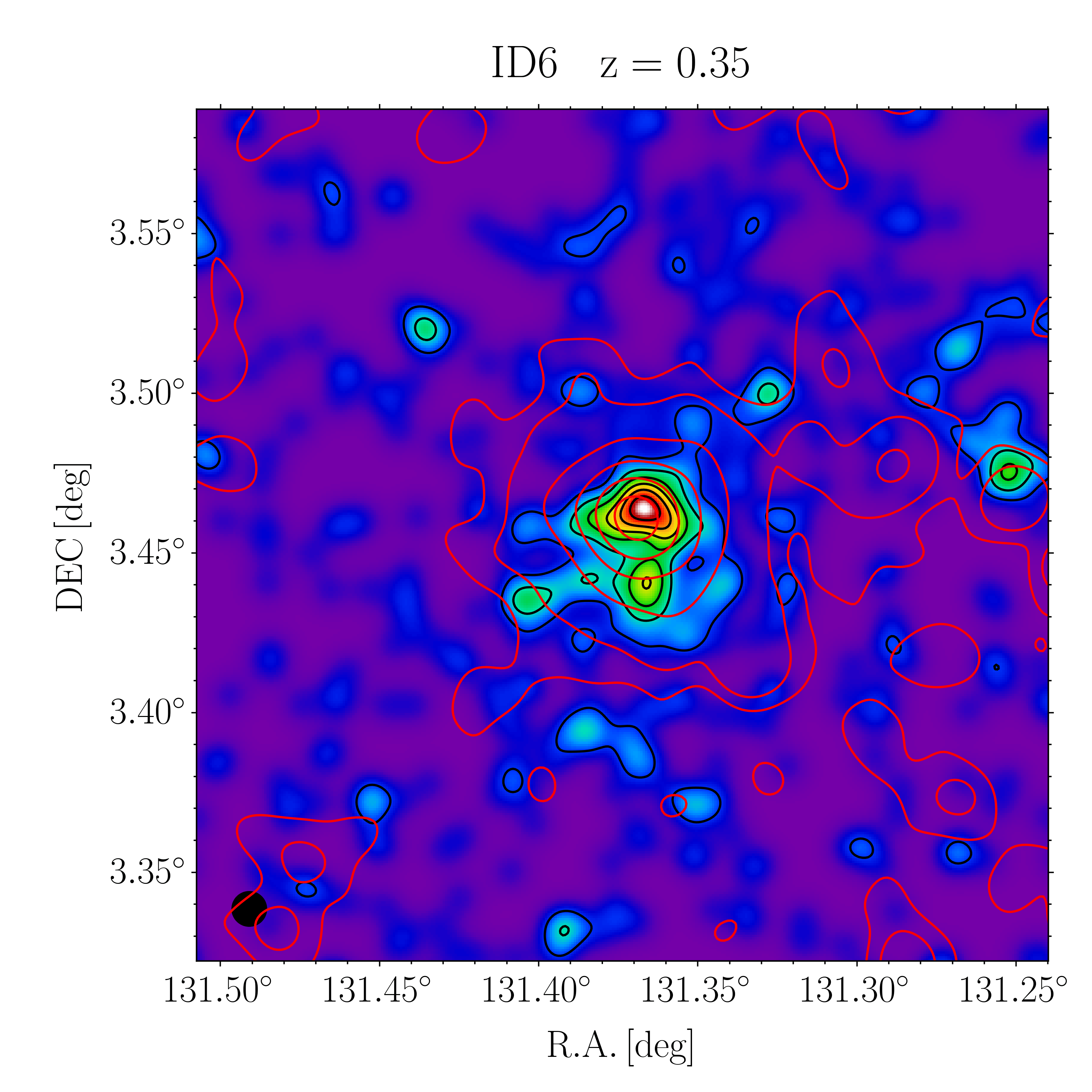}}
     \subfloat[][]{\includegraphics[width=0.24\columnwidth]{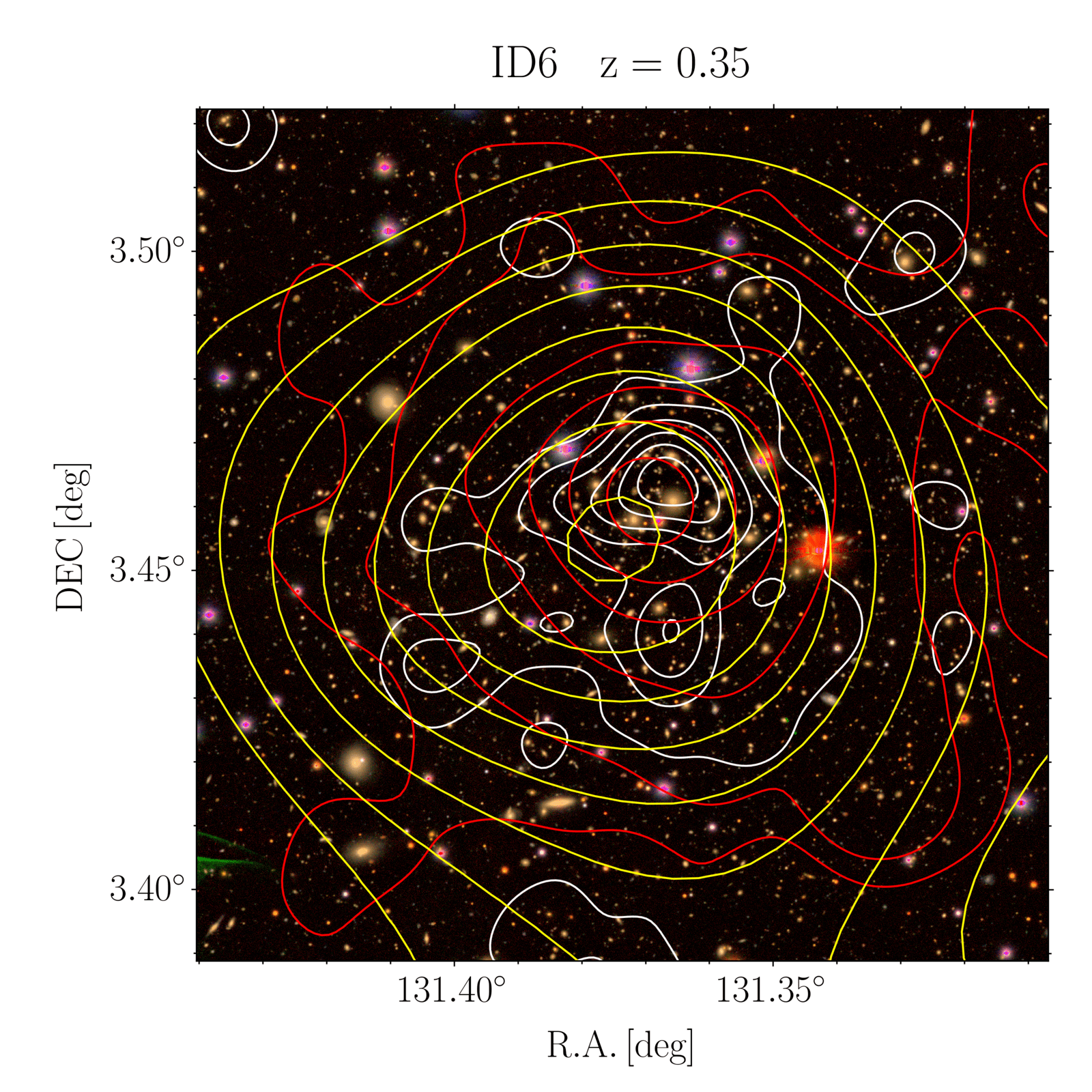}}
     \hspace{0.4cm}               
     \subfloat[][]{\includegraphics[width=0.24\columnwidth]{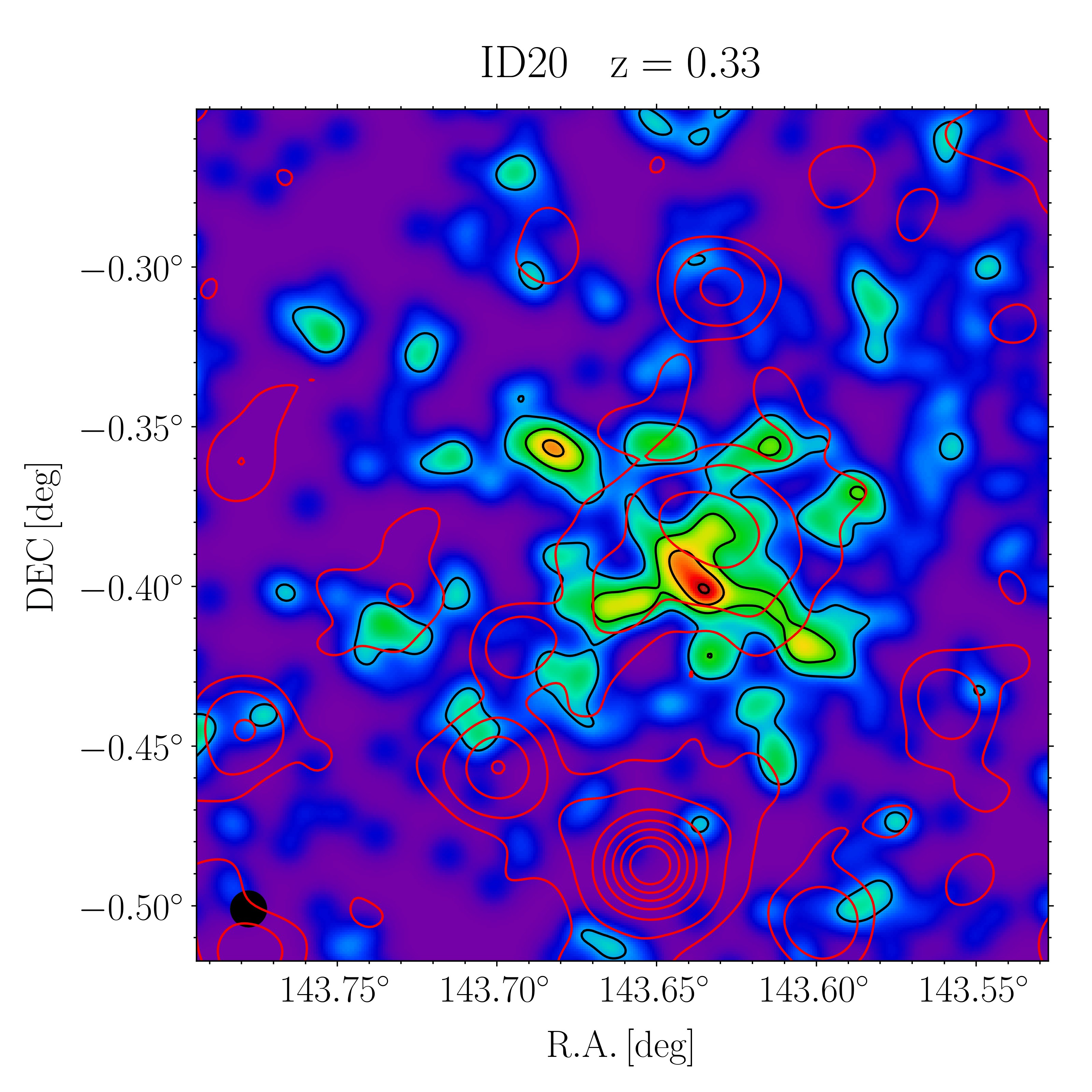}}
     \subfloat[][]{\includegraphics[width=0.24\columnwidth]{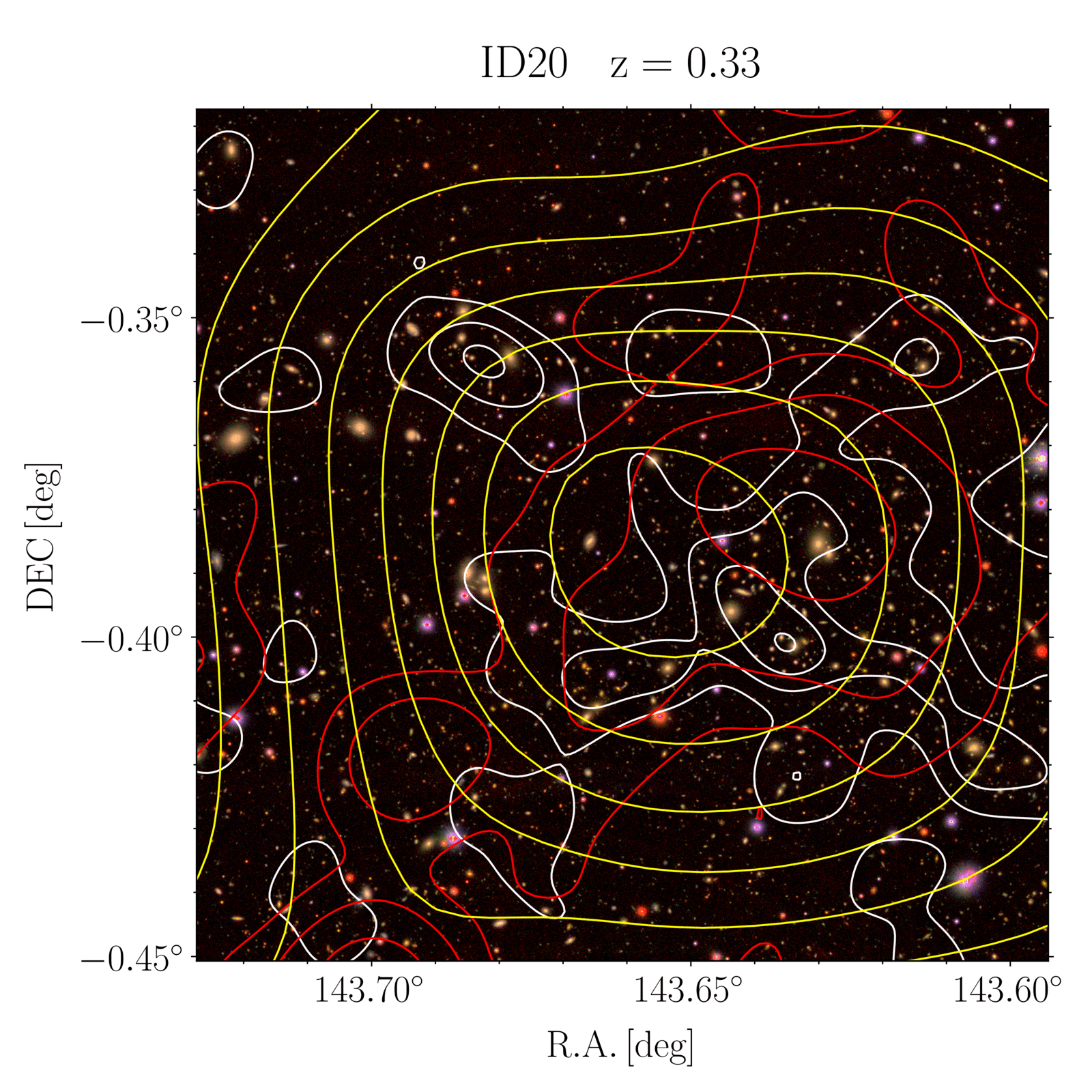}}\\[-5ex]
               
     \subfloat[][]{\includegraphics[width=0.24\columnwidth]{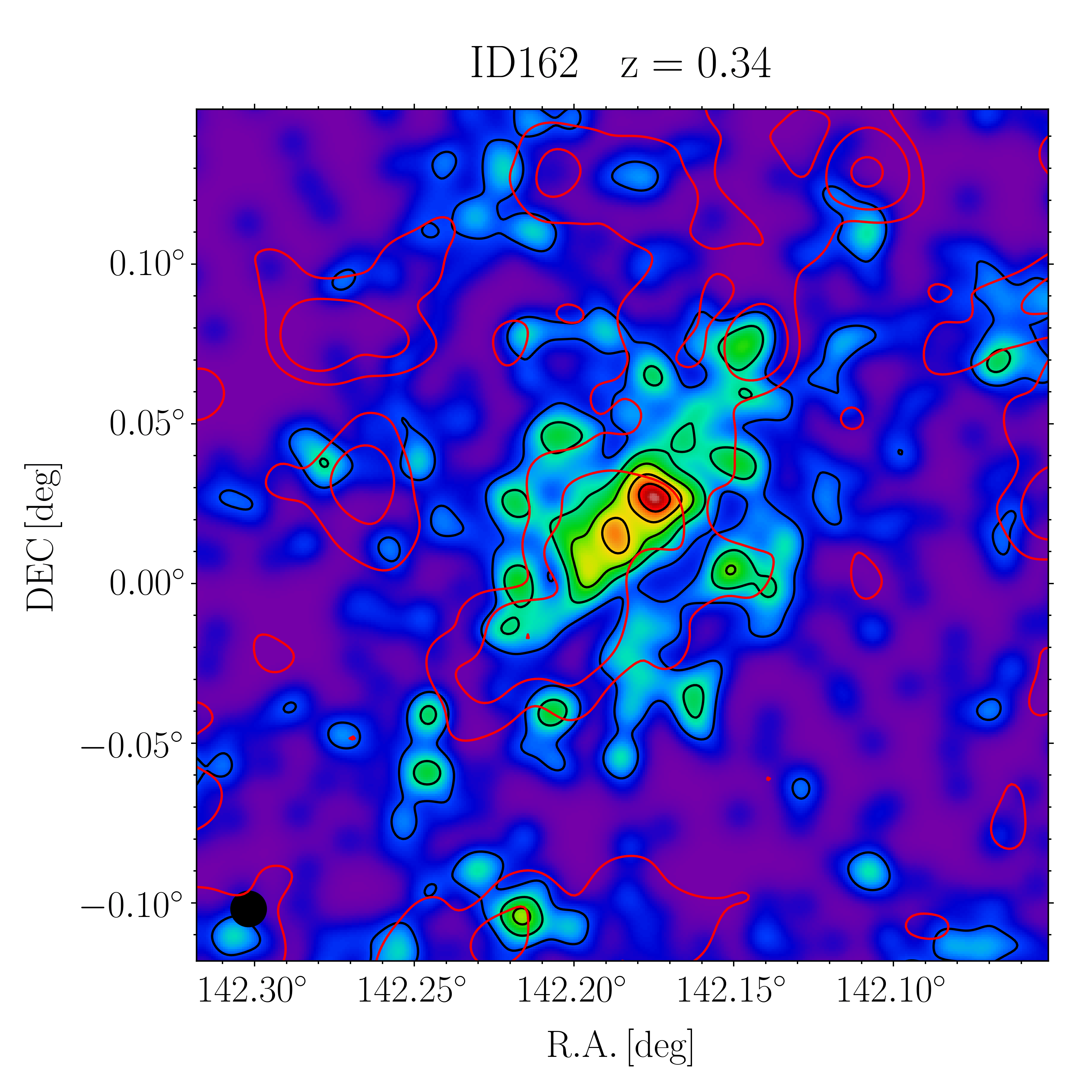}}
     \subfloat[][]{\includegraphics[width=0.24\columnwidth]{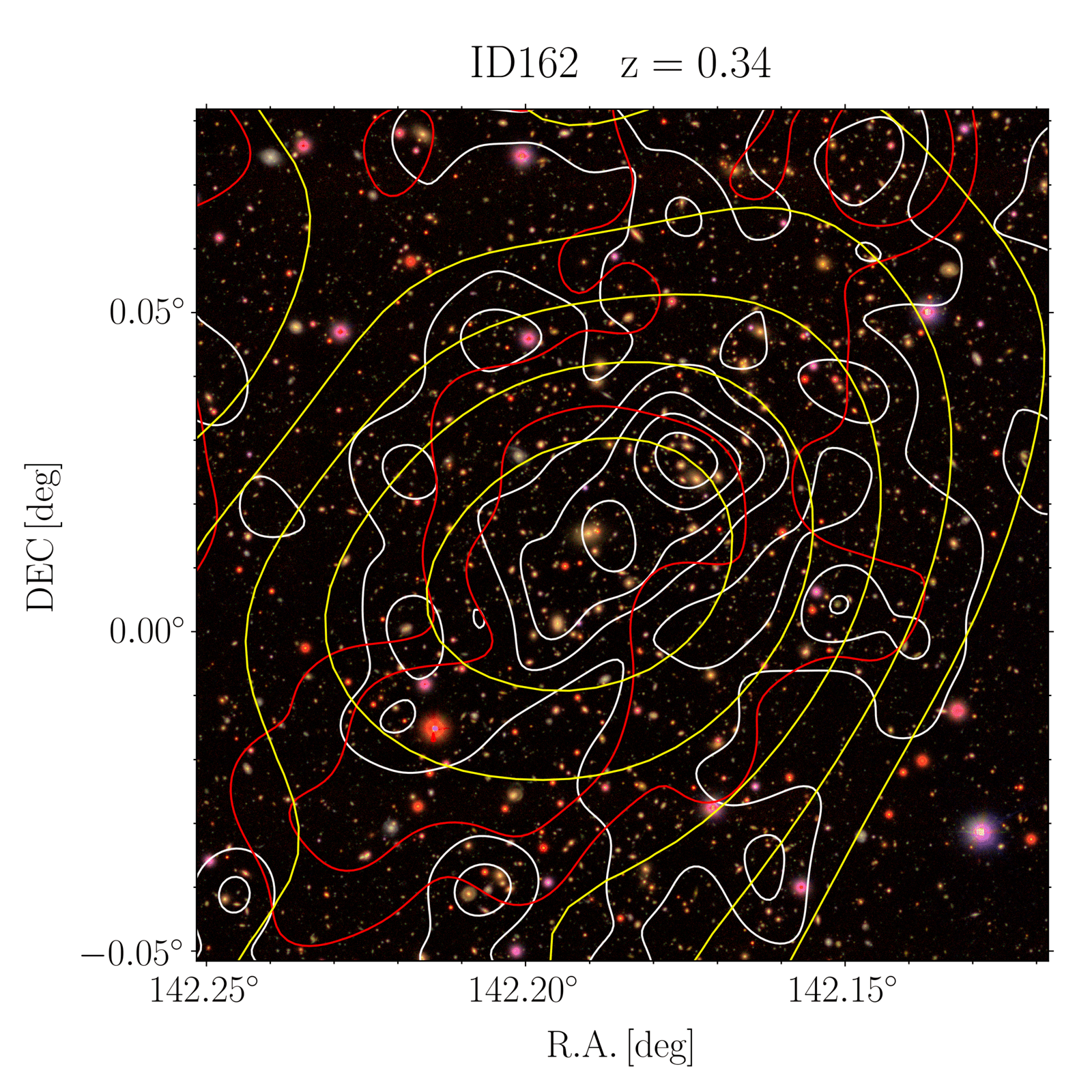}}
     \hspace{0.4cm}               
     \subfloat[][]{\includegraphics[width=0.24\columnwidth]{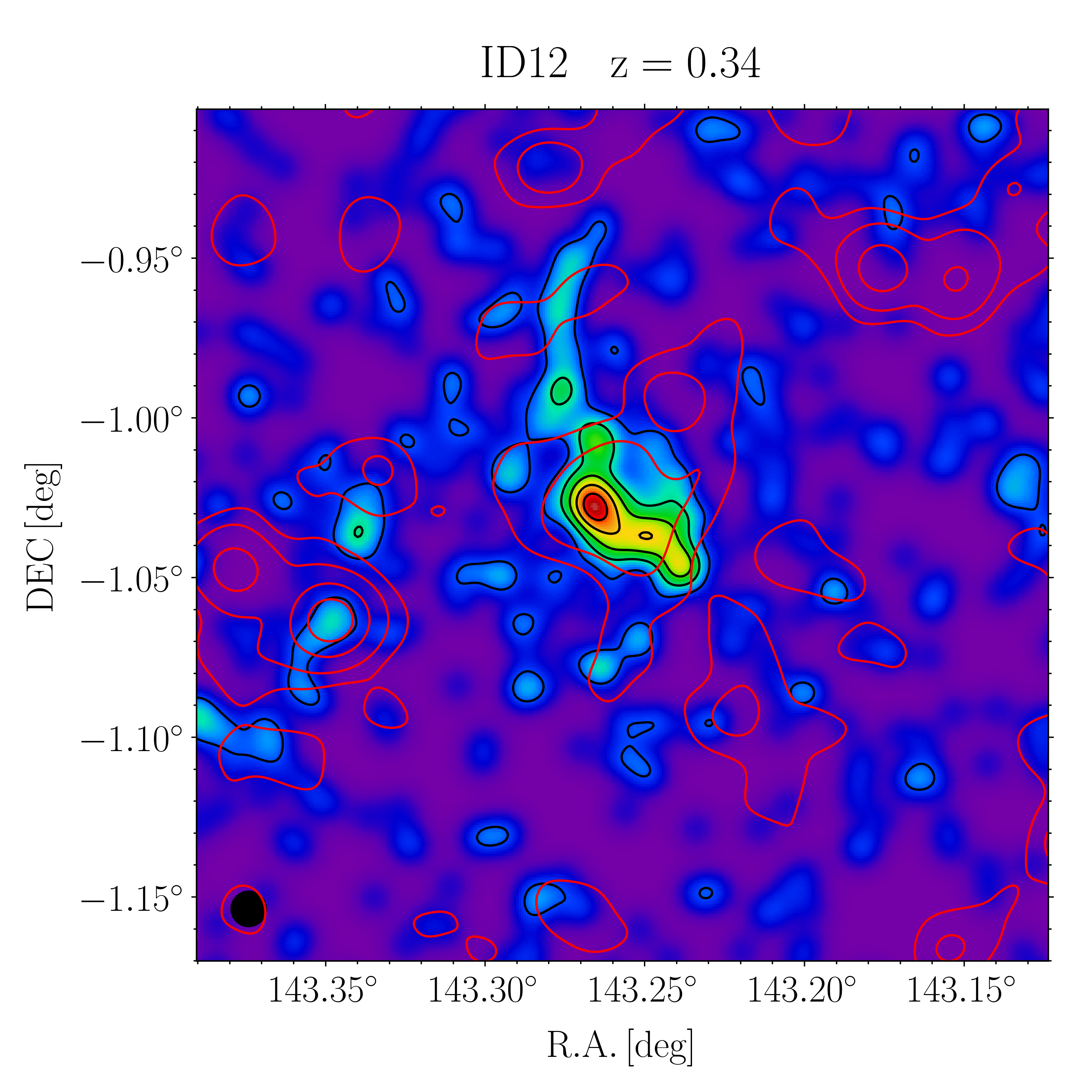}}
     \subfloat[][]{\includegraphics[width=0.24\columnwidth]{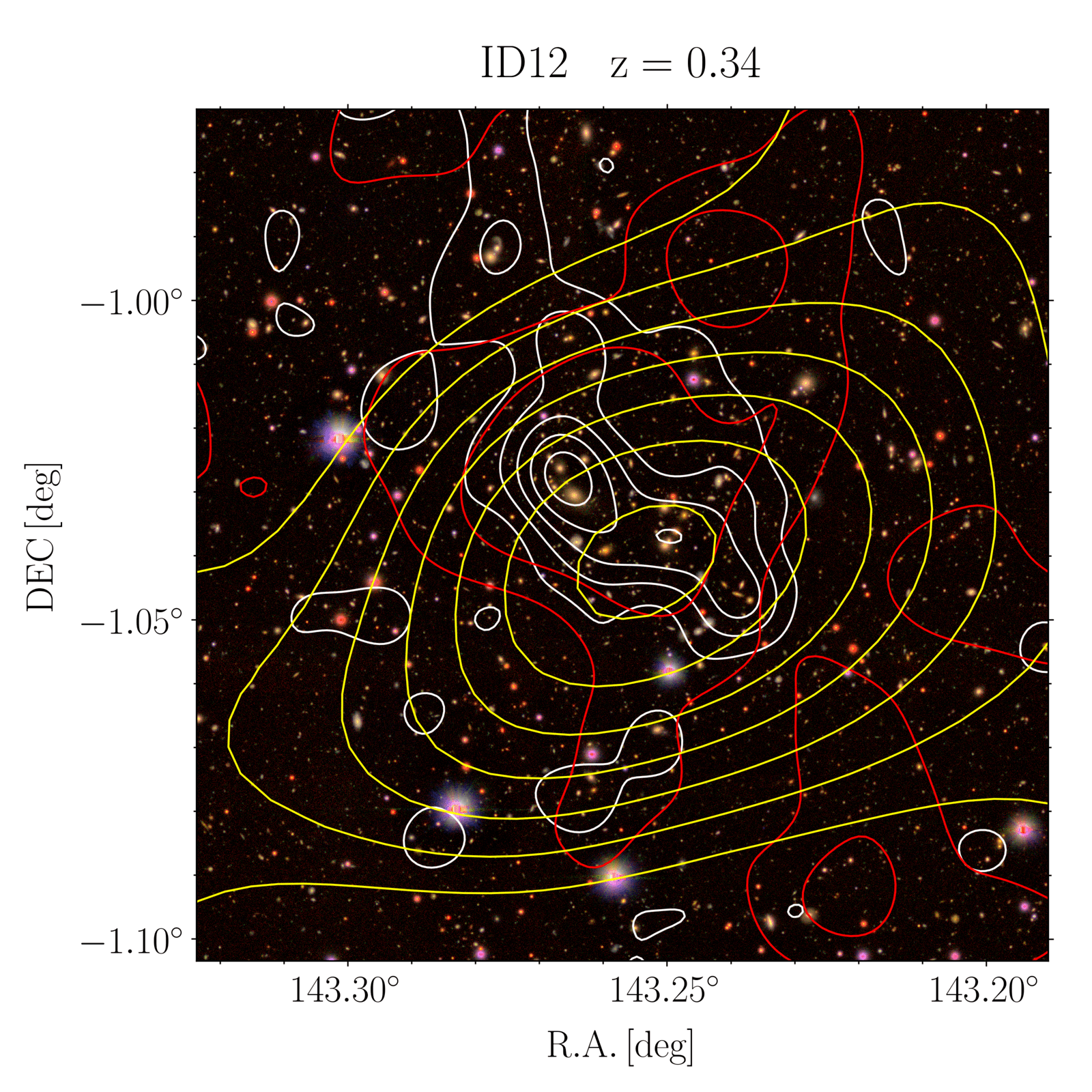}}\\[-5ex]
               
     \subfloat[][]{\includegraphics[width=0.24\columnwidth]{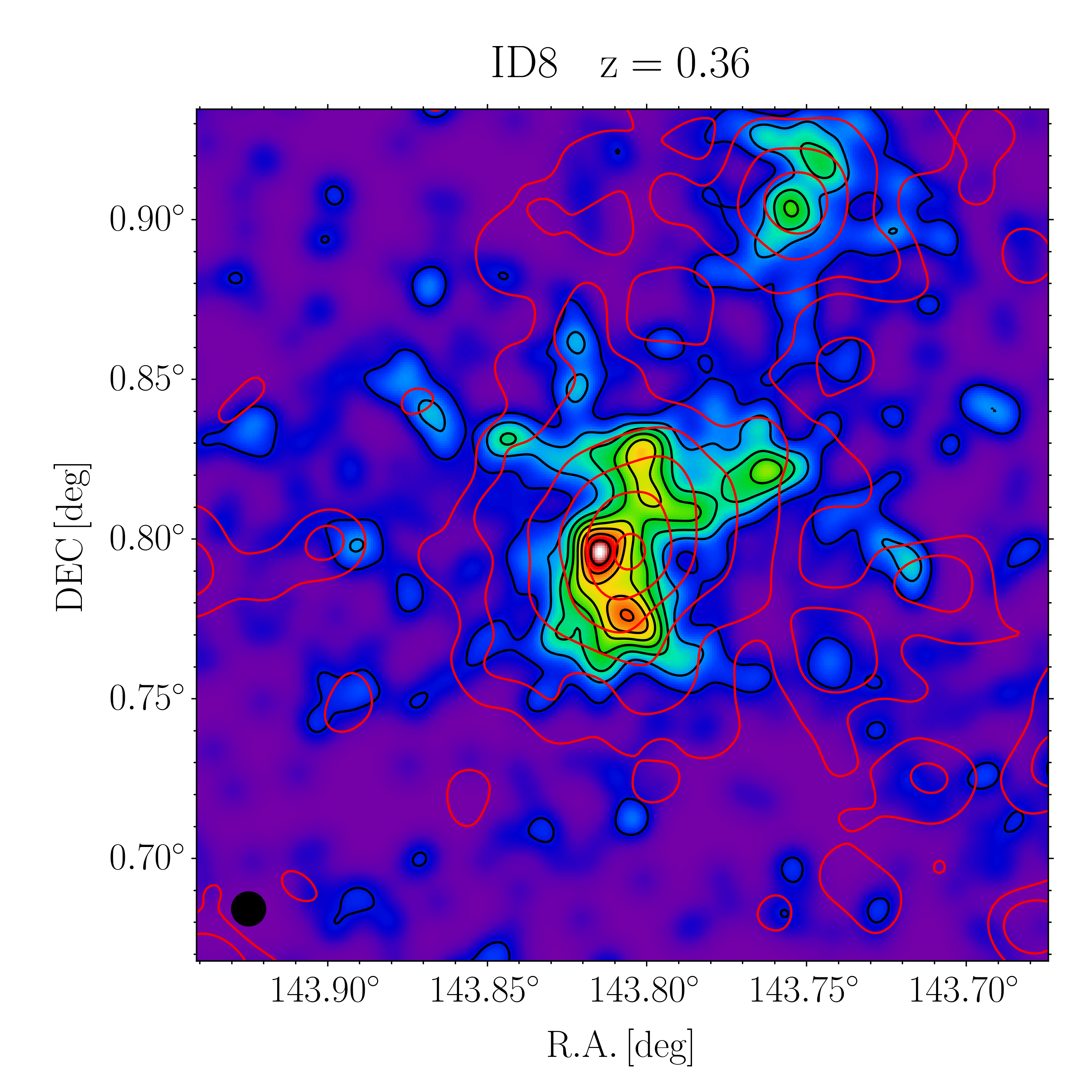}}
     \subfloat[][]{\includegraphics[width=0.24\columnwidth]{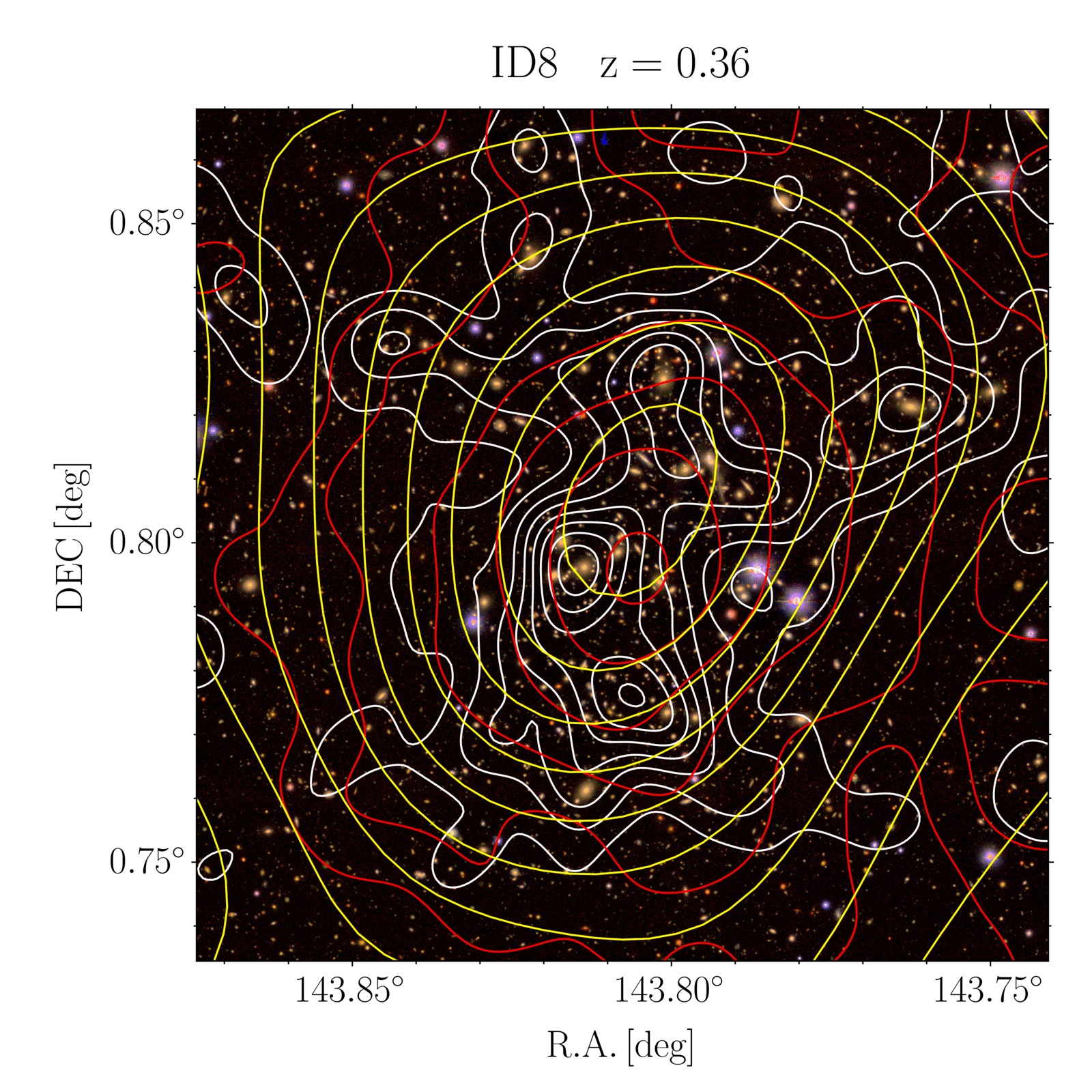}}
     \hspace{0.4cm}               
     \subfloat[][]{\includegraphics[width=0.24\columnwidth]{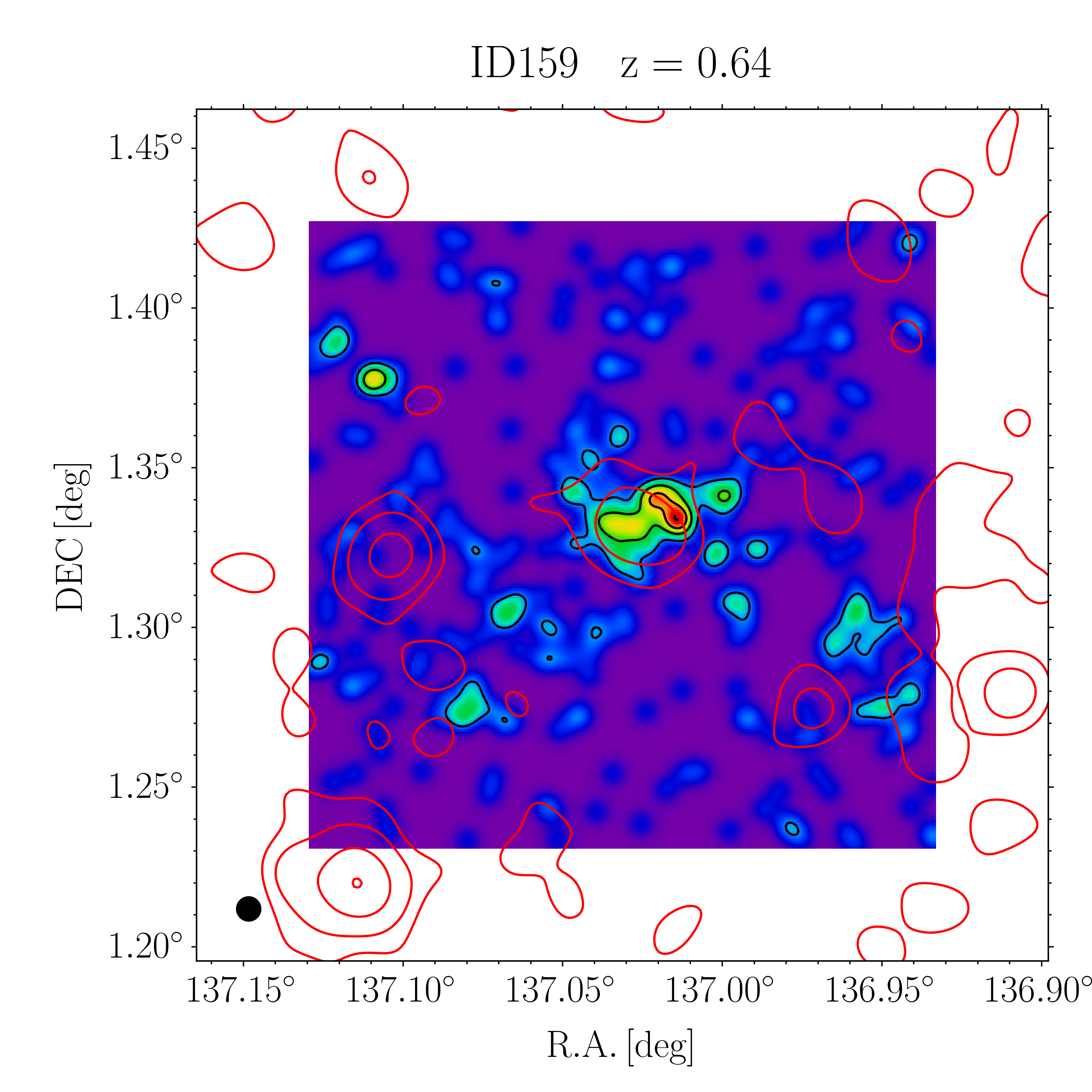}}
     \subfloat[][]{\includegraphics[width=0.24\columnwidth]{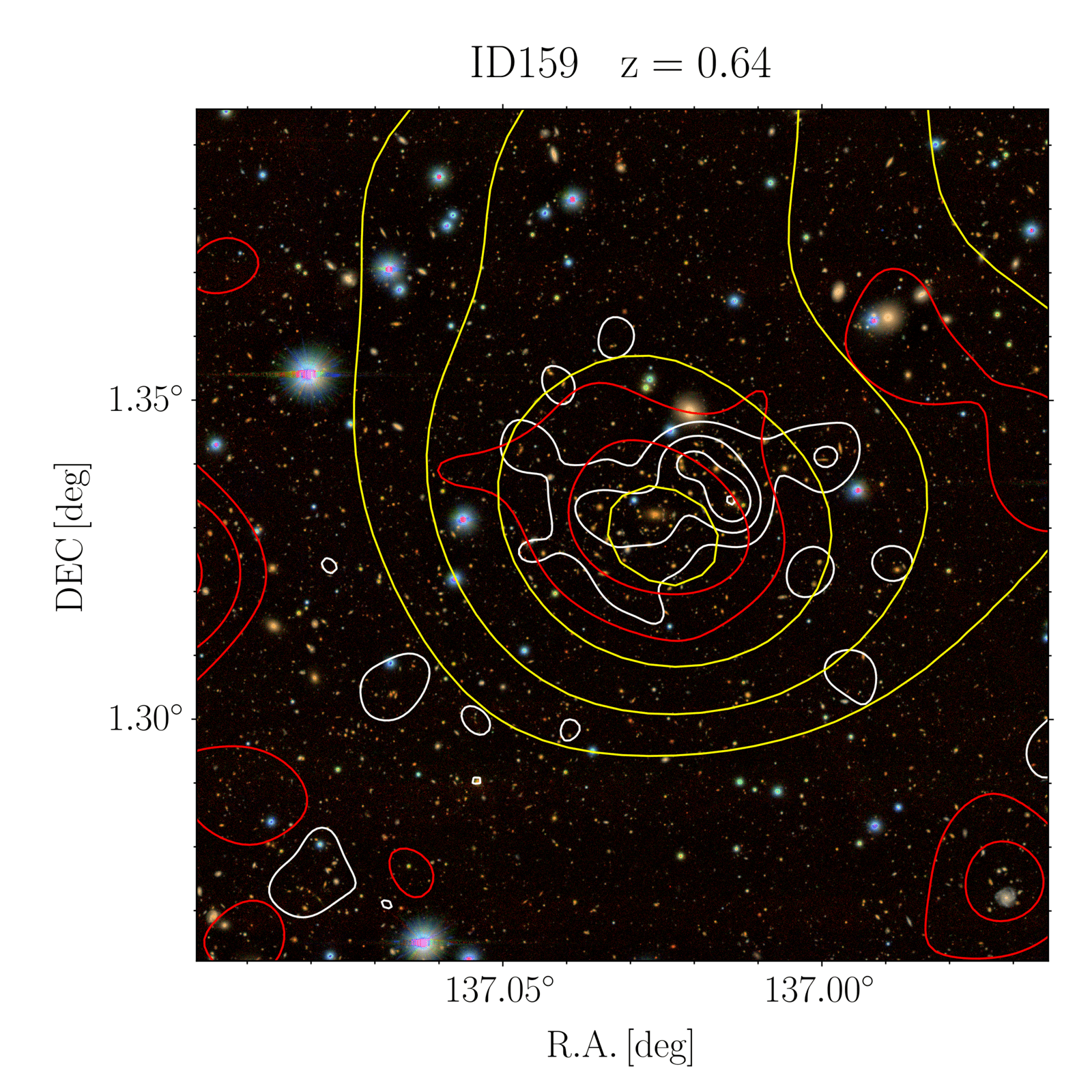}}\\[-5ex]
     \caption{Continued.}
 \end{figure*}
               
 \begin{figure*}[t]    
     \centering
     \includegraphics[width=0.24\columnwidth]{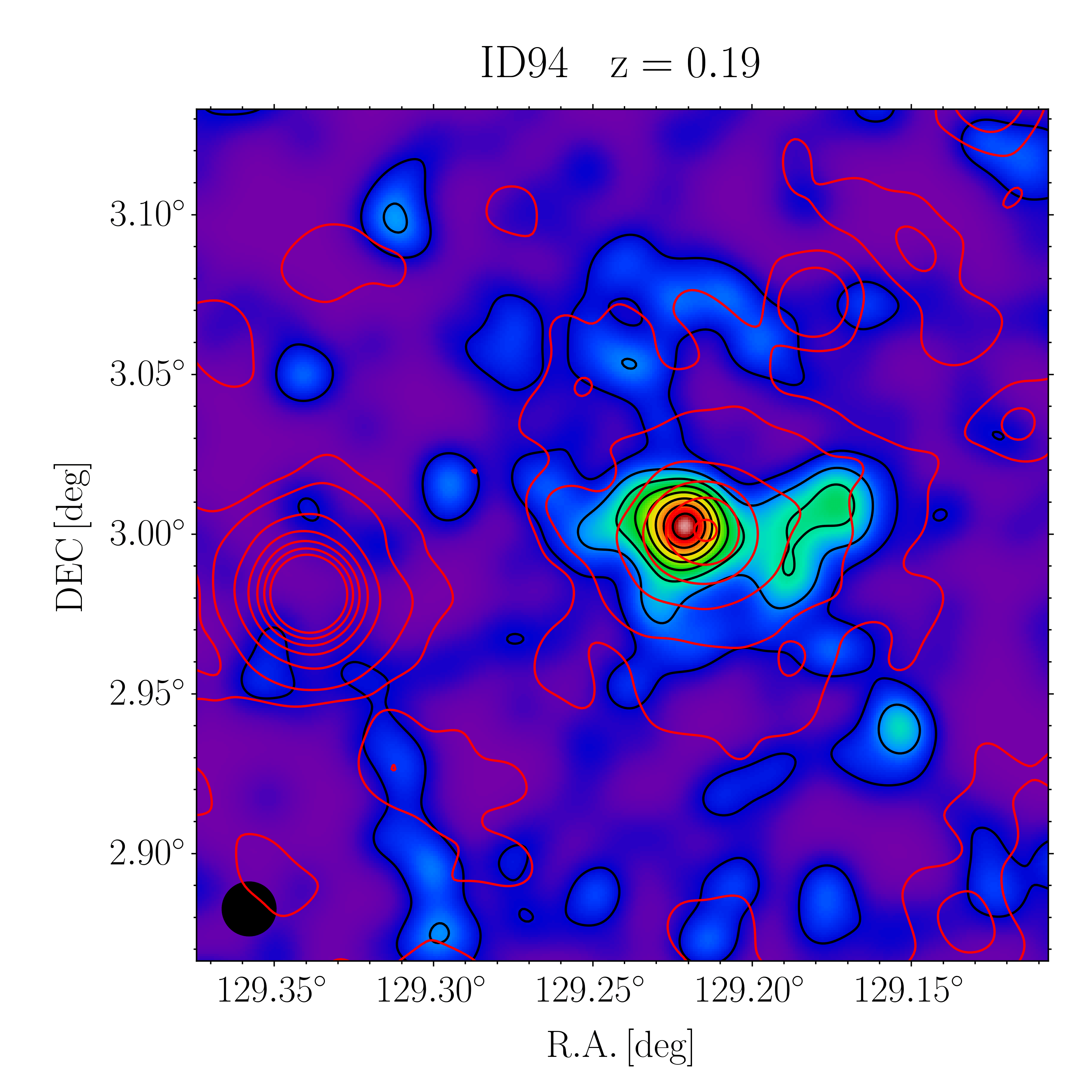}
     \includegraphics[width=0.24\columnwidth]{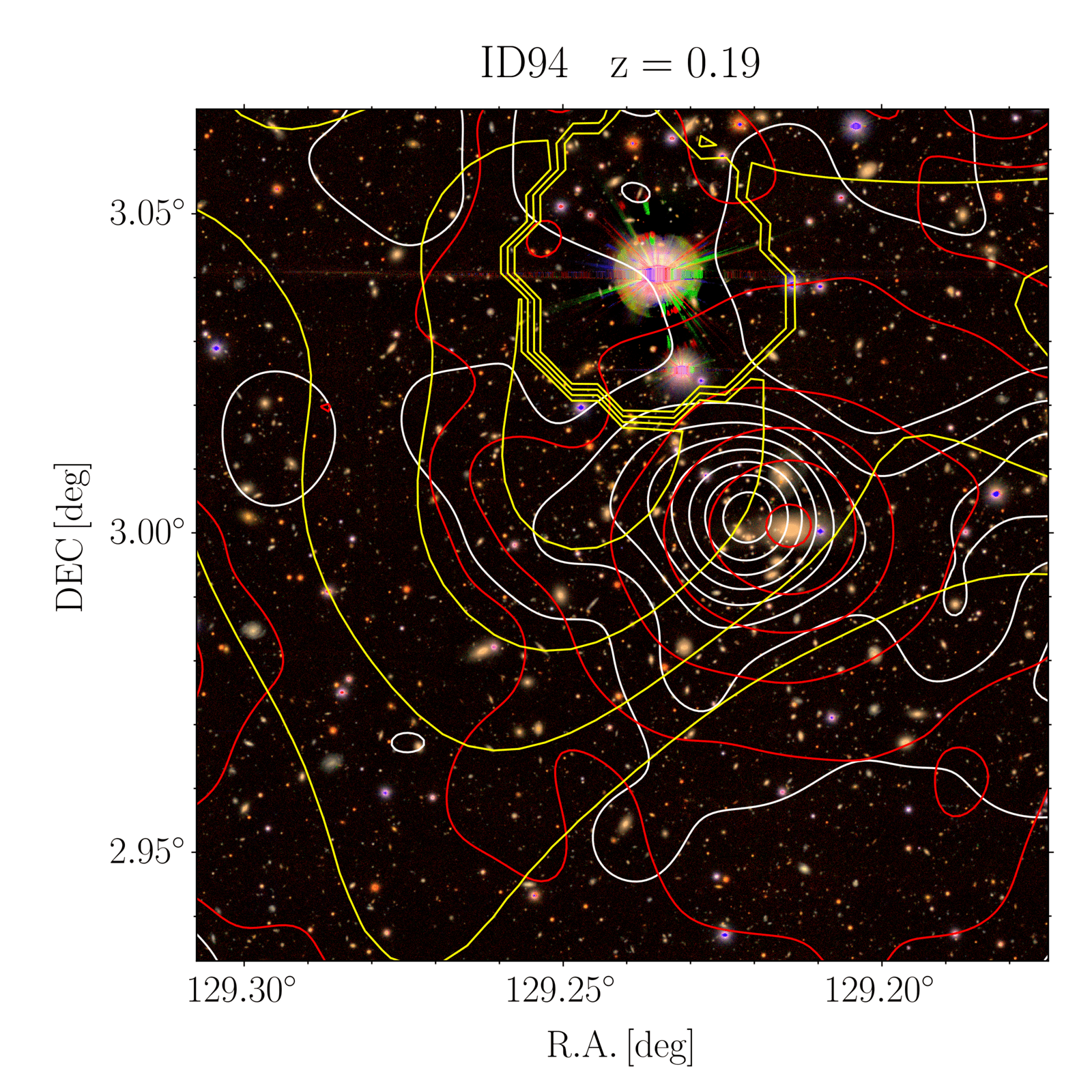}
     \hspace{0.4cm}               
     \includegraphics[width=0.24\columnwidth]{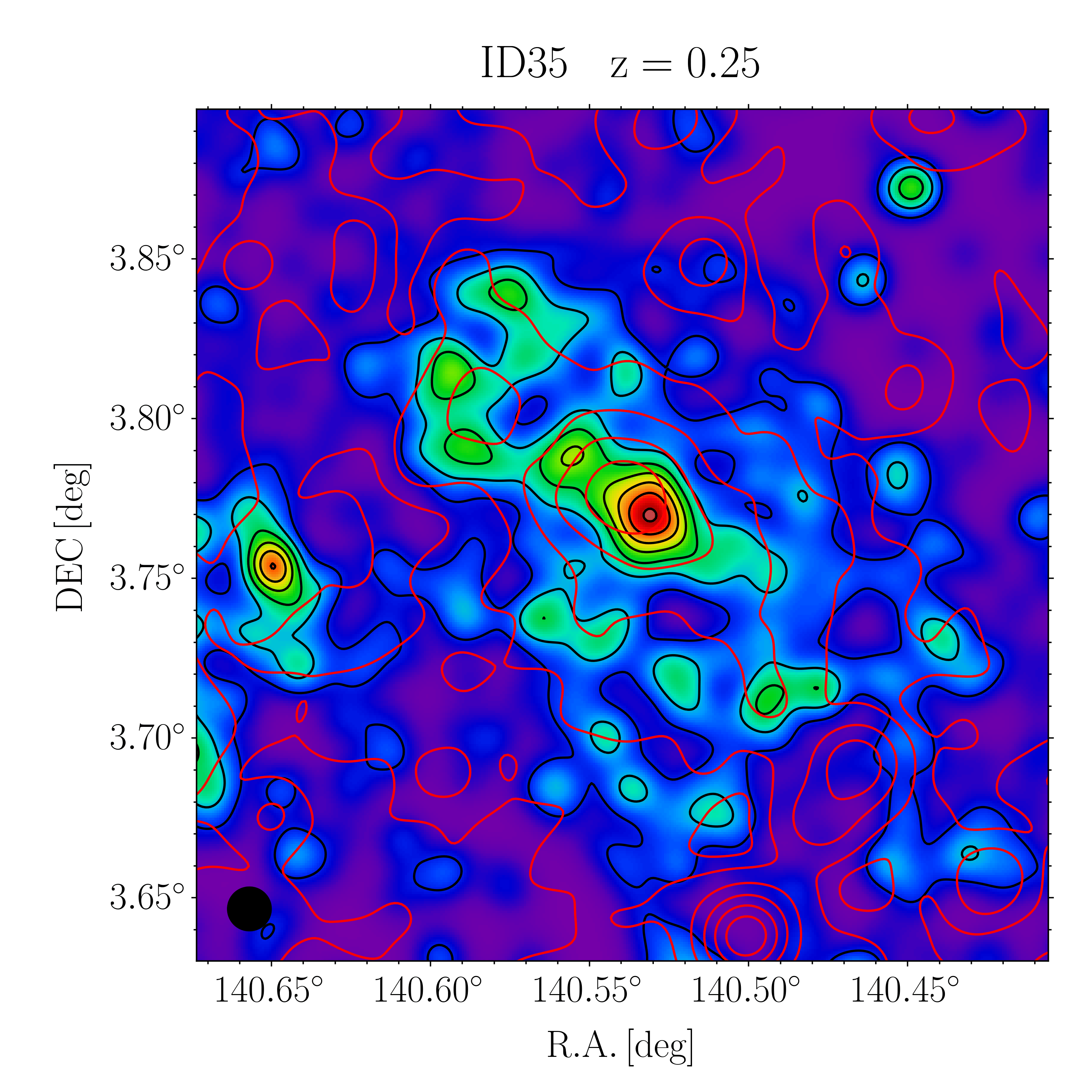}
     \includegraphics[width=0.24\columnwidth]{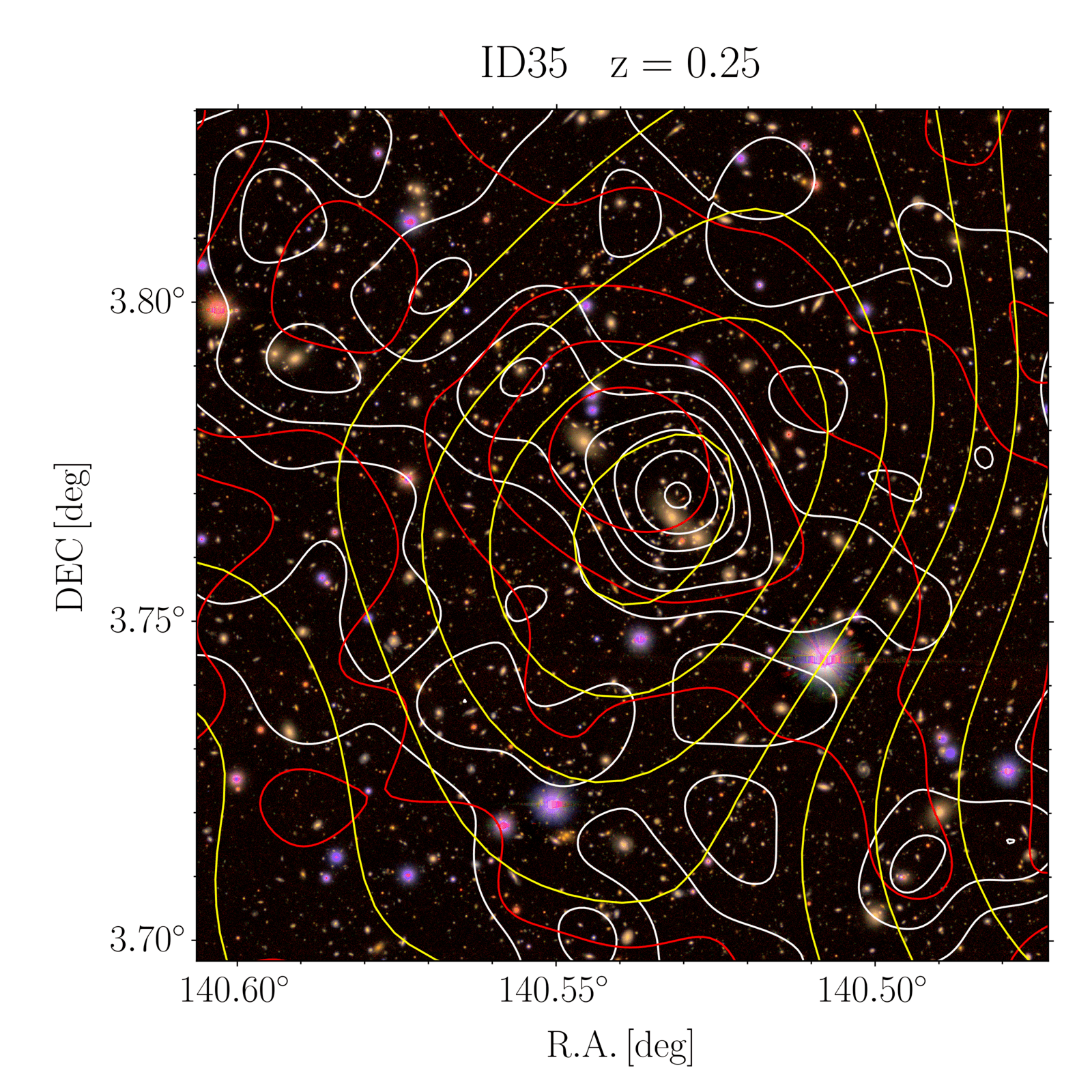}
               
     \includegraphics[width=0.24\columnwidth]{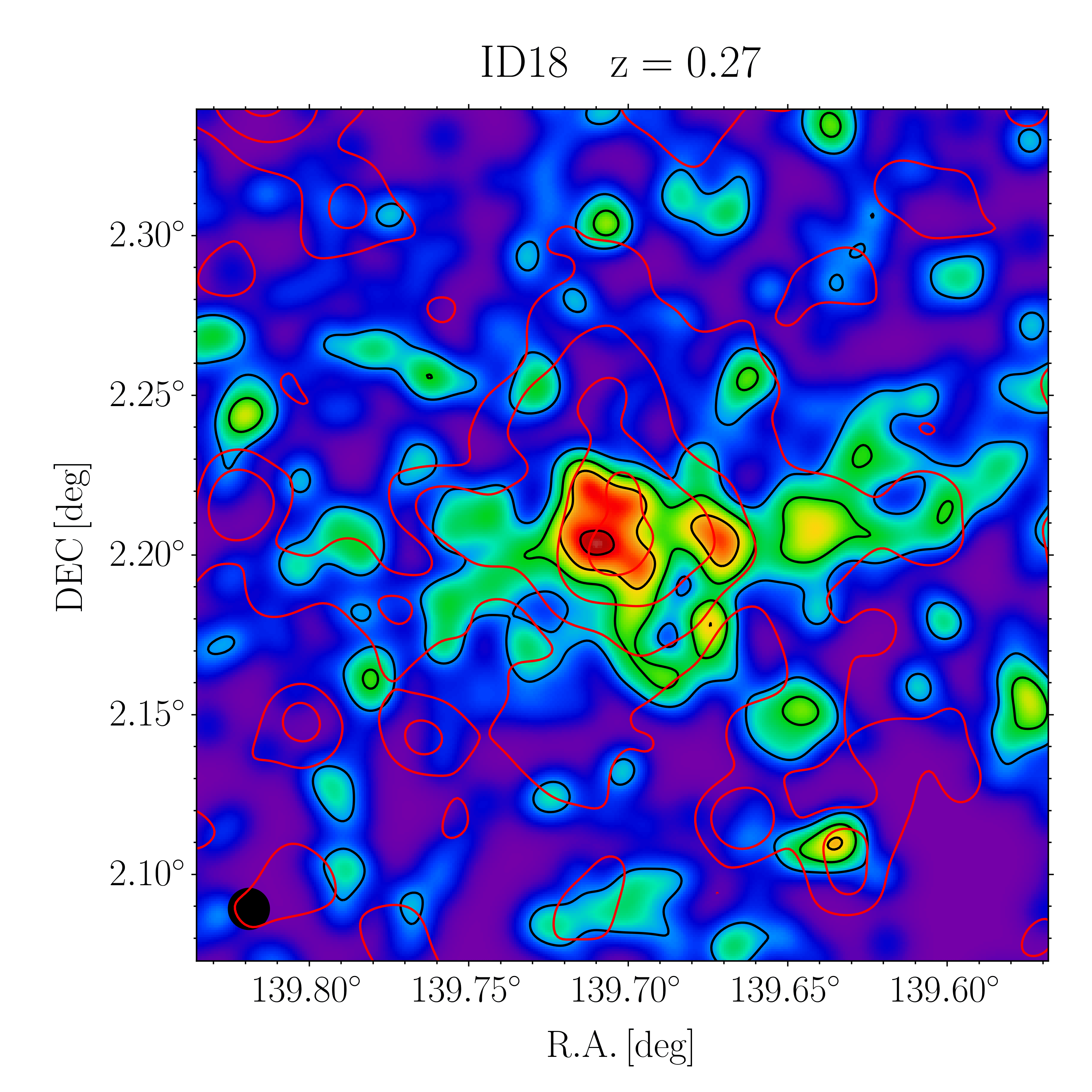}
     \includegraphics[width=0.24\columnwidth]{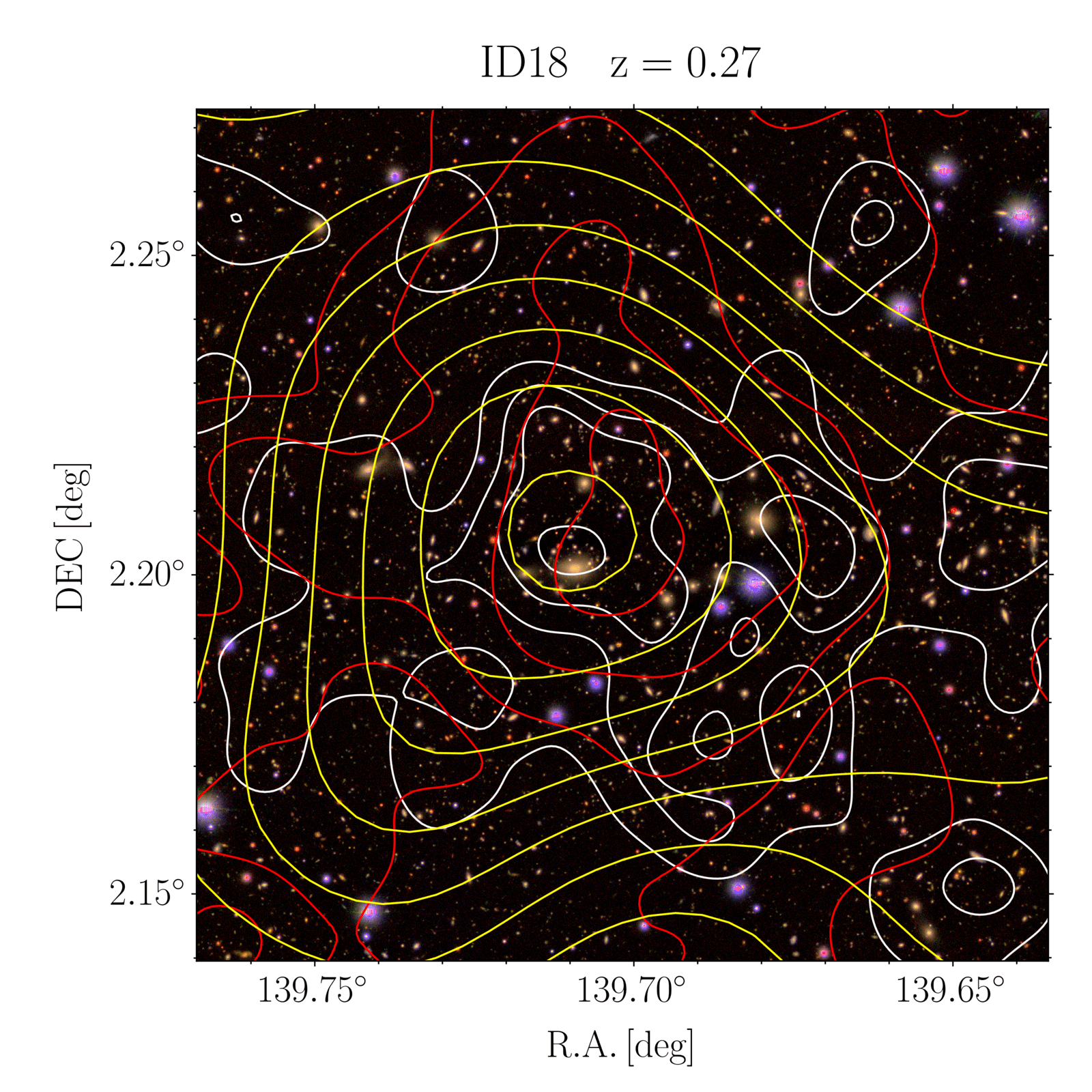}
     \hspace{0.4cm}                    
     \includegraphics[width=0.24\columnwidth]{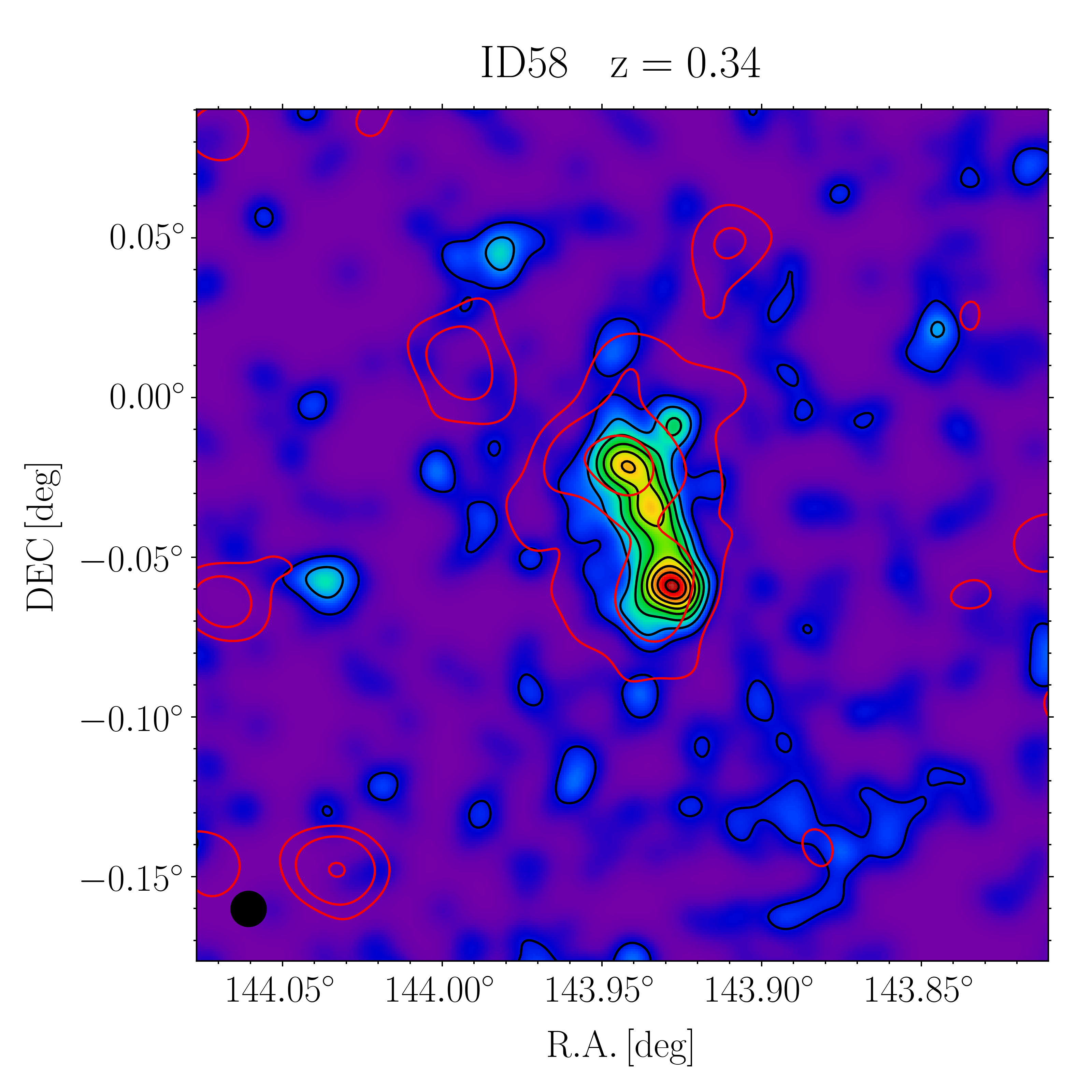}
     \includegraphics[width=0.24\columnwidth]{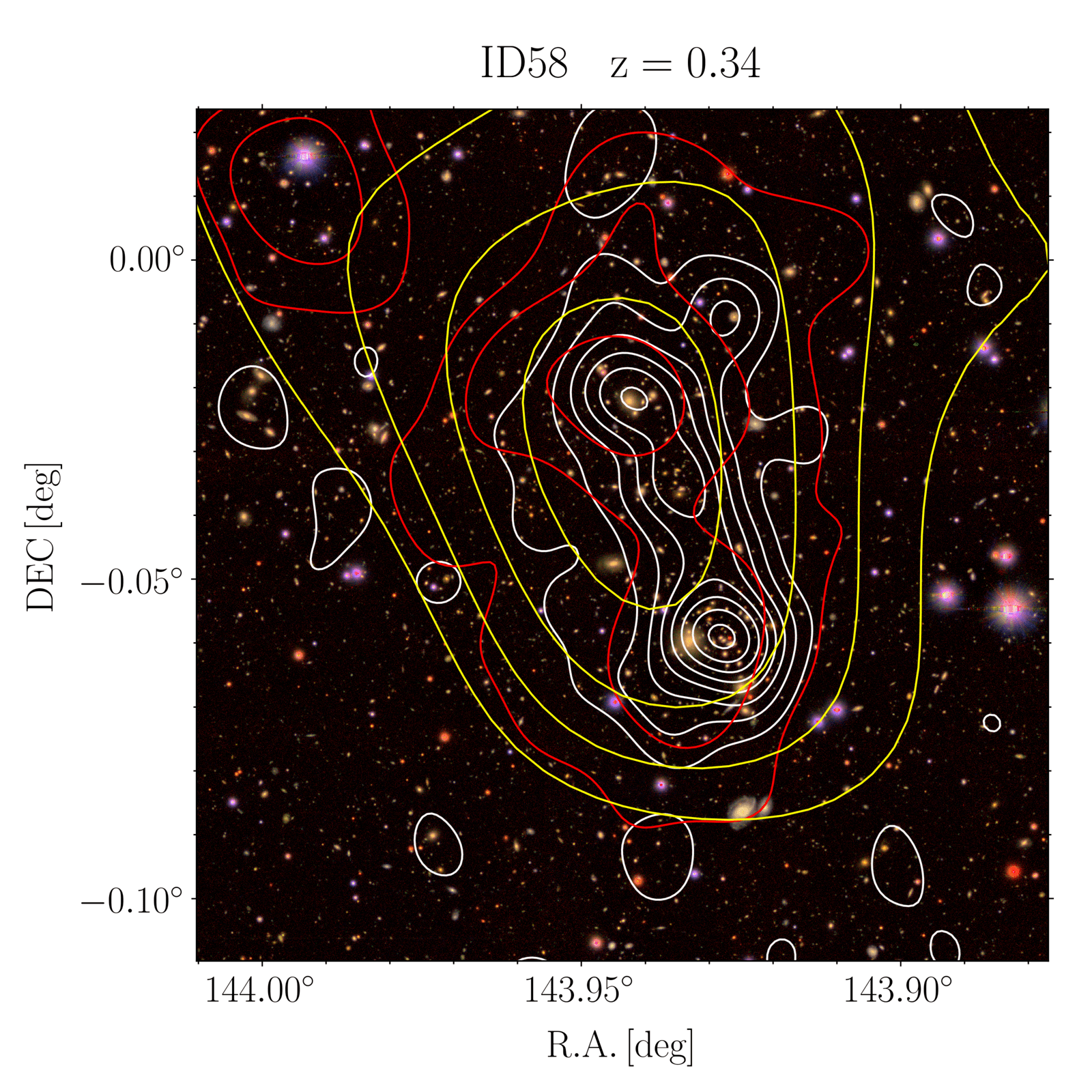}
                    
     \includegraphics[width=0.24\columnwidth]{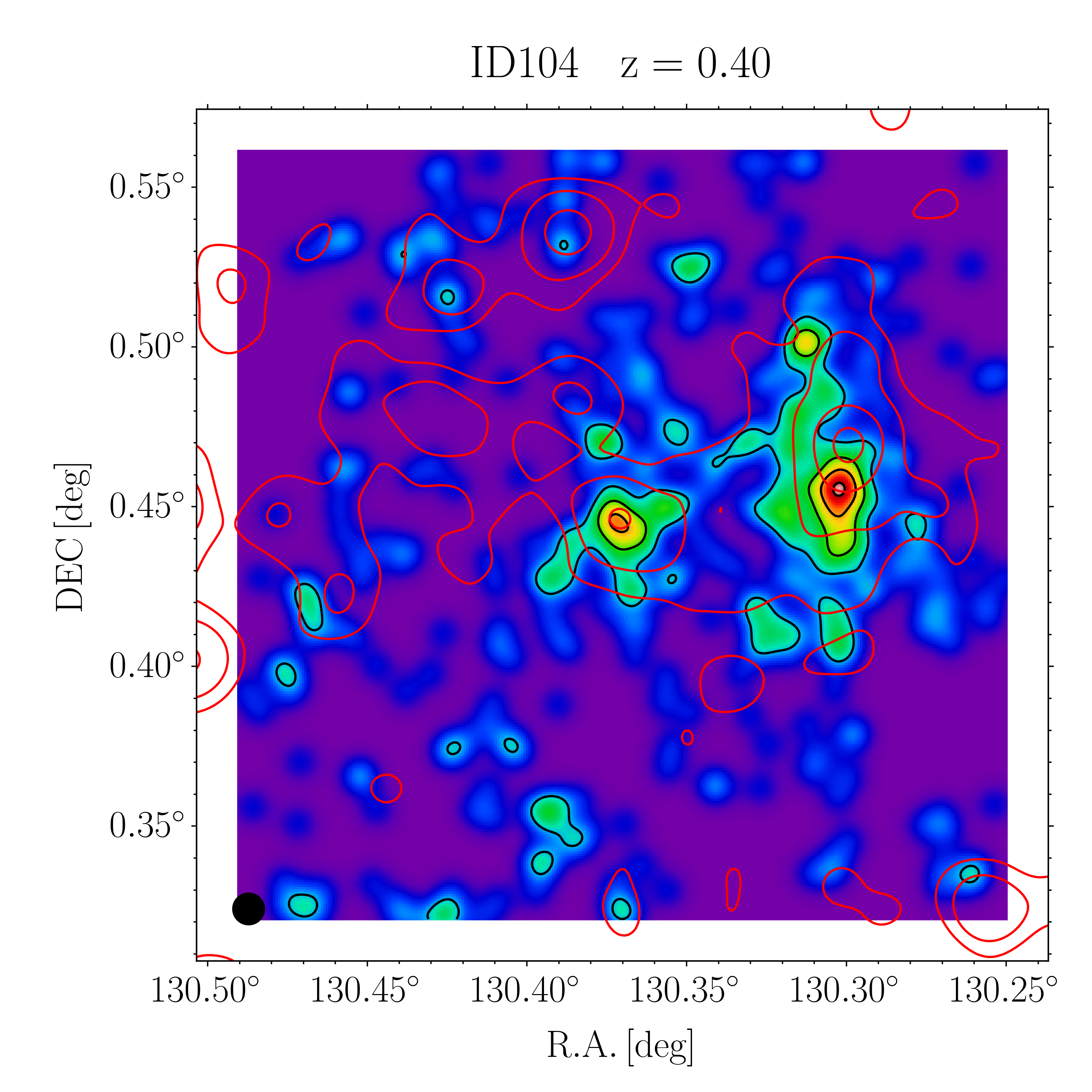}
     \includegraphics[width=0.24\columnwidth]{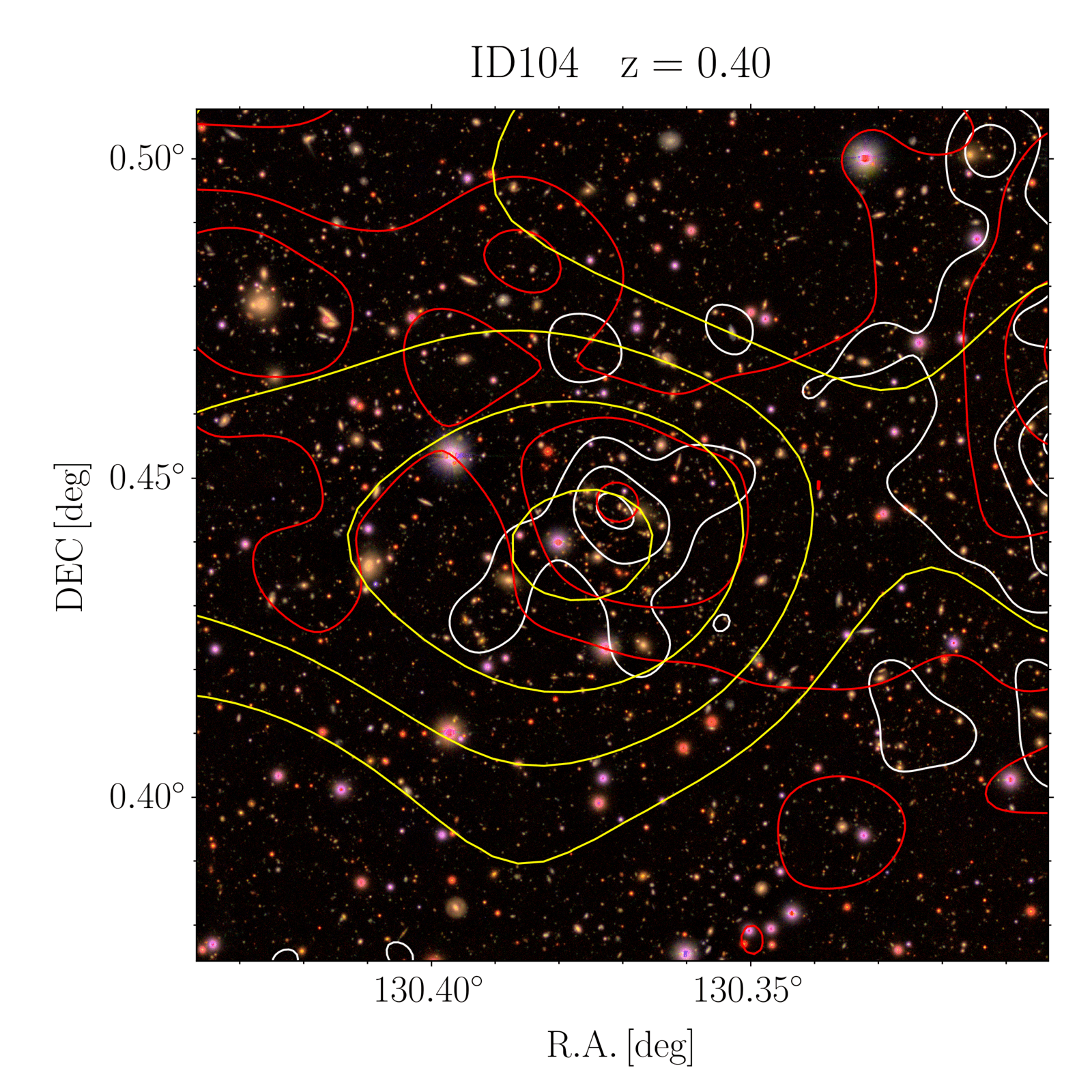}\\  
     \caption{{\it Left}: Galaxy density maps ($16\times16$ arcmin) centred on the HSC shear-selected cluster positions with primary eFEDS match. Overlaid in red are X-ray contours, which were obtained by smoothing the raw X-ray image in the $0.5-2.0$~keV energy band with a Gaussian of $24$~arcsec. The black circles in the lower-left corners show the smoothing scale, FWHM\ =\ $200$~kpc. {\it Right}: HSC-SSP optical images centred on the HSC shear-selected cluster positions. The $8\times8$ arcmin optical images of the central region are created using the $z$, $i$, and $r$ bands. X-ray contours are shown in red, galaxy density contours in white (they are the same contours as the black ones in the corresponding galaxy density maps), and weak-lensing mass contours in yellow.}
     \label{fig:primarymatchesxrayoptical}
 \end{figure*}

\clearpage
\section{X-ray morphological parameters distribution}
\label{app:C}

\begin{figure*}[ht]
    \centering
    \includegraphics[width=0.95\textwidth, trim=80 80 80 120,clip]{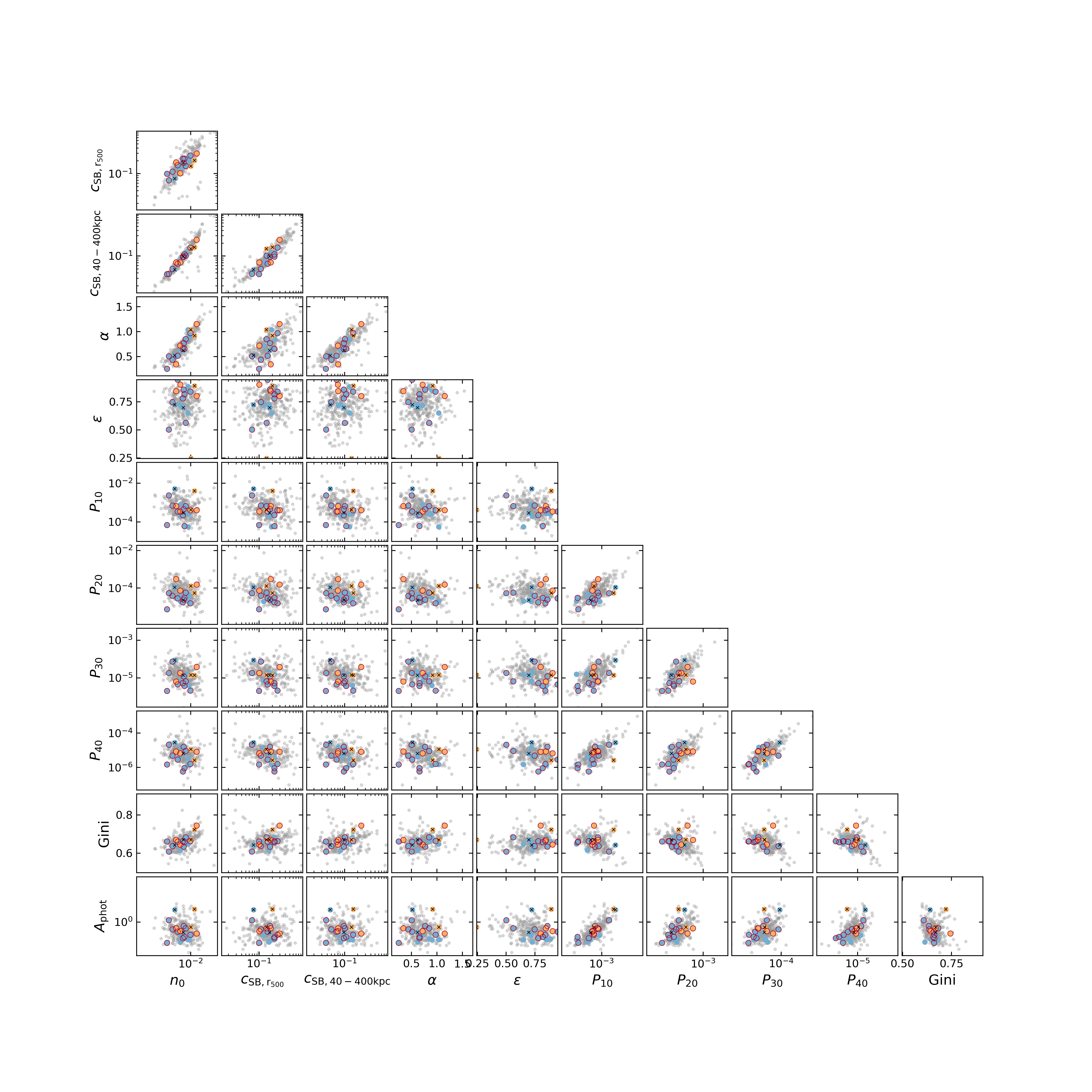}
    \caption{$L_{\rm X}$ and $z$ corrected X-ray morphological parameters obtained within $r_{500}$ in the parameter-parameter planes. Grey points represent the eFEDS sample discussed in \cite{Ghirardini2021b}. Blue-filled circles show the shear-selected clusters with a unique eFEDS match; filled-orange circles display weak-lensing peaks with a primary eFEDS counterpart. Points with a red ring around have multiple-peaks, which were classified as such using the peak-finding method described in section~\ref{sect:peakfinding}. Shear-selected clusters that fall out from the \cite{Ghirardini2021b} selection are marked with a black cross. Errors are omitted for clarity purposes. Please refer to \cite{Ghirardini2021b} for the error visualisation.}
    \label{fig:morphoparam}
\end{figure*}

The results of the X-ray morphological analysis by \cite{Ghirardini2021b} is summarised in Fig.~\ref{fig:morphoparam}, which shows the distribution of the different parameters presented in section~\ref{sect:morphoparam} of $325$ eFEDS clusters (grey points). Shear-selected clusters with an eFEDS counterpart are highlighted by blue-filled circles, clusters with a unique eFEDS match, and filled-orange circles, peaks with a primary eFEDS counterpart. 

\end{document}